\documentclass[twocolumn,trackchanges,twocolappendix]{aastex701}

\usepackage{amsmath,amssymb}
\usepackage{xspace}
\usepackage{enumitem}

\makeatletter 
  \patchcmd{\NAT@citex}
    {\@citea\NAT@hyper@{%
      \NAT@nmfmt{\NAT@nm}%
      \hyper@natlinkbreak{\NAT@aysep\NAT@spacechar}{\@citeb\@extra@b@citeb}%
      \NAT@date}}
    {\@citea\NAT@nmfmt{\NAT@nm}%
    \NAT@aysep\NAT@spacechar\NAT@hyper@{\NAT@date}}{}{}

  \patchcmd{\NAT@citex}
    {\@citea\NAT@hyper@{%
      \NAT@nmfmt{\NAT@nm}%
      \hyper@natlinkbreak{\NAT@spacechar\NAT@@open\if*#1*\else#1\NAT@spacechar\fi}%
        {\@citeb\@extra@b@citeb}%
      \NAT@date}}
    {\@citea\NAT@nmfmt{\NAT@nm}%
    \NAT@spacechar\NAT@@open\if*#1*\else#1\NAT@spacechar\fi\NAT@hyper@{\NAT@date}}
    {}{}
\makeatother

\DeclareUnicodeCharacter{03B1}{\ensuremath{\alpha}}
\newcommand{\kms}{\,\mathrm{km} \, \mathrm{s}^{-1}}

\newcommand{\pcm}{\, \mathrm{cm}^{-3}}
\newcommand{\Msolpc}{\, \mathrm{M}_{\sun} \, \mathrm{pc}^{-2}}
\newcommand{\ergpcm}{\, \mathrm{erg} \, \mathrm{cm}^{-3}}
\newcommand{\ar}{a_{\mathrm{R}}}

\newcommand{\kappaP}{\kappa_{\mathrm{P}}}

\newcommand{\Tg}{T_{\mathrm{gas}}}
\newcommand{\Td}{T_{\mathrm{dust}}}

\newcommand{\Msun}{\mathrm{M}_{\sun}}

\newcommand{\HII}{\ensuremath{\mathrm{H\,II}}}
\newcommand{\CII}{\ensuremath{\mathrm{C\,II}}}
\newcommand{\CI}{\ensuremath{\mathrm{C\,I}}}
\newcommand{\OI}{\ensuremath{\mathrm{O\,I}}}
\newcommand{\OII}{\ensuremath{\mathrm{O\,II}}}
\newcommand{\HI}{\ensuremath{\mathrm{H\,I}}}
\newcommand{\Htwo}{\ensuremath{\mathrm{H_2}}}
\newcommand{\Hplus}{\ensuremath{\mathrm{H^+}}}
\def\H{\rm H}
\def\elec{\rm e^{-}}

\def\Zsun{\rm Z_{\sun}}

\usepackage{chemformula}
\setchemformula{
    arrow-offset = 0em,
    arrow-min-length = 0pt
}
\newcommand{\dbDefault}{}

\begin{document}


\title{Non-equilibrium chemistry and radiation hydrodynamics in metal-poor and metal-rich interstellar gas}

\author[0000-0002-0311-2206]{Shyam H. Menon}
\email{smenon@flatironinstitute.org}
\affiliation{Center for Computational Astrophysics, Flatiron Institute, 162 5th Avenue, New York, NY 10010, USA}
\affiliation{Department of Physics and Astronomy, Rutgers University, 136 Frelinghuysen Road, Piscataway, NJ 08854, USA}

\author[0000-0003-3347-7094]{Piyush Sharda}
\email{sharda@strw.leidenuniv.nl}
\affiliation{Leiden Observatory, Leiden University, P.O. Box 9513, 2300 RA Leiden, The Netherlands}
\affiliation{Center for Computational Astrophysics, Flatiron Institute, 162 5th Avenue, New York, NY 10010, USA}

\author[0000-0001-5817-5944]{Blakesley Burkhart}
\email{b.burkhart@physics.rutgers.edu}
\affiliation{Department of Physics and Astronomy, Rutgers University, 136 Frelinghuysen Road, Piscataway, NJ 08854, USA}
\affiliation{Center for Computational Astrophysics, Flatiron Institute, 162 5th Avenue, New York, NY 10010, USA}

\author[0000-0001-5817-5944]{Tommaso Grassi}
\email{tgrassi@mpe.mpg.de}
\affiliation{Max-Planck-Institut für Extraterrestrische Physik, Giessenbachstraße 1, 85748 Garching, Germany}
\affiliation{Exzellenzcluster Origins, Boltzmannstr. 2, 85748 Garching, Germany}

\begin{abstract}

    We present a flexible implementation of non-equilibrium chemical networks with the \texttt{KROME} framework, coupled to a nonlocal variable Eddington tensor-closed radiation transport scheme (\texttt{VETTAM}). Our fiducial network is designed to explicitly capture the formation pathways of important molecular species such as CO, OH, and $\rm H_2O$, intermediaries like $\rm H_3^+$, and the destruction of these species through photon and cosmic-ray dissociation. We validate our scheme through benchmark tests, and demonstrate its success in 1D and 3D problems covering a broad range of metallicities ($0 \leq Z \lesssim \Zsun$) and densities ($n \lesssim 10^{17} \pcm$). We confirm earlier works that suggest that OH and $\rm H_2O$ are the key molecular species in dense gas ($n \gtrsim 10^{5} \pcm$) for $Z\lesssim 0.01 \Zsun$, dominate the cooling at these densities, and demonstrate that our scheme can reproduce their abundances with similar accuracy as dedicated static photo-dissociation region (PDR) codes. We also find that the dominant fine-structure coolant in dense ($n \gtrsim 10^3 \, \pcm$) PDRs transitions from $\CI$ to $\CII$ to $\OI$ at increasingly lower $Z$. Finally, we find that in turbulent media, rapid mixing at PDR interfaces renders the chemical species out of equilibrium, and show that this could potentially alleviate the long-standing problem of (static) PDR models producing excess $\CI$ in the diffuse ISM over UV absorption-line measurements. We conclude by presenting and comparing with alternative implementations of reduced but faster chemical networks in our scheme, discussing the problem-dependent performance-accuracy tradeoffs and future prospects.  

\end{abstract}


\section{Introduction} 

The neutral gas phases of the Interstellar Medium (ISM) contain several atoms and molecules that govern its thermal and dynamical state, and through a myriad of emission/absorption line mechanisms, and act as observational tracers of the ISM. The thermochemical state of the gas is controlled by the coupled interplay of heating, cooling, and stiff chemical reactions, all of which are also sensitive to the local ultraviolet (UV) radiation and cosmic-ray (CR) fields. This has led to the development of dedicated photoionization/photodissociation region (PDR) models that capture this complexity \citep{Tielens_Hollenbach_1985,Black_vanDishoeck_1987,vanDishoeck_Black_1988,Sternberg_Dalgarno_1989,Sternberg_Dalgarno_1995,Lee_1996,Storzer_1996,Ferland_1998,  Kamp_Bertoldi_2000,Kamp_vanZadelhoff_2001,Bensch_2003,Bell_2005,Meijerink_Spaans_2005,LePetit_2006,2014MNRAS.440.3349R,2014MNRAS.442.2780R}. However, these models are too computationally expensive to couple to hydrodynamic codes, and therefore their usage fails to capture the complex physical structure/geometry of gas and any (potential) time-dependent effects that turbulent gas dynamics in the ISM impose on the PDR structure. 

These considerations have led to several efforts to model thermochemistry coupled to hydrodynamic codes, with differing levels of accuracy and completeness depending on the problems they are designed to address. The simplest approaches only consider hydrogen species, capturing the formation of $\Htwo$ molecules mediated by dust grains, the photodissociation/ionization of $\HI$/$\Htwo$ \citep{Glover_MacLow_2007,kim2023_tigress,2025MNRAS.543..891P}, and for codes developed for  primordial/low-metallicity ($Z$) applications, the gas-phase formation pathways and intermediary primordial species involved in them \citep[e.g.,][]{Glover_Abel_2008,smith_grackle_2017}. In these methods, the chemistry of metal species is considered to be in equilibrium, and associated cooling rates are calculated from the equilibrium abundances and/or pre-tabulated with codes like \texttt{CLOUDY} \citep{Gunasekera_2025_cloudy}. While simple, these approaches are able to capture the thermal evolution of the ISM at $\sim \rm \Zsun$ metallicity and the $\HI$-to-$\Htwo$ transition reasonably, but cannot accurately represent the time-evolving nature of chemical species like $\CII$, $\CI$, $\OI$ and CO -- which act as important coolants of diffuse/dense neutral gas and as critical tracers of the atomic/molecular ISM.

On this front, several chemical networks have been designed to capture the critical reaction pathways for the formation and destruction of CO -- motivated by its high abundance and capability to trace $\Htwo$ \citep[][]{1999RvMP...71..173H,2013ARA&A..51..207B} -- ignoring species not involved in these pathways. For instance, \citet{Nelson_1997} consider CO formation through intermediate hydrocarbon radicals (e.g., CH, $\rm CH_2$) and ignoring other important intermediaries (e.g., $\CI$) and hydroxyl-mediated (e.g., OH, $\H_2O$) CO formation pathways; however, these simplifications make it computationally efficient, facilitating its extensive applications in numerical simulations of ISM patch and isolated galaxy simulations \citep{Smith_2014_CO,Walch_2015,Hu_2016,Safranek-Shrader_2017,Seifried_2017,Jeffreson_2020,Smith_2020_cloudfactory,2021ApJ...920...44H}. The tradeoff is that they tend to be less accurate, for instance by overestimating CO \citep{glover_approximations_2012} and $\Htwo$ abundances \citep{Ploeckinger_2025} compared to extended networks that include additional CO formation and destruction pathways. Networks considering $\CI$ and OH-mediated CO formation pathways have also been developed \citep{Nelson_1999,Gong_2017}, with some recent numerical simulations adopting them \citep[e.g.,][]{Hunter_2023,Whitworth_2023,Ho_2026}. While these capture CO formation to reasonable levels of accuracy, they cannot explicitly evolve OH and $\rm H_2O$ abundances, which can be important molecular tracers and coolants, particularly at low metallicities ($\lesssim 0.01 \rm \Zsun$) where they replace CO as the prominent molecular tracer in dense gas \citep{Glover_2010,Bialy_2015} and dominate its cooling to prestellar densities \citep{Omukai_2005,Omukai_2010}. 

On the other hand, the dawn of astrochemistry and discovery of complex organic molecules has also ushered in an era of exhaustive chemical networks (often consisting of 1000s of gas phase chemical reactions as well as two- and three-phase gas-grain reactions -- \citealt{2004MNRAS.354.1141V,umist12,2015ascl.soft07010M,2017AJ....154...38H,umist24}). These networks can reliably predict the abundances of several other molecules that have been routinely detected in the ISM, but they often assume a constant gas/dust temperature, cannot do thermochemistry, and remain infeasible to be embedded into hydrodynamical simulations (see, however, \citealt{2021A&A...653A..76H,2022A&A...668A.139G,2023MNRAS.518.5718B,2025MLS&T...6b5069V,2025arXiv251110558V} for recent progress in this area with neutral networks, emulators, and auto-differentiable solvers).

It is of equal importance to evaluate the approaches with which the (UV) radiation field is solved for and coupled to the thermochemistry in the implementations above. This is because UV photon-facilitated photoionization and dissociation are often the dominant destruction mechanisms for atoms/molecules, and also act as the primary heating mechanism in the diffuse ISM and PDRs through the photoelectric effect and direct photoionization. To include radiation, most implementations use approximate six-ray or tree-based methods \citep{Clark_2012,wunsch_tree-based_2018} to attenuate the FUV radiation field \citep[e.g.,][]{Smith_2014_CO,Walch_2015,Hu_2016}. While they are suitable for considering background radiation, they suffer from inaccurate solutions for point sources, particularly at larger distances, and ignore the time-dependence of the radiation field -- important, for instance to the transport of dust-reprocessed IR photons that facilitate cooling in dense dusty media. Other implementations use time-dependent moment-based methods, however, using a purely local Moment-1 ($M_1$) \citep{Minerbo_1978, Levermore_1984}, which is known to produce unphysical discontinuities and spurious radiation fluxes when the radiation from multiple sources interacts \citep[see, for e.g.,][Figure 13]{Jiang_2012,Menon_2022_VETTAM} with dynamical consequences for launching radiation pressure-driven winds \citep{Smith_2020}. More recently, photochemistry implementations coupled to accurate long-characteristics ray-tracing approaches have been developed \citep{Baczynski_2015,kim2023_tigress}, although their chemical networks only consider non-equilibrium $\HI$-$\Htwo$ and highly simplified, steady state treatments for CO. Moreover, the radiation transport (RT) in these schemes can become prohibitively expensive in the presence of a large number of radiating sources (e.g., a star cluster) as the cost scales linearly with the number of sources. 

In this paper, we present an implementation that addresses some of these limitations. We compile a detailed, yet computationally tractable, chemical network that we show can accurately produce abundances of molecules like CO, OH, $\rm H_2O$, several important intermediaries in their formation pathways that act as tracers of ISM physics and chemistry ($\rm H_3^+$, $\rm CH$), and the thermal evolution of gas across a broad range of densities ($0.1 \lesssim n \lesssim 10^{\rm 15}\, \rm cm^{-3}$) and metallicities ($0 \leq Z \lesssim \Zsun$). We couple the thermochemistry to \texttt{VETTAM}\footnote{\url{https://github.com/shm-1996/VETTAM}} \citep{Menon_2022_VETTAM}, a variable Eddington tensor (VET) closed-moment-based RT method that combines the accuracy of ray-tracing approaches with retaining the time-dependence of the RT and a cost that scales independently of the number of sources  \citep{Buntemeyer_2016}. In addition to our fiducial network, we also couple the RT to simpler but faster networks from the literature \citep{Gong_2017,Gong_2023,kim2023_tigress}, driven by the philosophy that the \textit{cost-accuracy trade-off} can be different depending on the scientific question. We implement and evolve the thermochemistry networks within the flexible \texttt{KROME} framework \citep{grassi_krome_2014}, and couple our implementation to the \texttt{FLASH} radiation-magnetohydrodynamics (MHD) code \citep{Fryxell_2000,Dubey_2008}.

In Section~\ref{sec:methods}, we describe our networks and their coupling to \texttt{VETTAM}. In Section~\ref{sec:OneZoneTests}, we present one-zone demonstrations of the working/accuracy of our thermochemistry, focusing on the improvements over existing implementations. In Section~\ref{sec:3DDemonstrations}, we present 3D radiation hydrodynamic simulations of turbulent PDRs with background/point UV sources as a proof-of-concept of the working of our implementation. In Section~\ref{sec:Discussion}, we compare the different network implementations, discuss the cost-accuracy tradeoffs, and conclude with a summary and outlook in Section~\ref{sec:Summary}.

\section{Methods}
\label{sec:methods}

Throughout this work we adopt the protosolar abundances \citep{Asplund_2009} for elements in the gas-phase at Solar metallicity ($\Zsun$), scaled appropriately for lower metallicities. We also assume that the dust-to-gas ratio scales linearly with the metallicity, adopting a dust-to-gas mass ratio of the Solar neighborhood \citep[0.0081][]{Weingartner_2001b} as the reference value at $\Zsun$.

\subsection{Chemical network}
\label{s:chemicalnetwork}
Our aim is to design a chemical network that can be reliably used across primordial to Solar metallicities, and remains computationally tractable to be embedded into hydrodynamical simulations involving a wide range of gas densities. We present our fiducial network that satisfies these conditions below in Section \ref{s:fiducial_network}, supplemented by alternate chemical networks that have been widely used in the literature in Section \ref{sec:AlternativeNetworks}.

\subsubsection{Fiducial network}
\label{s:fiducial_network}
We construct our fiducial chemical network in a hierarchical fashion, starting from the \textit{reduced} network of \citet[][]{Omukai_2005} and augmenting it with the primordial (including deuterium) chemistry used in the \texttt{POPSICLE} radiation-MHD simulations \citep[][]{2025MNRAS.540.1745S,2025MNRAS.541L...1S} that was adopted from \citet[][]{Galli_1998,Omukai_2000,grassi_krome_2014}. We then progressively add additional species and reactions (with the most up-to-date reaction rates), retaining only those that are found to play an important role in the thermochemical evolution at any metallicity. In particular, we add important chemical reactions from \citet[][]{Glover_Abel_2008,Glover_2010} that become relevant at low metallicities but are not considered by \citet[][]{Omukai_2005}, as well as high-metallicity reactions from \citet[][]{Gong_2017} and high-temperature reactions from \citet[][]{Hunter_2023}. We adopt the dust temperature-mediated H$_2$ formation rate on dust grains from \citet{Hollenbach_1979}. To assess the importance of each reaction and chemical species, we perform a suite of validation tests (described in the Appendix) spanning equilibrium, non-equilibrium, and time-dependent conditions. We find that the significance of individual species and reactions often depends on the specific test considered, with some becoming important only in equilibrium calculations and others only in non-equilibrium or time-dependent evolution. Further, we include some chemical species which are not relevant for thermochemistry, but their presence ensures that appropriate loss route exists for important species at all metallicities (0 to $\rm{Z_{\odot}}$) and densities ($0.01 - 10^6\,\rm{cm^{-3}}$ for ISM tests and $0.01 - 10^{17}\,\rm{cm^{-3}}$ for cloud collapse tests). Our results demonstrate that constructing a robust and accurate chemical network requires validation across all three regimes.

Our fiducial network consists of 37 chemical species and 206 chemical reactions. The chemical species we consider are: e$^-$, H$^-$, H, H$^+$, H$_2$, H$_2^+$, H$_3^+$, HD, D, D$^-$, D$^+$, HD$^+$, He, He$^+$, He$^{++}$, C$^+$, C, CH, CO, CO$^+$, CH$_2$, CH$_3$, CH$_4$, CH$_2^+$, CH$^+$, CH$_3^+$, CO$_2$, HCO$^+$, O$^+$, O, OH, H$_2$O, O$_2$, OH$^+$, H$_2$O$^+$, H$_3$O$^+$, O$_2^+$. We include 30 photochemical reactions and 22 CR-induced reactions within the network (see Appendix~\ref{sec:appendix_chemistry} for details).

\subsubsection{Alternative chemical networks}
\label{sec:AlternativeNetworks}
In addition to the fiducial chemical network above, we also implement two reduced networks from the literature, designed to capture the essential aspects of thermochemistry at progressively lower computational cost. This was done with the philosophy of the implementation in mind: i.e., a common flexible framework for thermochemistry which leaves cost/accuracy trade-off choices to the user, depending on the problem and the quantities they aim to capture. 

The first network is described in \citet{Gong_2017}, which includes 18 independent species, including the use of pseudo-species that combine the abundances of several species involved in CO formation pathways; for instance, hydrocarbon species such as $\rm CH$, $\rm CH_2$, $\rm CH^+$, $\rm CH_2^+$ and $\rm CH_3^+$ are combined into a single (pseudo-)species $\rm CH_{x}$. Similarly, the pseudo-species $\rm OH_{x}$ represents the combination of the hydroxyl radicals $\rm OH$, $\rm H_2O$, $\rm OH^+$, $\rm H_2O^+$, and $\rm H_3O^+$. In addition to the reactions described in \citet{Gong_2017}, we include the modifications and additional reactions described in \citet{Hunter_2023}, which are required to prevent unphysical results in hot gas. In Appendix~\ref{sec:GOWComparison}, we verify that our implementation can match the fiducial PDR test outcomes presented in \citet{Gong_2017}, reproducible from their publicly available PDR code\footnote{https://github.com/munan/pdr}. 

Second, we implement the minimal chemical network described in \citet{kim2023_tigress} used in the TIGRESS-NCR simulations, which evolves the dominant H species in non-steady state\footnote{Both reduced networks are designed for higher metallicity environments, and therefore do not include the gas-phase formation pathways of \Htwo.}. For metals, the species $\CI$, $\CII$ and $\OI$, $\OII$ are tracked, with the $\CI$-$\CII$ assumed to be in ionization equilibrium and the $\OI-\OII$ ionization balance to be dictated by charge balance with $\HII$ ions. Finally, the CO abundance is obtained from an analytical fit to PDR models in \citet{Gong_2017} at every step of the thermochemistry updates. 

The coupling to the radiation is identical to that described for our extensive network, with the exception that the \citet{kim2023_tigress} network simply uses the dust-attenuated Lyman-Werner (LW) radiation field in the fitting function for the CO abundance, and does not require accounting for self-shielding and cross-shielding by CO and $\Htwo$ molecules along the line-of-sight. The heating/cooling mechanisms are also almost identical, only omitting $\rm H_2O$ and OH cooling, which are not explicitly tracked by the networks.

\subsection{Thermal processes}

Our implementation considers all important heating/cooling mechanisms relevant to the evolution of gas across a broad density range $0.1 \lesssim n \lesssim 10^{17} \, \rm cm^{-3}$ -- i.e., from the diffuse ISM to the densities achieved in protostellar cores. This necessitates capturing heating from the photoelectric effect, CRs, photoionization, $\Htwo$ Lyman-Werner pumping/photodissociation, and exothermic chemical reactions (most notably $\Htwo$ formation on dust grains), in addition to cooling from a comprehensive set of atomic, molecular and dust-related processes: collisional excitation/ionization and recombination of the primordial species, free-free emission, fine-structure line cooling from $\CI$, $\CII$ and $\OI$, rovibrational/rotational line cooling from $\Htwo$, HD, CO, OH and $\rm H_2O$, Compton cooling, dust-continuum cooling, grain-assisted recombination cooling, and gas-dust collisional coupling. 

Since our chemical network is designed to explicitly resolve the cold/warm neutral phases of the ISM rather than the detailed ionization structure of photoionized gas, we account for nebular cooling from highly ionized metal species not explicitly tracked by our network (i.e. beyond $\CII$, $\OII$ and other elements) using the simplified fitting function presented by \citet{kim2023_tigress}, which is calibrated against a fixed, characteristic ionization-state composition for O, N, and S; this approximation reproduces the broad thermal properties of \HII\ regions but cannot capture the detailed dependence of the nebular cooling rate on the local hardness of the radiation field and non-equilibrium ionization states \citep[see,][for efforts in this direction]{Sarkar_2021,Katz_2022,Chan_2026}. At temperatures $\Tg \gtrsim 2\times10^4 \, \rm K$, where neither our chemical network nor this nebular treatment remain appropriate, we smoothly transition to a tabulated cooling rate assuming collisional ionization equilibrium (CIE). 

We compute the dust temperature self-consistently at every step by balancing dust heating (from absorbed UV/IR/CMB radiation and gas-dust collisions) against dust thermal (IR) emission, accounting for the dust-to-gas ratio scaling with metallicity throughout. The full set of heating and cooling terms, their functional forms, and the associated references are detailed in Appendix~\ref{sec:appendix_thermodynamics}.

\subsection{Radiation Hydrodynamics and coupling to thermochemistry}

We use the \texttt{VETTAM} radiation hydrodynamics (RHD) scheme \citep{Menon_2022_VETTAM} implemented in the hydrodynamics code \texttt{FLASH} \citep{Fryxell_2000,Dubey_2008} to solve for the radiation field and the associated rates for photon-mediated thermochemistry. \texttt{VETTAM} solves the non-relativistic, frequency-integrated, two-moment equations of radiation hydrodynamics in the mixed frame formulation \citep{Mihalas_1982}. It uses a non-local VET closure computed with a hybrid characteristics ray-tracing approach that guarantees accurate solutions in media that contain a mix of optically thick and thin gas/dust, and/or multiple point/diffuse sources of radiation. These are situations where the entirely local $M_1$ closure used in most existing approaches \citep[see][for a review]{Wunsch_2024} tend to yield incorrect solutions. In addition, \texttt{VETTAM} uses an implicit backward-Euler temporal update to treat the hierarchy of timescales in RHD -- distinct from reduced speed of light approaches that face feasibility concerns at high optical depths \citep{Skinner_2013}. For this, \texttt{VETTAM} utilizes the sparse matrix solver capabilities of the open-source \texttt{PETSc} library \citep{PetscConf,PetscRef}. 

\texttt{VETTAM} provides the option of solving for the radiation transfer in up to five characteristic frequency bands, designed to capture the ionization/dissociation of important species tracked by our chemical networks and/or heating mechanisms relevant to the diffuse/dense ISM. The bands we follow are 
\begin{enumerate}
    \item $h\nu \geq 15.42 \mathrm{eV}$: This band can ionize $\Htwo$ and $\H$.
    \item $13.6 \mathrm{eV} \leq h\nu < 15.42 \mathrm{eV}$: Can ionize $\H$, but not $\H_2$.
    \item $11.2 \mathrm{eV} \leq h\nu < 13.6 \mathrm{eV}$: Lyman Werner (LW) band; can dissociate several molecules (e.g., $\Htwo$ and CO molecules) and photoionize $\CI$. 
    \item $6 \mathrm{eV} \leq h\nu < 11.2 \mathrm{eV}$: Photoelectric (PE) band, primarily responsible for the heating of neutral gas through the photoelectric effect ($\Gamma_{\rm PE}$). 
    \item Infrared band: Dust-reprocessed infrared (IR) radiation emitted by warm dust grains heated by the absorption of UV radiation, and relevant to the thermal balance of dust/gas.
\end{enumerate}

Additional frequency bands can be incorporated in a straightforward fashion (e.g., $\rm He$ ionizing bands).

\subsubsection{Solving for the radiation field}

For each discrete radiation band $\nu_i$, \texttt{VETTAM} solves the equations

\begin{gather}
    \frac{\partial E_{r,\nu_i}}{\partial t} +  \nabla \cdot \boldsymbol{F}_{r,\nu_i} =  j_{*,\nu_i} - cG^0_{\nu_i} \label{eq:EradUV} \\
    \frac{\partial \boldsymbol{F}_{r,\nu_i}}{\partial t} + \nabla \cdot (c^2\mathbf{T}_{r,\nu_i} E_{r,\nu_i}) =  -c^2\boldsymbol{G}_{\nu,i} \label{eq:FradUV} \\
    \frac{\partial I_{r,\nu_i}}{\partial s} = \frac{j_{*,\nu_i} + j_{\nu_i}}{4\pi} - \rho \kappa_{F,\nu_i} I_{r,\nu_i},
\end{gather}
where $E_{r,\nu_i}$, $\boldsymbol{F}_{r,\nu_i}$ and $\mathbf{P}_{r,\nu_i}$ are the radiation energy density, flux, and pressure tensor in that frequency band, $j_{*,\nu_i}$ and $j_{\nu_i}$ is the (isotropic) contribution from point sources (stars/AGN) and diffuse sources, $\rho$ is the gas density, $c$ the speed of light, $\kappa_{F,\nu_i}$ is the flux-mean absorption opacity, and $\mathbf{T}_{\nu_i}$ is the Eddington tensor given by the ratio of the second and zeroth (discrete) angular moments of the specific radiation intensity $I_{r,\nu_i}$,
\begin{equation}
    \mathbf{T}_{\nu_i} = \frac{\int d \Omega\, \hat{\boldsymbol{n}}_k \hat{\boldsymbol{n}}_k\,I_{r,\nu_i}.}{\int d \Omega\, I_{r,\nu_i}},
\end{equation}
obtained from the solution of the time-independent radiative transfer equation using the hybrid characteristics approach outlined in \citet{Buntemeyer_2016}. The source terms $G^0_{\nu_i}$ and $\boldsymbol{G}_{\nu_i}$ are given by
\begin{equation}
    \label{eq:G0}
    \begin{aligned}
    G^0_{\nu_i} =\;& \rho (\kappa_{\rm E,\nu_i}E_{r,\nu_i} - j_{\nu_i}) + \rho \left(\kappa_{\mathrm{F},\nu_i} - 2\kappa_{\mathrm{E},\nu_i} \right) \frac{\boldsymbol{v} \cdot \boldsymbol{F}_{\nu_i}}{c^2} \\
    &+ \rho \left( \kappa_{\mathrm{E},\nu_i} - \kappa_{\mathrm{F},\nu_i} \right) \left[\frac{v^2}{c^2}E_{r,\nu_i} + \frac{\boldsymbol{v}\boldsymbol{v}}{c^2} :\mathbf{P}_{r,\nu_i}\right],
\end{aligned}
\end{equation}
and
\begin{equation}
\label{eq:G}
\boldsymbol{G}_{\nu_i} = \rho \kappa_{\mathrm{F},\nu_i} \frac{\boldsymbol{F}_{\nu_i}}{c} - \rho \kappa_{\mathrm{F},\nu_i} E_{r,\nu_i}\frac{\boldsymbol{v}}{c}  - \kappa_{\mathrm{F},\nu_i} \frac{\boldsymbol{v}}{c} \cdot \mathbf{P}_{\nu_i},
\end{equation}
are the gas-radiation coupling terms to leading order in all regimes of RHD \citep{Krumholz_2007a} for the radiation energy and flux, respectively. $\kappa_{E,\nu_i}$ is the continuum energy-mean and flux-mean absorption opacity\footnote{We do not consider scattering (e.g., due to dust or electrons) for any of the bands in our implementation.}. For the purposes of this paper $j_{\rm IR} = \kappa_{\rm P,IR}a_{\rm R}\Td^4$ where $\kappa_{\rm P,IR}$ is the Planck-mean opacity (i.e. dust IR emission) and $j_{\nu_i} = 0$ for all the UV bands, although, in principle it will be straightforward to add any diffuse sources of UV radiation, such as Lyman continuum  (LyC) photons emitted by recombinations to the ground state relevant for scenarios where hydrogen case-B recombination is not a good approximation.

\subsubsection{Gas and dust opacities}

$\kappa_{E,\nu_i}$ and $\kappa_{F,\nu_i}$ include differing contributions from gas ($\kappa_{\rm g,\nu_i}$) and dust ($\kappa_{\rm d, \nu_i}$) depending on the band. For the purposes of this paper, dust acts as a source of (continuum) opacity in all bands, and the gas only for the LyC bands ($h\nu > 13.6 \, \rm {eV}$) where $\kappa_{\rm g,LyC}$ includes contributions from the ionization cross sections of $\H$ and $\Htwo$ to these photons. The energy- and flux-averaging technically depend on the local radiation field seen by the dust/gas mixture; we simplify this somewhat for the UV bands by assuming an unattenuated source spectrum (implying $\kappa_{E,\nu_i} = \kappa_{F,\nu_i} $)  -- such as from a fully IMF-sampled stellar population provided by \texttt{STARBURST99} \citep{Leitherer_1999} or \texttt{BPASS} \citep{2017PASA...34...58E}, or a stellar spectrum -- and pre-tabulate dust and gas opacities as a function of age, mass, metallicity and other properties. If there are multiple point sources with distinct spectral properties (such as a resolved star cluster with massive/low-mass stars), we also provide the option to perform a unattenuated flux-weighted opacity averaged over the sources\footnote{In practice, this is implemented in a completely modular fashion; \texttt{VETTAM} simply requires an average dust/gas opacity for the band for each cell, irrespective of how it is averaged, and more complicated prescriptions can be adopted.}. Our default implementation uses the \citet{Weingartner_2001b} dust opacity model for UV dust opacities and ionization cross sections from \citet{Osterbrock_1989}. In the IR we use the approximation $\kappa_{E,\rm IR} = \kappa_{\rm P,IR}(T_{\rm d})$ and $\kappa_{ F, \rm IR} = \kappa_{\rm R, IR}(T_{\rm d})$ where $\kappa_{\rm P,IR}(T_{\rm d})$ and $\kappa_{\rm R,IR}(T_{\rm d})$ are the Planck- and Rosseland-mean opacities for dust at temperature $T_{\rm d}$ obtained from the \citet{Semenov_2003} model. We refer the reader to \citet{Menon_2022_VETTAM} on the caveats associated with this assumption. 

Our approach for the LW band differs somewhat from the other bands, as there exist sources of gas opacity driven by a discrete band of lines rather than a continuum opacity. These lines correspond to thousands of discrete transitions of rovibrational levels between the ground and excited electronic states of $\Htwo$ and CO, rendering a fully consistent solution prohibitively expensive/complicated for hydrodynamic simulations. We instead resort to pre-computed shielding factors to treat these sources of opacity, which depend on the column densities of the chemical species from each source to the target cell. We use the hybrid characteristic ray-tracer \citep{Buntemeyer_2016} to compute the column densities of H, $\Htwo$, CO, and $\CI$ from each point source of radiation to each cell. We describe how these quantities are then used to compute dissociation/ionization rates in the following section. 

\subsubsection{Coupling to thermochemistry}

The ionization rates for $\Htwo$ ($\mathcal{I}_{\rm H_2}$) and $\H$ ($\mathcal{I}_{\rm H}$) are computed from the total absorption rates of LyC radiation energy density $cG^0_{{\rm LyC,15.4}}$ and $cG^0_{{\rm LyC,13.6}}$ as
\begin{gather}
    \mathcal{I}_{\rm H_2} = f_{\rm H_2,15.4}\frac{ cG^0_{{\rm LyC,15.4}}}{n_{\rm H_2}\langle h \nu \rangle_{\rm LyC,15.4}} \\
    \mathcal{I}_{\rm H} = f_{\rm H,15.4} \frac{ cG^0_{{\rm LyC,15.4}}}{n_{\rm H}\langle h \nu \rangle_{\rm LyC,15.4}} + f_{\rm H,13.6} \frac{ cG^0_{{\rm LyC,13.6}}}{n_{\rm H}\langle h \nu \rangle_{\rm LyC,13.6}}.
\end{gather}
In the expressions above, $\langle h \nu \rangle_{i}$ is the average photon energy in the corresponding band, which is incident-spectrum dependent, similar to the opacities discussed above. $f_{\rm H_2,15.4}$ and $f_{\rm H,15.4}$ are the fraction of total radiation energy absorbed by $\Htwo$ and $\H$ in the $\Htwo$ ionizing LyC band, and  and $f_{\rm H,13.6}$ is the fraction for the $\H$ ionizing LyC band. The fractions are obtained from the ionizing and dust optical depths per unit length in the given band as
\begin{gather}
    f_{\rm H_2,15.4} = \frac{n_{\rm H_2}\sigma_{\rm H_2,15.4}}{n_{\rm H_2}\sigma_{\rm H_2,15.4} + n_{\rm H} \sigma_{\rm H,15.4} + n_{\rm Hnuclei}\sigma_{\rm d,15.4}},
\end{gather}
where $\sigma_{\rm H_2,15.4}$, $\sigma_{\rm H,15.4}$, and $\sigma_{\rm d,15.4}$ are the ionization cross sections for $\Htwo$/$\H$ and the dust cross section respectively, and $n_{\rm Hnuclei}$ is the total number density of H-species. A similar expression can be made for the $\H$ ionizing LyC band, with the exception that $f_{\rm H_2,13.6} = 0$. For certain applications, one may wish to use a single LyC band, in which case the approach is identical to the one for the $\Htwo$ ionizing band, but with distinct values for the cross sections, appropriately averaged over the new band energy limits. 

For the rates associated with the LW band, we use the column densities of $\H$ ($N_{\rm H}$), $\Htwo$ ($N_{\rm H_2}$), CO ($N_{\rm CO}$) and $\CI$ ($N_{\rm CI}$) from every source to the target cell and accumulate the dissociation rates of $\Htwo$ ($\mathcal{D}_{\rm H_2}$) and CO ($\mathcal{D}_{\rm CO}$) and the ionization rates of $\CI$ ($\mathcal{I}_{\rm CI}$) as
\begin{align}
    \nonumber \mathcal{D}_{\rm H_2} = \frac{\mathcal{D}_{\rm H_2,ISRF
    }}{4 \pi J_{\rm LW,ISRF}}\sum_i \frac{L_{i,\rm LW}}{4 \pi r_i^2} \exp({- \tau_{\rm d,LW,i}) } \times \\
    f_{\rm sh,\rm H_2} (N_{\rm H_2,i}) f_{\rm sh,\rm H} (N_{\rm H,i}) \\
    \nonumber \mathcal{D}_{\rm CO} = \frac{\mathcal{D}_{\rm CO,ISRF
    }}{4 \pi J_{\rm LW,ISRF}}\sum_i \frac{L_{i,\rm LW}}{4 \pi r_i^2} \exp({- \tau_{\rm d,LW,i}) } \times \\
    f_{\rm sh,\rm CO} (N_{\rm H_2,i},N_{\rm CO,i})\\
    \nonumber \mathcal{I}_{\rm CI} = \frac{\mathcal{I}_{\rm CI,ISRF
    }}{4 \pi J_{\rm LW,ISRF}}\sum_i \frac{L_{i,\rm LW}}{4 \pi r_i^2} \exp({- \tau_{\rm d,LW,i}) } \times \\
    f_{\rm sh,\rm CI} (N_{\rm H_2,i},N_{\rm CI,i}),
\end{align}
where $\mathcal{D_{\rm \Htwo,ISRF}},\mathcal{D_{\rm CO,ISRF}}$ and $\mathcal{I_{\rm CI,ISRF}}$ are the free-space dissociation and ionization rates corresponding to the \citet{Draine_1978} interstellar radiation field background (ISRF) fitting function, and $J_{\rm LW,ISRF}$ the corresponding mean LW-band intensity for the ISRF. The summation is over all radiation sources/sinks $i$, each of which has a luminosity in the LW band ($L_{\rm LW,i}$), distance from the given cell $r_i$ and accounting for extinction by the dust optical depth $\tau_{\rm d, LW,i}$ and line-shielding using the columns to the source. $f_{\rm sh,\rm H_2}$ accounts for the self-shielding by $\Htwo$ molecules, for which we use the fitting function provided by \citet{Wolcott-Green_2011}\footnote{Their expression is very similar to the seminal result by \citet{Draine_Bertoldi_1996}, with a modification that provides more accurate solutions for a broader range of gas temperatures due to their accounting of its effects on level populations of the $\Htwo$ molecule, an effect that \citet{Draine_Bertoldi_1996} ignored.}. $f_{\rm sh,\rm H}$ accounts for the absorption of photons in the LW band which overlap with electronic transitions of $\H$ (Lyman series) that can become important at relatively high column densities ($N_{\rm H} \gtrsim 10^{23} \, \rm cm^{-2}$, \citealt{Draine_Bertoldi_1996,2026ApJ..1008..253C}); we use the fitting function provided by \citet{Wolcott-Green_2011b} for $f_{\rm sh,\rm H}$. We adopt the treatment in \citet{Tielens_Hollenbach_1985} for $f_{\rm sh,\rm CI}$, which considers self-shielding by $\CI$ and cross-shielding by the more abundant $\Htwo$ molecules; for $f_{\rm sh,\rm CO}$ we use the tabulated data (Table 5) in \citet{Visser_2009} assuming $b(\rm CO) = 0.3\, \kms$, $T_{\rm ex} (\rm CO) = 5 \, \rm K$ and $N(^{12}\rm CO)/N(^{13}\rm CO) = 69$.

We account for photoelectric heating on dust grains by accumulating the absorbed radiation energy density in the LW ($cG^0_{\rm LW}dt$) and PE ($cG^0_{\rm PE} dt$) bands to calculate the dimensionless FUV intensity ($\chi_{\rm FUV}$) with respect to the ISRF. This is given by 
\begin{equation}
    \chi_{\rm FUV} =\frac{c}{4\pi} \frac{E_{r,\rm LW} + E_{r,\rm PE}}{J_{\rm FUV,ISRF}},
\end{equation}
where $J_{\rm FUV,ISRF}$ is again obtained from the \citet{Draine_1978} ISRF. $\chi_{\rm FUV}$ is then used to obtain the photoelectric heating rate ($\Gamma_{\rm PE}$) using the expression given by Equation~\ref{eq:PE_heatingrate}. The absorbed FUV radiation by dust is also used to set $\dot{u}_{\rm r,UV}$, the UV heating term in the equation for the dust temperature (Equation~\ref{eq:dusttemp}) as
    $\dot{u}_{\rm r,UV} = cG^0_{\rm LW} + cG^0_{\rm PE}$.

\section{One-zone demonstrations}
\label{sec:OneZoneTests}

\subsection{ISM Phase structure}
\label{sec:ismphase}

\begin{figure*}
\centering
\includegraphics[width=\textwidth]{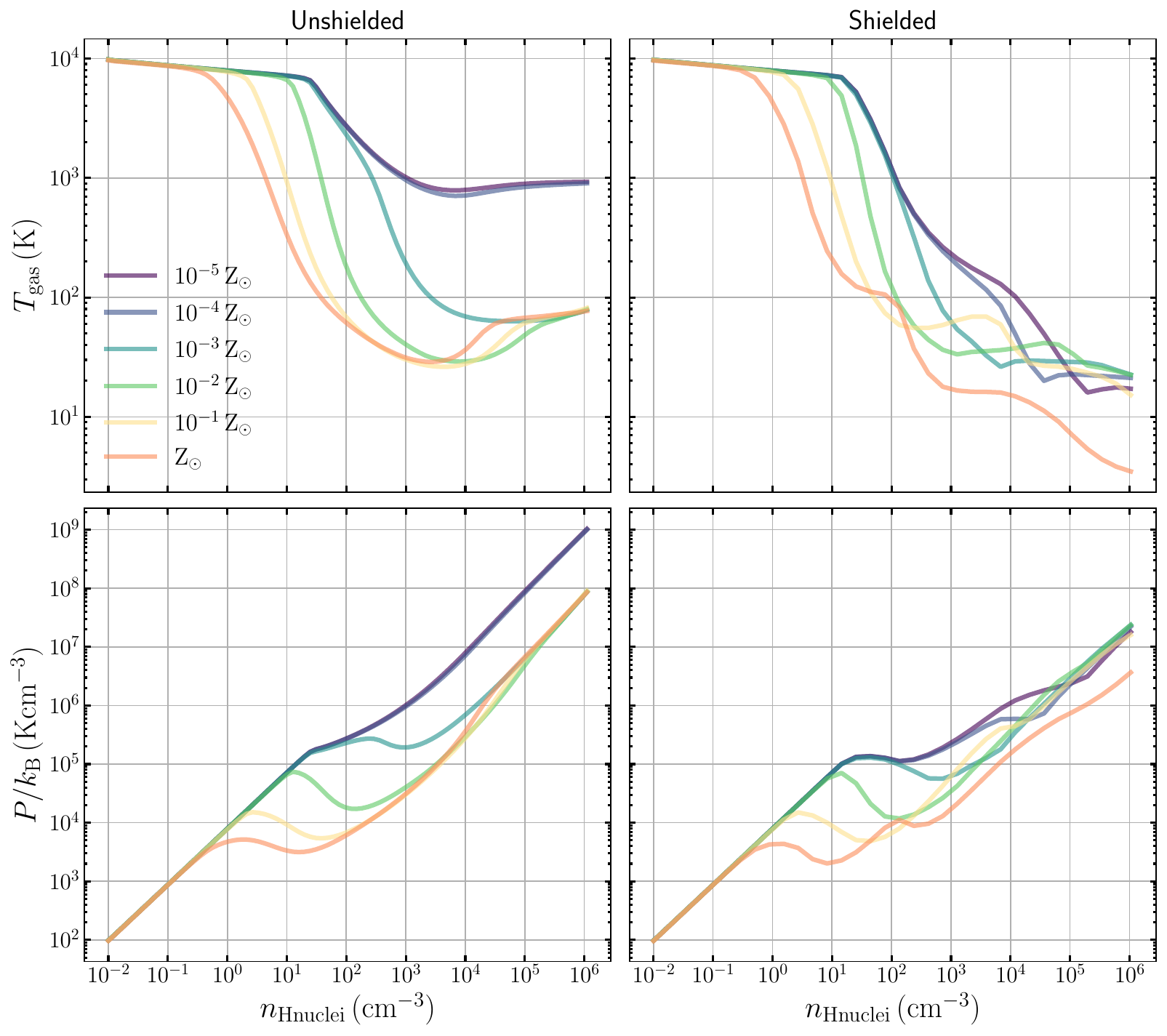}
\caption{Equilibrium phase structure obtained with our scheme for gas at different $Z$ exposed to a background FUV radiation field of $\chi_{\rm FUV} =1$ and CR background $\xi_{\rm H} = 10^{-16} \, \rm s^{-1}$. Left and right columns correspond to cases with parameterized dust+line-shielding, respectively. Top and bottom rows show the gas temperature and thermal pressure, respectively.}
\label{fig:ism}
\end{figure*}

We first test the ability of our implementation to capture the (equilibrium) multiphase behavior of the ISM. We do this by adopting a given fixed total number density of hydrogen nuclei ($n$), background FUV radiation field ($\chi_{\rm FUV}$), CR ionization rate of HI $\xi_{\rm H}$, and metallicity ($\Zsun$) and evolving the thermochemistry for a sufficiently long time such that equilibrium is achieved, defined as the point where the relative change in the abundances and temperature are $<10^{-8}$. We do this for $10^{-2} \lesssim n \lesssim 10^6 \, \rm cm^{-3}$ and $10^{-5} \lesssim \rm Z \lesssim 1 \, \rm \Zsun$. 

We perform two versions of this calculation. An unshielded case exposed to the free-space radiation/CR background and a shielded case that attenuates due to dust absorption and line overlap (self/cross-shielding), and considers an approximation for ionization losses for CRs. This shielding reduces the effective UV and CR flux the gas is exposed to, lowering the total heating as a result, and also permitting the formation of molecules unhindered by photodissociation and CR ionization. The amount of shielding column is a non-local quantity, and depends non-trivially on the turbulent density structure of gas/dust. While we expect denser gas to have higher shielding, in a turbulent medium the correlation is weaker than one naively expects, with significant scatter. Nevertheless, for testing purposes, we adopt the relation between the shielding length and $n$ used in \citet{kim2023_tigress}, estimated from ray-tracing calculations of the shielding column in numerical simulations of turbulent star-forming regions. The shielding length 
\begin{equation}
    L_{\rm shield} = 5 (n/100)^{-0.7} \, \rm pc,
\end{equation} 
corresponding to a shielding column
\begin{equation}
    N_{\rm shield} = n L_{\rm shield} = 3.87 \times 10^{20} n^{0.3} \, \rm cm^{-2}.
\end{equation}
This prescription leads to a V-band extinction $A_{\rm V }\sim 1$ at $n\sim 190 \, \rm cm^{-3}$. The shielding column is used along with the LW and PE dust cross sections to apply dust shielding to attenuate the LW/PE radiation fluxes, and the shielding lengths along with the number density of the species ($\Htwo$, $\CI$, CO) are used to compute self-shielding. The CR flux is also attenuated through a $\propto N_{\rm shield}^{-1}$ for $N_{\rm shield} > 9.35 \times 10^{20} \, \rm cm^{-2}$ \citep{Neufeld_2017} to account for energy losses via ionization and Coulomb scattering by thermal gas. More sophisticated treatments for CR attenuation can be readily incorporated within our numerical scheme \citep[e.g.,][]{padovani18a,Padovani2024,2022A&A...658A.151G}. 

In the left panel of Figure~\ref{fig:ism}, we show the equilibrium temperature (top; $\Tg$) and pressure (bottom; $P$) for the unshielded case. The $n$-$P$ relation at metallicities $Z \gtrsim 10^{-3} \, \Zsun$ shows the behavior expected for the two neutral phases of the ISM in pressure equilibrium \citep{Field_1969}. For $Z=\Zsun$, we obtain this stable multiphase behaviour over the range of pressures $3000 \lesssim P/k_{\rm B} \lesssim 6000 \, \rm K\,cm^{-3}$ corresponding to the upper and lower limits of the warm neutral medium (WNM) and cold neutral medium (CNM) respectively characterised by densities (temperatures) $n\lesssim 1 \, \rm cm^{-3}$ ($T_{\rm gas} \gtrsim 6000 \, \rm K$) and $n \gtrsim 20 \, \rm cm^{-3}$ ($T_{\rm gas} \lesssim 100 \, \rm K$) respectively. These values are consistent with previous estimates for these conditions \citep[e.g.,][]{Wolfire_2003,2017ApJ...839...90B,Bialy_Sternberg_2019}. We also find that the equilibrium pressure ranges shift to higher values at lower $Z$, and the multiphase region narrows, with less clear separation between the maximum and minimum pressures. This continues until eventually for $Z \lesssim 10^{-4} \, \Zsun$, the multiphase disappears completely, and the temperatures are always $\gtrsim 800 \, \rm K$. This metallicity dependence is consistent with the findings of \citet{Bialy_Sternberg_2019} where this loss of multiphase structure at low $Z$ was emphasized\footnote{We find slight qualitative differences for the trends with metallicity at $\sim 0.1 \Zsun$, which we ascribe to the differences in our scaling of dust abundance with metallicity: we have adopted a linear scaling whereas \citet{Bialy_Sternberg_2019} use the steeper, non-linear scaling at low metallicities from \citet{remy-ruyer_gas--dust_2014}.}. 

In the right panel of Figure~\ref{fig:ism}, we show the corresponding behavior with dust- and self-shielding for the same range of $Z$ as the unshielded test. The gas shows similar multiphase behavior as earlier, but at lower temperatures and pressures in the CNM, as would be expected with the lower PE and CR heating imposed by attenuation. In addition, the formation of molecules facilitated by shielding allows further cooling down to $\lesssim 100 \, \rm K$ for all $Z$, and $\lesssim 10 \, \rm K$ for gas at $\Zsun$. Even at the lowest $Z$, we observe multiphase behavior and gas temperatures approaching $\sim 20 \, \rm K$ at the highest densities. However, it is important to note that these represent the state of gas in thermochemical equilibrium, and at the lowest $Z$, the thermochemical timescales are typically much longer than dynamical timescales, and therefore the equilibrium states presented here may not be achieved in a dynamically evolving metal-poor ISM. Nevertheless, these curves demonstrate that our implementation achieves physically consistent ISM phase structure in equilibrium across a wide range of $Z$.

\subsection{Prestellar collapse}
\label{sec:collapse}

\begin{figure}
\centering
\includegraphics[width=0.49\textwidth]{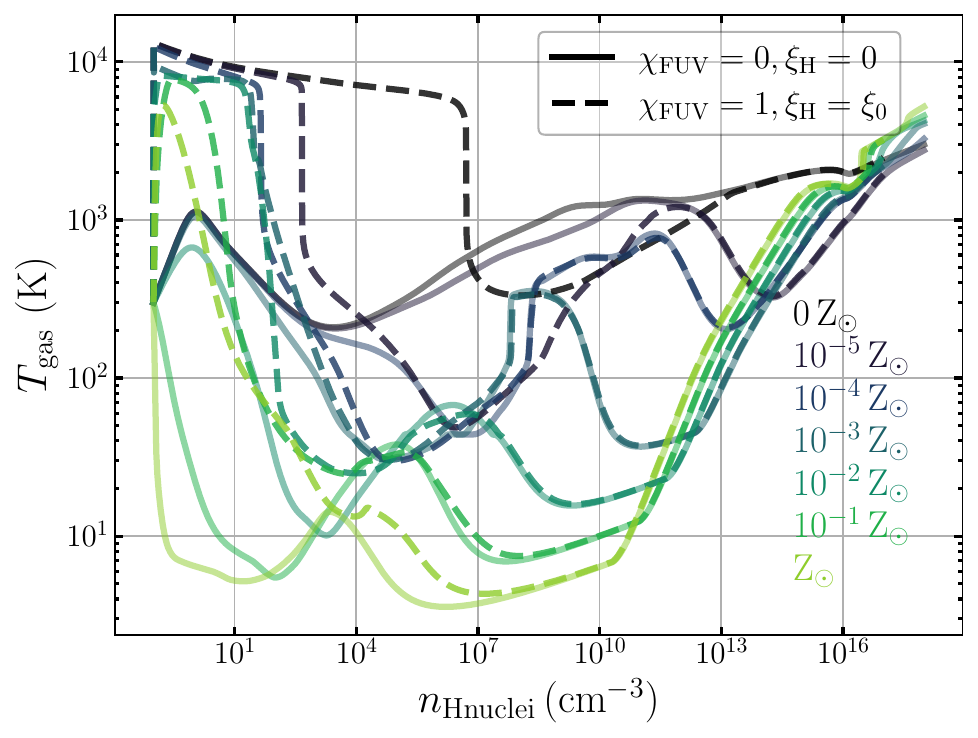}
\caption{The evolution of central gas temperature with density for the free-fall collapse of prestellar gas at different metallicities, and initial $n_{\rm Hnuclei} = 0.1 \, \pcm$ ($\rho_0 = n_0m_{\rm H}$) and temperature $\Tg = 300 \, \rm K$. Solid and dashed lines indicate cases without and with a radiation/CR background, respectively.}
\label{fig:CollapseTest}
\end{figure}

We model here the thermochemical evolution of prestellar gas undergoing free-fall gravitational collapse for a wide range of metallicities, to test the ability of our scheme to model very high density gas ($\sim 10^{15} \, \pcm$) beyond the range considered in Section~\ref{sec:ismphase}. The test follows the approach outlined in \citet{Omukai_2000}, where we consider the one-zone thermochemical evolution of the gas, following the evolution of the central density of a spherically-symmetric collapsing core. The central density, $\rho$, in this case, follows the free-fall relation
\begin{equation}
    \frac{d\rho}{dt} = \frac{\rho}{t_{\rm ff}},
\end{equation}
where $t_{\rm ff} = \sqrt{3\pi/32G\rho}$ is the instantaneous free-fall time at that density. We also evolve the coupled ODEs for the chemical rates and thermal evolution as the density/time increases. We initialize the gas with number density $n_0 = 0.1 \, \pcm$ ($\rho_0 = n_0m_{\rm H}$) and temperature $T_0 = 300 \, \rm K$, assuming a fully atomic medium, and all C (O) in its singly ionized (neutral) phase. We compute the evolution for gas metallicities of $0 \lesssim Z \lesssim 1 \, \Zsun$, assuming that the dust-to-gas ratio scales linearly with $Z$. We perform calculations with/without a background radiation field and CRs; when included, we assume a FUV flux of $\chi_{\rm FUV} = 1$ and CR ionization rate of $\xi_{\rm H} = 3 \times 10^{-17} \, \rm s^{-1}$. Following \citet{Omukai_2000} we assume that the core is surrounded by gas with identical density $\rho$ out to its Jeans length $\lambda_{\rm J} = \pi c_{\rm s}/\sqrt{G \rho}$ for computing a shielding column (for external radiation) and/or optical depth corrections to line/dust cooling; this assumption is founded on the expectation of a flat central region for a Penston-Larson self-similar collapse solution \citep{Penston_1969,Larson_1969}. CRs are not attenuated at high column densities in this calculation. 

\begin{figure*}
    \centering
    \includegraphics[width=\textwidth]{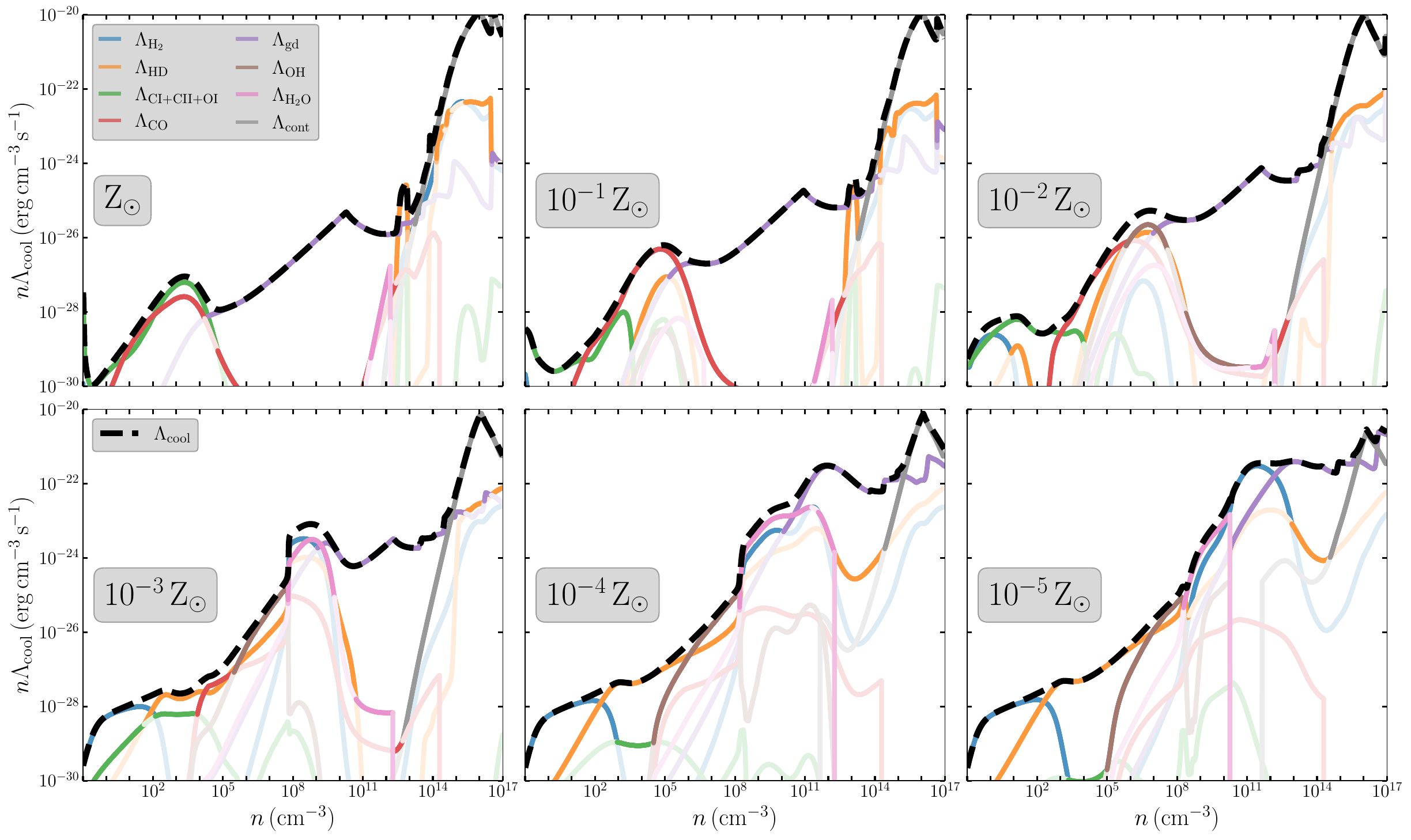}
    \caption{The most important heating/cooling mechanisms driving the evolution of gas temperature for a one-zone prestellar collapse calculation with our scheme at various metallicities (without a radiation/CR background) described in Section~\ref{sec:collapse}. The two most important coolants at all densities are shown with others made transparent for clarity. At all $Z$, dust-gas collisions ($\Lambda_{\rm gd}$) and continuum cooling processes ($\Lambda_{\rm cont}$) dominate at high $n$, and line cooling from chemical species at lower $n$. The species critical for cooling are CI/CII/OI fine-structure ($\Lambda_{\rm CI+CII+OI}$) and CO ($\Lambda_{\rm CO}$) at higher metallicities ($\gtrsim 0.1 \Zsun$) whereas OH and $\rm H_2O$ cooling becomes critical at low $Z \lesssim 0.01 \Zsun$ in dense gas. At the lowest $Z \lesssim 10^{-5} \Zsun$, there is a significant contribution from primordial species ($\Lambda_{\rm H_{2}}$, $\Lambda_{\rm HD}$). The transition from line to dust/gas continuum cooling occurs at increasingly higher $n$ at lower $Z$, implying that the gas temperature is sensitive to the chemical composition of the gas at low $Z$.}
    \label{fig:CollapsePhysics}
\end{figure*}

In Figure~\ref{fig:CollapseTest}, we show the evolution of the central gas temperature as a function of density obtained with our scheme, and in Figure~\ref{fig:CollapsePhysics} we highlight the most important heating/cooling terms at different densities and metallicities. We find excellent agreement with the evolution found in previous calculations without CR/radiation backgrounds \citep{Omukai_2000,Omukai_2005} and reproduce several key features in the temperature evolution, which we discuss below. 

Initially, after a short transient phase, the gas cools with the general trend of cooler temperatures at higher $Z$ at low to intermediate $n$, facilitated by fine-structure and molecular line cooling at higher $Z$ and $\Htwo$/HD line cooling at lower $Z$. This cooling phase is followed by a clear jump in temperatures, occurring at higher $n$ for lower $Z$, due to the heat released via $\Htwo$ formation on the surface of dust grains. The temperature rise is eventually reversed as molecules are formed and the associated ro-vibrational line emission cools the gas; the molecular cooling is dominated by CO at $Z \gtrsim 10^{-1} \Zsun$, with increasing contributions from OH and H$_2$O cooling\footnote{There are sharp drops in OH and $\rm H_2O$ cooling at $n \gtrsim 10^{11} \, \pcm$, which signify the limits of the tabulated cooling rates we adopt from \citet{Omukai_2010}. This doesn't affect the thermal evolution, however, as dust-gas cooling dominates at these $n$.} at lower metallicities ($\lesssim 10^{-2} \, \Zsun$), due to the higher gas temperatures encountered in dense gas at lower $Z$ ($\gtrsim 200 \, \rm K$), which increase both the formation rates of H$_2$O and OH and the ability to excite their ro-vibrational transitions \citep{Omukai_2005}. At the lowest metallicities ($Z \lesssim 10^{-5}\Zsun$), there is significant cooling from primordial species, i.e. $\Htwo$ and HD line cooling. Eventually, at sufficiently high densities, gas-dust collisions become frequent enough to dominate cooling, even for the $Z = 10^{-5} \Zsun$ case, keeping the gas almost isothermal in this phase. As the density increases further, the dust becomes optically thick to its IR continuum radiation, saturating the cooling rates, which lead to an adiabatic heating phase as the gas continues to collapse -- this phase is related to the formation of the first Larson core \citep{Larson_1969}. There is a slight, but noticeable change in the slope of the adiabatic phase, due to a change in specific heats of the gas facilitated by the excitation of rotation/vibrational degrees of freedom of $\Htwo$ molecules. The dominant cooling in this regime is the collision-induced continuum emission by $\Htwo$, which, however, is insufficient to keep up with compressional heating, and is metallicity independent. This explains the convergence in evolution with $Z$ on arriving at this phase. As the gas continues to collapse, temperatures reach $\sim 2000 \, \rm K$, which dissociate $\Htwo$ molecules (an endothermic process) and briefly cools the gas. This is followed by the second adiabatic collapse at $n\gtrsim 10^{17} \rm cm^{-3}$ to increasingly higher $\Tg$, until eventually nuclear fusion is initiated in the core. The case with radiation/CRs is heated via CR/photoelectric heating at lower $n$, leading to higher temperatures, until $n$ and the associated cooling rates are high enough, after which it evolves identically to the case without radiation/CRs.

Through this test, we demonstrate the ability of our scheme to capture the evolution of the temperature through the distinct chemistry-dependent cooling pathways (especially at low $Z$) out to densities approaching protostellar densities ($n\sim 10^{17} \pcm$). 

\subsection{PDR Benchmarks}
\label{sec:RolligPDR}

\begin{figure*}
    \centering
    \includegraphics[width=0.9 \textwidth]{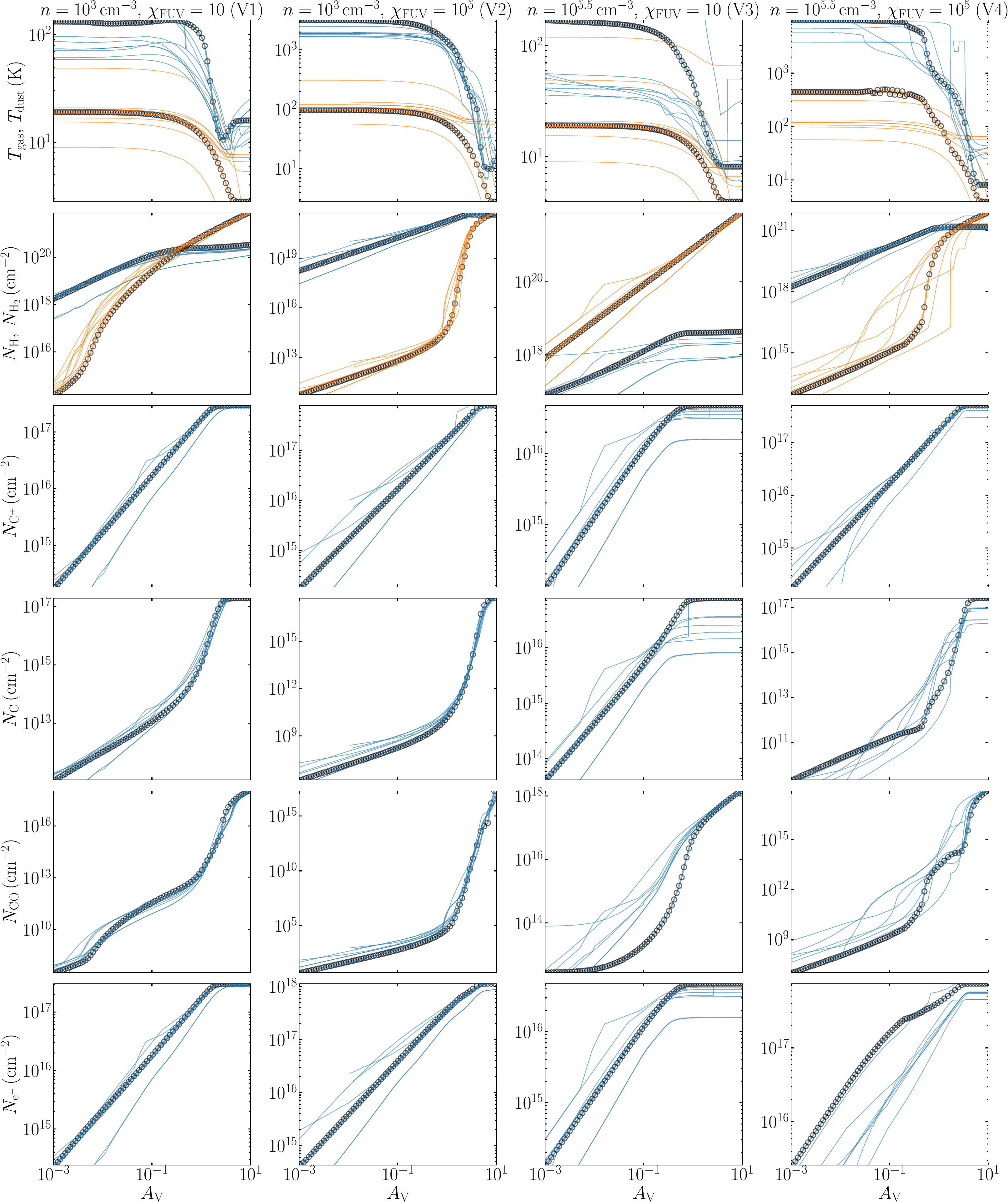}
    \caption{Column densities of various species as a function of $A_{\rm V}$ for the four PDR benchmark tests described in \citet{Rollig_2007}. Rows denote (from top to bottom): gas and dust temperatures, HI and $\Htwo$, $\CII$, $\CI$, CO and $\elec$ columns. The markers denote the values we obtain, and the lines in the background are results from PDR codes that participated in the benchmark: namely \texttt{Cloudy} \citep{Ferland_1998}, \texttt{COSTAR} \citep{Kamp_Bertoldi_2000,Kamp_vanZadelhoff_2001}, \texttt{HTBKW} \citep{Tielens_Hollenbach_1985}, \texttt{KOSMA-$\tau$} \citep{Storzer_1996,Bensch_2003}, \texttt{Lee96mod} \citep{Lee_1996}, \texttt{Leiden} \citep{Black_vanDishoeck_1987,vanDishoeck_Black_1988}, \texttt{Meijerink} \citep{Meijerink_Spaans_2005}, \texttt{Meudon} \citep{LePetit_2006}, \texttt{Sternberg} \citep{Sternberg_Dalgarno_1989,Sternberg_Dalgarno_1995}, and \texttt{UCL\_PDR} \citep{Bell_2005}.}
    \label{fig:RolligTesta}
\end{figure*}

\begin{figure*}
    \centering
    \includegraphics[width=0.9 \textwidth]{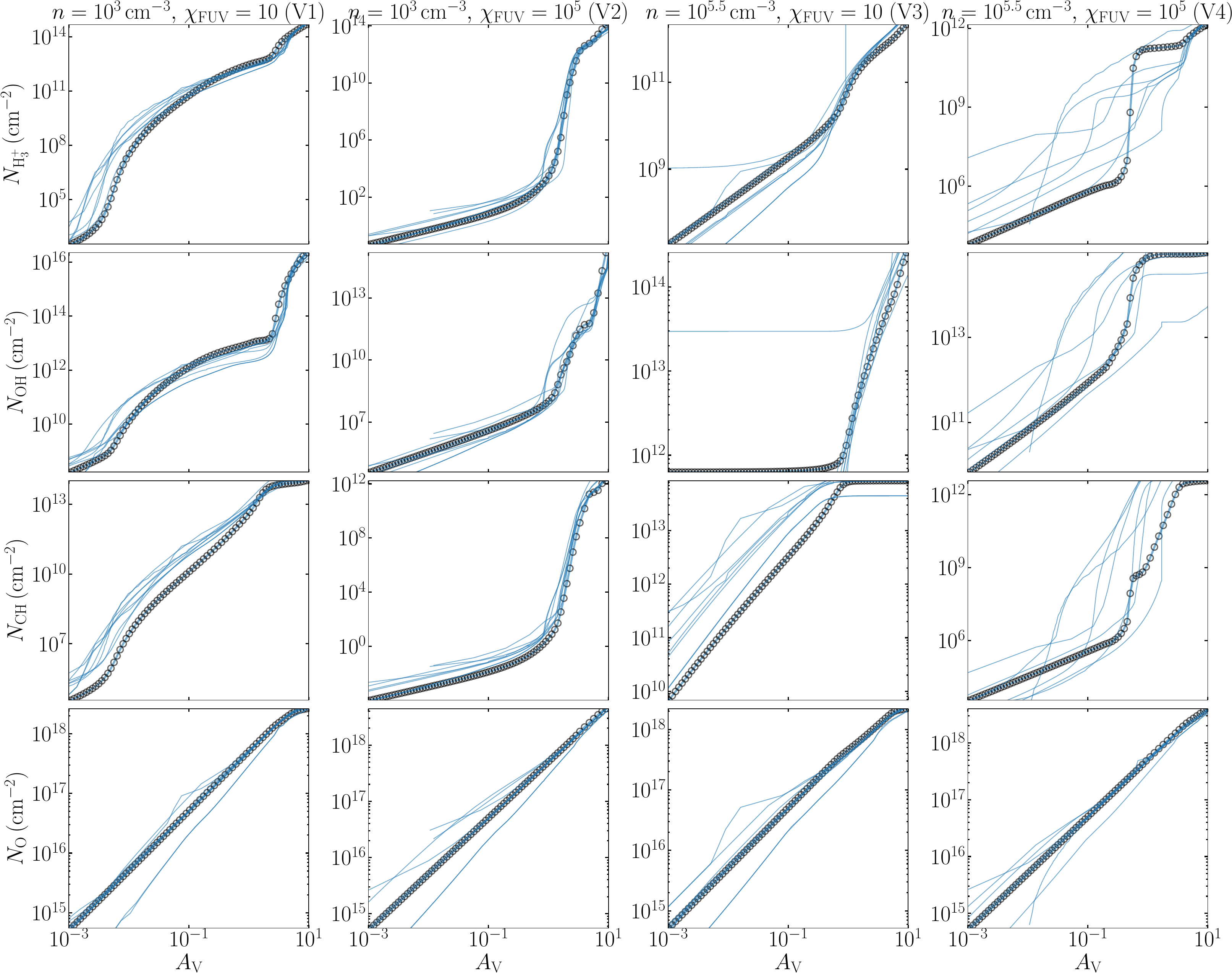}
    \caption{Continuation of Figure~\ref{fig:RolligTesta}. Rows denote (from top to bottom): column density of $\rm H_3^+$, OH, CH, and $\OI$.}
    \label{fig:RolligTestb}
\end{figure*}

We next test our ability to capture the coupled chemical and thermal structure of a PDR illuminated by an incident FUV radiation field of strength $\chi_{\rm FUV}$ and a CR ionization rate $\xi_{\rm H}$ at its semi-infinite edge. We assess whether our implementation reproduces accurate PDR transitions (the HI-to-$\Htwo$ transition and the CII$\to$CI$\to$CO transition), their characteristic depths, and the abundances of other species/molecules involved in CO formation pathways (e.g., $\rm H_3^+,CH$) and/or cooling at low-$Z$ (OH). We test this by running the benchmark models of \citet{Rollig_2007} and comparing against the ensemble of PDR codes included in their code-comparison project\footnote{Code results published at \url{https://lecture.ph1.uni-koeln.de/pdr-comparison/results_data_post.htm}}. Some of these participating codes use substantially more detailed chemical networks than our implementation and therefore provide a stringent validation of our simplified approach.
We run the four \citet{Rollig_2007} benchmark models V1--V4, in which the gas and dust temperatures are determined self-consistently from the thermal balance rather than held fixed (as in the companion F1--F4 models of that study, which we do not consider here). These span two densities, $n=10^3 \, \pcm$ and $n=10^{5.5} \, \pcm$, combined with two radiation field strengths, $\chi_{\rm FUV}=10$ and $\chi_{\rm FUV}=10^5$: V1 ($n=10^3\,\pcm$, $\chi_{\rm FUV}=10$), V2 ($n=10^3\,\pcm$, $\chi_{\rm FUV}=10^5$), V3 ($n=10^{5.5}\,\pcm$, $\chi_{\rm FUV}=10$), and V4 ($n=10^{5.5}\,\pcm$, $\chi_{\rm FUV}=10^5$). Together, these four models bracket the widest range of PDR conditions (in particular $\chi_{\rm FUV}/n$) explored in the comparison project.

We adopt a spatially uniform, unattenuated CR ionization rate of $\xi_{\rm H} = 5 \times 10^{-17} \, \rm s^{-1}$, and set the FUV dust opacity to $\kappa_{\rm FUV} = 1134.8 \, \rm cm^{2} \, g^{-1}$. To be consistent with \citet{Rollig_2007}, we also adjust several of our default parameters: the H$_2$ formation rate on dust grains is set to $3 \times 10^{-18} \Tg^{1/2} \, \rm cm^{3} \, s^{-1}$, we use the \citet{Draine_Bertoldi_1996} $\Htwo$ self-shielding prescription with a Doppler parameter $b=1 \, \kms$, the V-band dust opacity such that $A_{\rm V} = 6.289 \times 10^{-22} N_{\rm Hnuclei}$, and the elemental C and O abundances to $10^{-4}$ and $3 \times 10^{-4}$, respectively.

We first assess whether our implementation reproduces the temperature structure by solving for thermal balance and comparing the resulting gas- and dust-temperature profiles to those from the specialized PDR codes. In the top row of Figure~\ref{fig:RolligTesta} we show these profiles as a function of cloud depth ($A_{\rm V}$) for all four models, together with the profiles reported by the participating codes. As noted by \citet{Rollig_2007}, the code-comparison results exhibit substantial scatter due to minor differences in the implementations of  heating/cooling process in the participating codes; our profiles lie well within the range spanned by the other models in all four cases. 

In Figures~\ref{fig:RolligTesta} and \ref{fig:RolligTestb}, we assess whether our code reproduces the chemical structure of PDRs. We show the column densities of HI and $\Htwo$, $\CII$, $\CI$, CO and $\elec$ (Figure~\ref{fig:RolligTesta}), and of $\rm H_3^+$, OH, CH, and $\OI$ (Figure~\ref{fig:RolligTestb}), as a function of $A_{\rm V}$ for the four benchmark models (columns). The abundances we obtain with our implementation are in overall excellent agreement across all four models with the PDR codes in the benchmark. We particularly highlight the agreement obtained for molecular intermediates (e.g., OH, CH) that participate in CO formation, not explicitly considered or tested in commonly used thermochemistry implementations coupled to hydrodynamical codes. As noted in Section~\ref{sec:collapse}, OH is a critical coolant in dense gas for $Z\lesssim 0.01 \Zsun$ and our ability to accurately reproduce its abundance is reassuring for planned applications to model the low-$Z$ ISM.

\section{Chemistry in 3D Turbulent Media}
\label{sec:3DDemonstrations}

In this section we show two example 3D applications of our scheme as a proof of concept of our implementation: i) a turbulent patch of the ISM exposed to an interstellar radiation field background at various gas-phase metallicities ($\Zsun$ to $10^{-4} \Zsun$), and ii) the expansion of a HII region/PDR driven by a massive UV radiation source into the turbulent ISM. These demonstrations are designed to illustrate the interplay of non-steady state chemistry and dust/self-shielding in the intermittent density/temperature structures imposed by turbulence. Our goal here is not an exhaustive parameter study, but rather to highlight the qualitative behavior of key species and thermodynamic phases, and to demonstrate numerical robustness when the chemistry, heating/cooling, and radiation field are evolved self-consistently. 

We list some common numerical choices for the simulations below: we use the split, five-wave HLLR Riemann solver \citep{Waagan_2011} with a CFL number of 0.5 for the hydrodynamics. We use a relative tolerance of $10^{-3}$ and an absolute tolerance of $10^{-10}$ for the implicit solver in the thermochemistry step, and discretize the ray-tracer with 48 HEALPix tessellations on the solid angle sphere. The dust opacities in the UV bands are assumed to be constant with the following values: $\kappa_{\rm d,LyC} = 1250 \, \rm cm^2 g^{-1}$, $\kappa_{\rm d,LW} = 900 \, \rm cm^2 g^{-1}$, $\kappa_{\rm d,PE} = 514 \, \rm cm^2 g^{-1}$ -- appropriate for dust exposed to a fully-sampled Chabrier IMF \citep{kim2023_tigress}. 

\subsection{Turbulent patch}
\label{sec:turbulent_sim}

\begin{figure*}
    \centering
    \includegraphics[width=0.9 \textwidth]{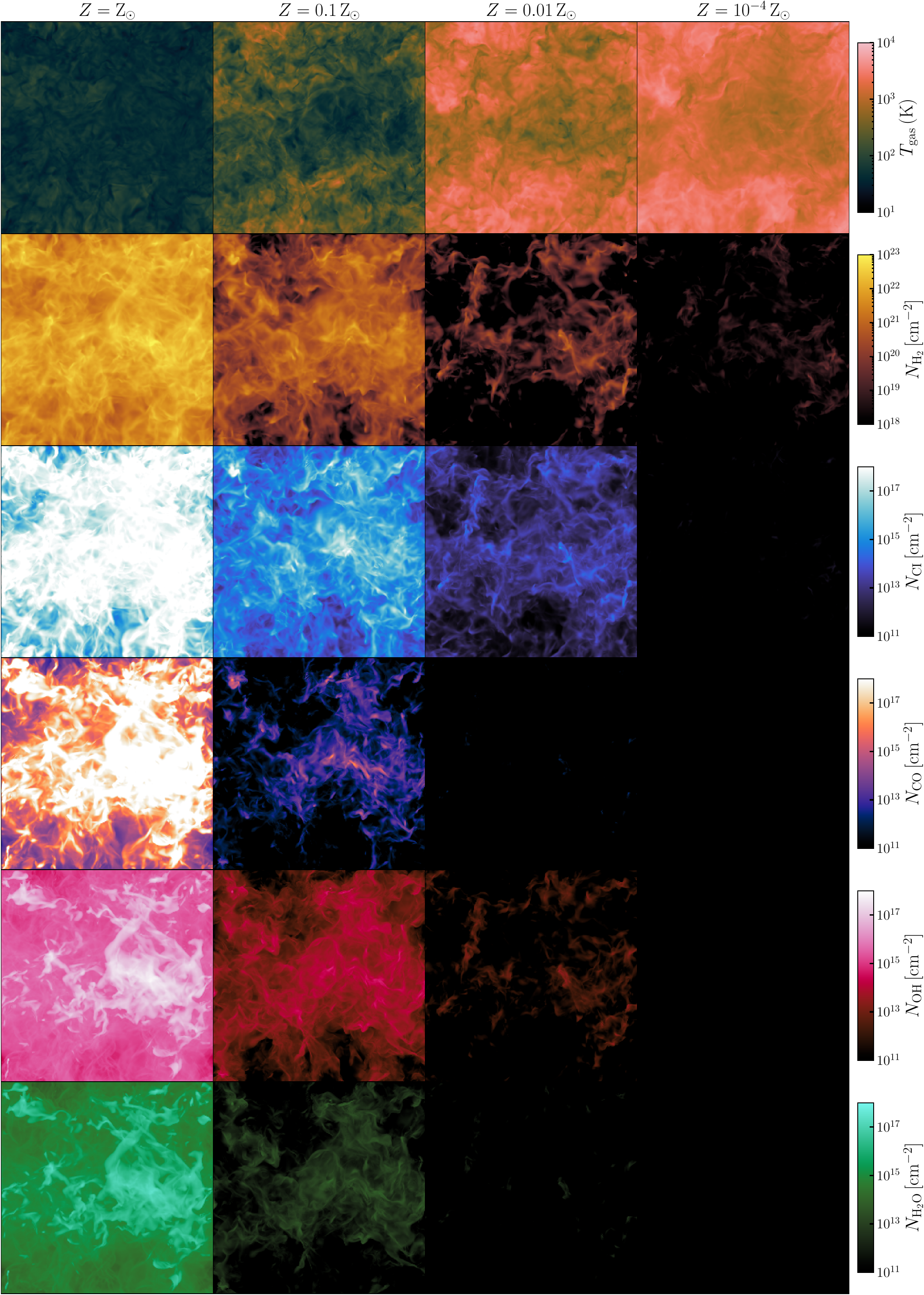}
    \caption{Thermochemical state of gas in supersonically turbulent patches at different metallicities (columns), as described in Section~\ref{sec:turbulent_sim}. The top row shows the mass-weighted average gas temperature ($\Tg$) and the following rows show column densities of $\Htwo$, CI, CO, OH, and $\rm H_2O$. We can see the expected evolution with $Z$ and highlight the presence of $\Htwo$ gas at low $Z$, which is better traced by CI and OH than CO.}
    \label{fig:TurbMetallicityAbundances}
\end{figure*}

We initialize a cubic domain of length $L = 100$ pc with average density $n_0 = 100 \, \pcm$, and consider four different cases with a spatially and temporally constant gas-phase metallicities of $\Zsun$, 0.1 $\Zsun$, 0.01 $\Zsun$ and $10^{-4 }\Zsun$. For each case, we set the temperature and chemical abundances to the one-zone steady state values appropriate for that density and metallicity. We expose the medium to a background UV radiation field of $\chi_{\rm FUV} =1$, CR ionization rate of $\xi_{\rm H} = 2 \times 10^{-16} \, \rm s^{-1}$, and use the ray-trace solution to account for dust and self-shielding for the radiation. Our choice of $n_0$ and $L$ corresponds to an $A_{\rm V,0} = 8 (Z/\Zsun)$ to the center of the domain for the different metallicities. 

We drive turbulent motions in the domain using a time-dependent Ornstein-Uhlenbeck process through the method outlined in \citet{Federrath_2010} and made publicly available in the \texttt{Turbulence Generator} module \citep{Federrath_2022} in the \texttt{FLASH} code. The turbulent forcing amplitude is controlled to ensure a 3D velocity dispersion $\sigma_{v,\rm 3D} = \sqrt{3}\sigma_{v,\rm 1D} = 13 \, \kms$, and follows a velocity power spectrum $E(k) \propto k^{-2}$ with a natural mixture of solenoidal and compressive modes \citep{Federrath_2013} over the range of scale $k/(2\pi/L) \in [2,64]$. The boundary conditions are set to periodic for the hydrodynamics, and the simulations are run for $8 t_{\rm dyn}$, where $t_{\rm dyn} = L/(2\sigma_{v,\rm 3D}) \approx 3.8 \, \rm Myr$ is the turbulent crossing timescale, by which point all relevant quantities achieve a statistical steady state.

Figure~\ref{fig:TurbMetallicityAbundances}, shows the gas structure and compare the gas temperatures ($\Tg$) and column densities of several important chemical species across the four metallicities. We can see that the gas columns are highly turbulent and filamentary, resulting in a highly turbulent temperature structure, and an expected trend of warmer temperatures at lower $Z$. We also see lower columns of species sensitive to dust-shielding from photodissociation -- i.e. $\Htwo$, $\CI$, CO, OH, and $\rm H_2O$ -- at lower $Z$ ($A_{\rm V}$). We also find that at $Z \lesssim 0.1 \Zsun$ there are noticeable reservoirs of $\Htwo$ gas, which, however, are not traced by CO -- i.e. CO-dark gas -- contrary to the $\Zsun$ case where CO faithfully traces the dense $\Htwo$ reservoir. As expected, we find that $\CI$ and OH are better tracers of $\Htwo$ at low $Z$ than CO. Nevertheless, the bulk of the H, C, and O at these $Z$ are in $\HI$, $\OI$, and $\CII$, respectively. 

\begin{figure*}
    \centering
    \includegraphics[width=\textwidth]{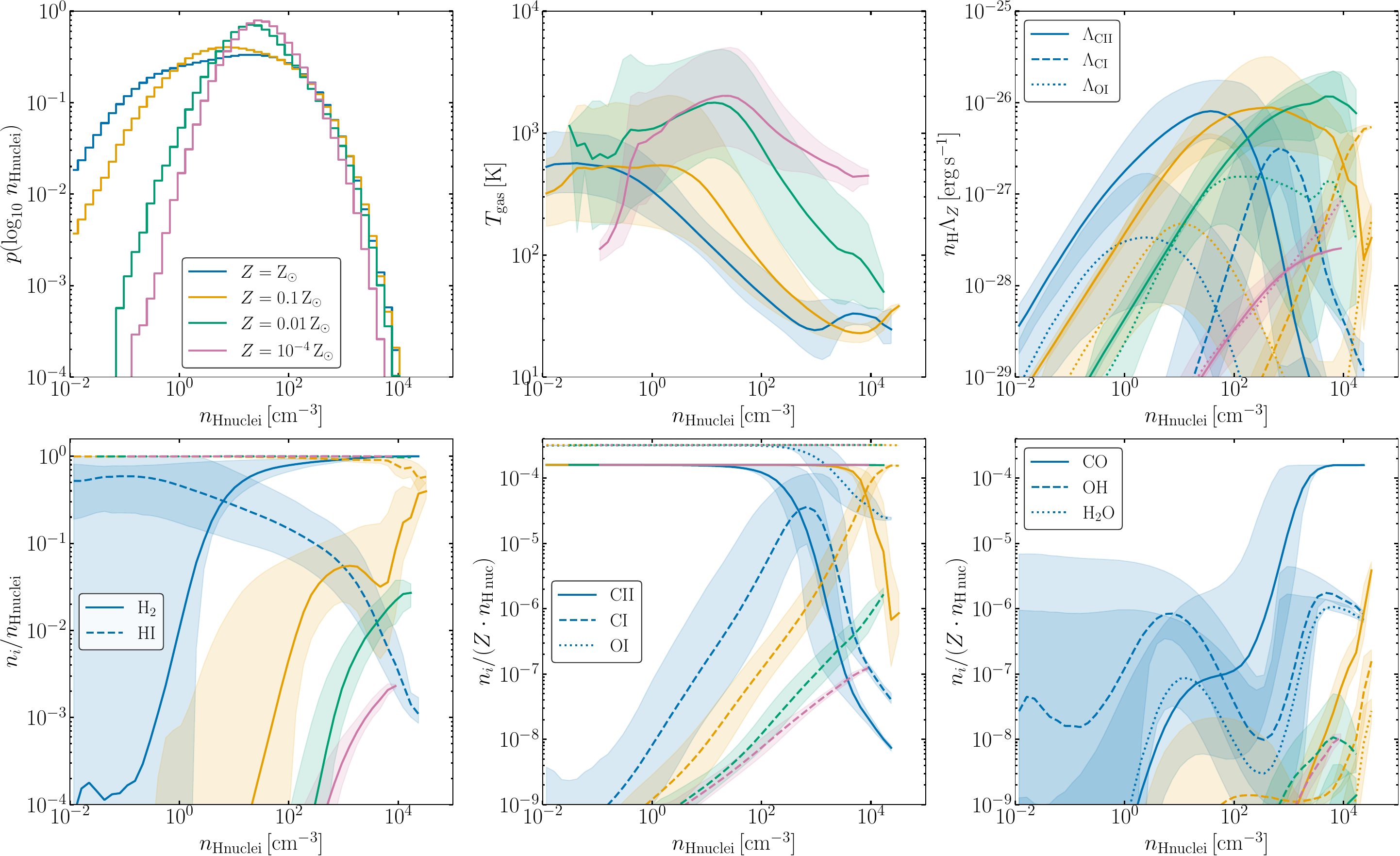}
    \caption{Average properties of the gas as a function of total hydrogen nuclei density ($n_{\rm Hnuclei}$) compared for the four $Z$, with shaded region showing the 1-$\sigma$ variation at that density. Panels show the volume-weighted PDFs (top-left), gas temperature ($\Tg$; top-middle), fine-structure cooling rates (top-right), abundances of $\HI$ and $\Htwo$ (bottom-left), of $\CI$, $\CII$ and $\OI$ (bottom-middle) and of molecular species CO, OH and $\rm H_2O$ (bottom-right).}
    \label{fig:TurbMetallicityPhase}
\end{figure*}

We quantify and compare the physical properties of the gas at various $Z$ in Figure~\ref{fig:TurbMetallicityPhase}. We show the average properties of the gas at different H nuclei density ($n_{\rm Hnuclei}$), with shaded regions indicating the 1-$\sigma$ variation. We can see that at lower $Z$ the gas is warmer and tends to show a more uniform $\Tg$ structure. This smooths out density fluctuations and strong shocks in the medium, most evidently seen in the density-PDFs, where there is a lack of very low-$n$ material in comparison to higher $Z$. We also note that the $\Tg$ values of gas we obtain at $n\lesssim 1 \rm cm^{-3}$ is much cooler than expected for the warm phase of the multiphase ISM (WNM; $\sim 10^4 \, \rm K$); indeed, in our static one-zone equilibrium calculations, we find $\Tg \sim 10^4 \, \rm K$ for gas at these $n$, as can be seen in Figure~\ref{fig:ism}. However, in our idealized stirred turbulent boxes, we find that the (thermal) equilibration timescales ($t_{\rm eq, \Tg}$) are longer than the dynamical/turbulent crossing timescale of the box ($t_{\rm turb}$), and therefore the gas at these $n$ are never allowed to reach their equilibrium values. We demonstrate this quantitatively in Appendix~\ref{sec:Turbulence_temperatures}, and demonstrate that at lower turbulent velocities where $t_{\rm eq, \Tg} \lesssim t_{\rm turb}$, we indeed recover temperatures approaching the expected equilibrium values. 

We also show in Figure~\ref{fig:TurbMetallicityPhase} that for the range of $n$ obtained in our turbulent media, the dominant carriers of H, C, and O nuclei change, consistent with expectations from static one-zone models: i.e., a larger abundance of species reliant on dust-shielding -- such as $\Htwo$, CI, and CO -- at higher $Z$. We also confirm that OH takes over as the dominant molecular species over CO at $Z \lesssim 0.01 \Zsun$. However, we find very large scatter for all species at a given $n$ due to turbulence-facilitated dynamical mixing at the interfaces of the PDR, combined with the time and space-varying radiation field seen by the gas due to the non-local shielding from the ISRF. In the top-right panel of Figure~\ref{fig:TurbMetallicityPhase} we also compare fine-structure cooling rates at different $Z$ finding that $\CII$ and $\CI$ cooling dominate at low- and high-$n$ respectively for $Z \gtrsim 0.1 \Zsun$, whereas $\OI$ cooling starts to become increasingly important at lower $Z$, with it dominating for the $Z \sim 10^{-4} \Zsun$ case in dense gas. While $\OI$ is the dominant carrier of O nuclei at all $Z$, only for $Z \lesssim 0.01 \Zsun$ does the gas achieve high enough temperatures ($\gtrsim 500 \, \rm K$) at $n \gtrsim 100 \pcm$ to excite the $\OI$ fine-structure lines. 

\subsection{Turbulent HII Region and PDR}
\label{sec:Turbulent_PDR}

\begin{figure*}
    \centering
    \includegraphics[width=0.9 \textwidth]{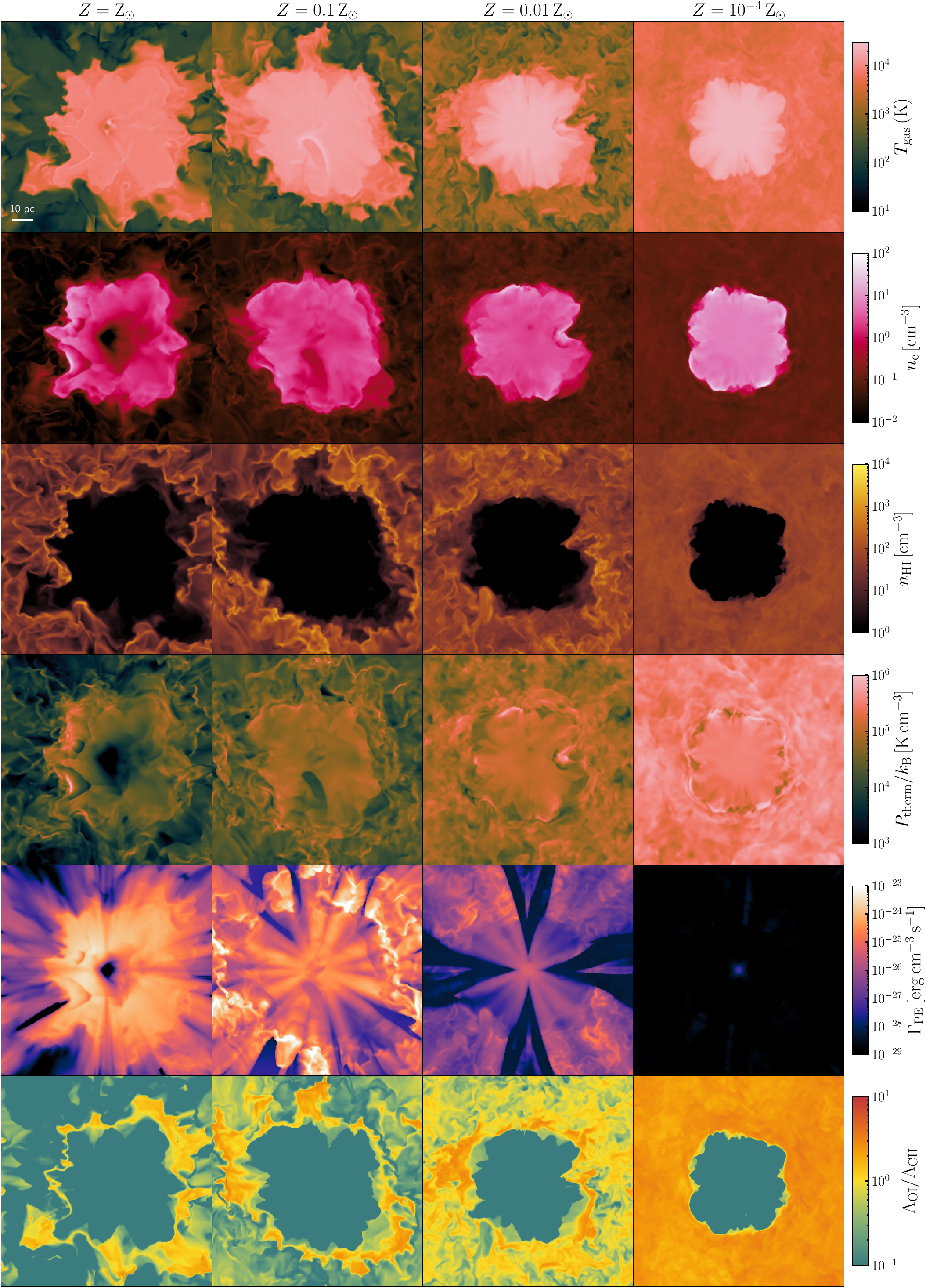}
    \caption{Relevant quantities compared for a HII regions/PDR in a $\sim 100 \, \rm pc$ turbulent patch at different gas-phase $Z$, driven by a UV source ($\sim 1000 \rm M_{\sun}$ star cluster). The rows show slices of gas temperature ($\Tg$), number densities of electrons ($n_{\rm e}$), $\HI$ ($n_{\rm HI}$), the gas thermal pressure ($P_{\rm therm}/k_{\rm B}$), the photoelectric heating rate ($\Gamma_{\rm PE}$) and the ratio of OI and CII fine-structure cooling rates. }
    \label{fig:TurbMetallicityRHD}
\end{figure*}

\begin{figure}
    \centering
    \includegraphics[width=\linewidth]{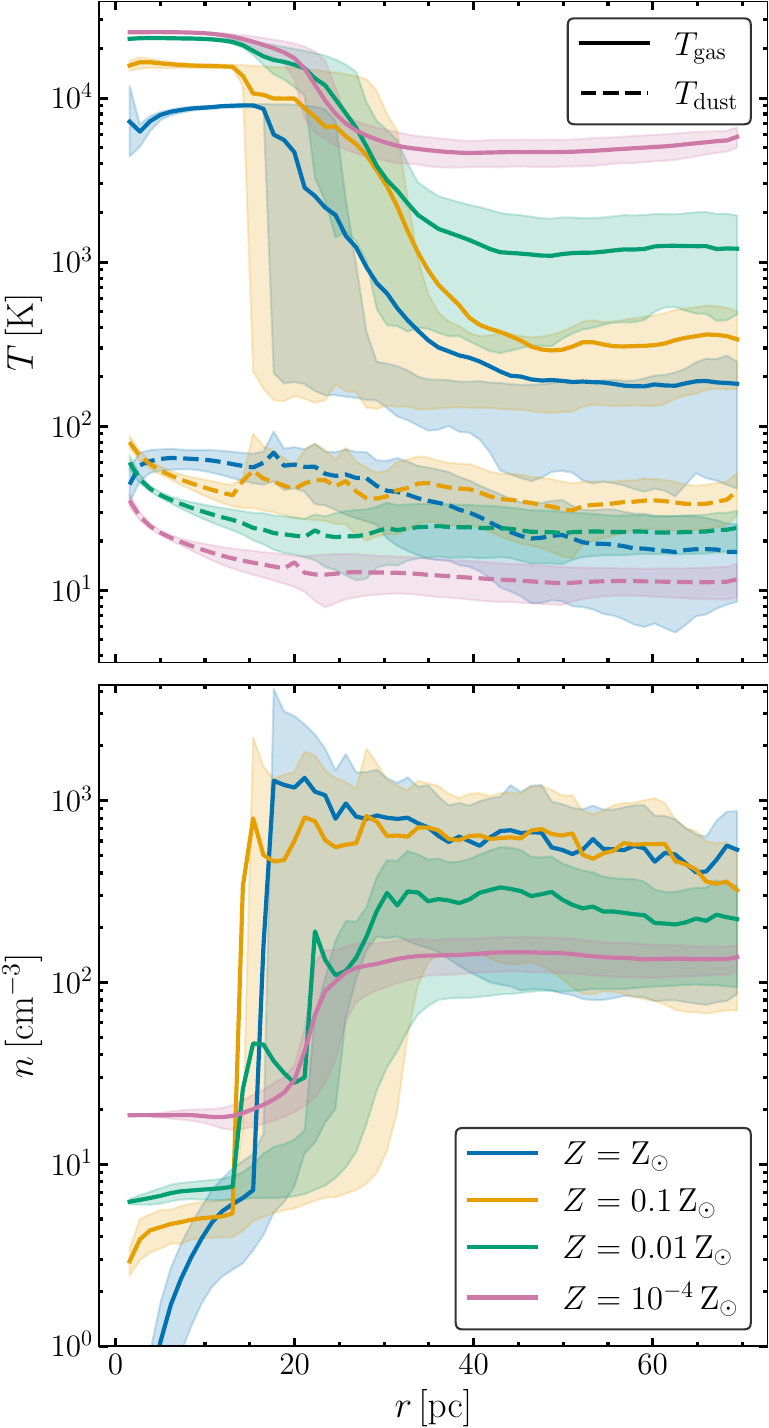}
    \caption{Radial profiles of gas temperature $\Tg$ (solid) and dust temperature $T_{\rm dust}$ (dashed, top panel), and total hydrogen nuclei number density $n$ (bottom panel), at $t\sim10\,\rm Myr$ for the turbulent \HII\ region/PDR calculation described in Section~\ref{sec:Turbulent_PDR}, for the four metallicities considered. Lines show the azimuthal median and shaded regions the $1\sigma$ scatter at fixed radius. The ionization front, marked by the sharp drop in $\Tg$ and the compressed density shell in the bottom panel, occurs at progressively larger and less well-defined radii as $Z$ decreases, as the associated density shell weakens and eventually disappears by $Z=10^{-4}\Zsun$, reflecting the reduced ability of the swept-up gas to cool and compress at low $Z$.}
    \label{fig:TurbHIIRadialProfile}
\end{figure}

In this section, we demonstrate the thermochemical evolution of the turbulent medium when a strong source of UV photons -- such as a massive star/cluster -- is placed in the turbulent medium, and compare the subsequent evolution of the HII region/PDR at the different $Z$. We use the final state of the stirred turbulent medium described in the previous section -- i.e. at $8t_{\rm dyn} \sim 30 \, \rm Myr$ -- where the system has reached a dynamical and thermochemical quasi-steady state. We restart the simulation by placing a source with the radiative properties of a $1000\rm M_{\odot}$ star cluster at the center of the domain, and modify the boundary conditions from periodic to outflow. The radiative properties of the source are assigned assuming the luminosity-to-mass of a fully IMF-sampled stellar population at $\Zsun$ obtained with \texttt{Starburst99}\footnote{These correspond roughly to ionizing, LW and PE luminosity-to-mass ratios of 500, 109 and 271 $\rm L_{\sun}/M_{\sun}$ respectively.}, and are kept the same for the runs with different $Z$. Although we expect lower $Z$ stellar populations to have different radiative outputs, we keep the source properties fixed here for a more controlled experiment to isolate the effects of $Z-$dependent thermochemistry on the evolution of the HII region/PDR. We evolve the system for 10 Myr.

We compare the properties of the HII region/PDR for the four $Z$ cases at the final time $\approx 10 \, \rm Myr$. In Figure~\ref{fig:TurbMetallicityRHD} we show slices of various relevant quantities, and in Figure~\ref{fig:TurbHIIRadialProfile}, where we compare the azimuthal-averaged radial profiles of temperatures ($\Tg$, $\Td$) and number density ($n$). We can see that the interior of \HII\ regions are evidently hotter at lower $Z$ , with characteristic interior temperatures rising from $\sim 8000 \, \rm K$ at $\Zsun$ to a few $\times10^4 \, \rm K$ at lower $Z$. This is consistent with the loss of collisionally-excited metal line cooling -- treated approximately in our scheme ($\Lambda_{\rm neb}$; (Appendix~\ref{sec:appendix_thermodynamics}) -- at lower $Z$, leading to higher equilibrium temperatures. 

However, despite this change in the interior temperature, the size of the ionized cavity -- traced by the sharp drop in $\Tg$ and the associated compressed density shell -- increases only slightly ($\lesssim 20\%)$ at lower $Z$ -- at odds with the zeroth-order expectation that higher ionized gas temperatures will facilitate
faster expansion of D-type ionization fronts. This can be understood from the fact that at lower $Z$ the neutral gas temperature is also significantly higher due to the loss of metal coolants, and therefore the pressure difference at the ionization front -- which drives its expansion -- is less pronounced at lower $Z$, as can be seen in the fourth row of Figure~\ref{fig:TurbMetallicityRHD}. This is also why at lower $Z$ the ionization gas pressure is unable to sweep up a dense, thin shell, and rather produces much broader, weaker density jumps than at higher $Z$ (lower panel, Figure~\ref{fig:TurbHIIRadialProfile}). The higher neutral gas temperatures also explain why the ionized gas and PDR are less clumpy and turbulent at the lowest $Z$ -- higher $\Tg$ lead to lower turbulent Mach numbers in the surrounding neutral gas, and therefore pressure fluctuations are promptly smoothed by sound waves. 

In the 5th row of Figure~\ref{fig:TurbMetallicityRHD} we show the photoelectric heating rate, $\Gamma_{\rm PE}$. We can see that $\Gamma_{\rm PE}$ is highest at the edge of the PDRs, tracing the dense filaments, globules and pillar-like structures at the edge of ionization front. We can also see the expected decrease with $Z$ (Equation~\ref{eq:PE_heatingrate}; $\Gamma_{\rm PE}\propto G_0 \, Z_d'$): even though the FUV field itself penetrates to somewhat larger $A_{\rm V}$ at low $Z$, the near-total loss of dust grains means there are simply too few absorbers left to convert that field into significant heating, and $\Gamma_{\rm PE}$ falls by more than five orders of magnitude between $\Zsun$ and $10^{-4}\Zsun$. We also note that the radial ``pinwheel'' pattern visible in $\Gamma_{\rm PE}$ is a numerical artifact of our finite angular sampling of the ray-tracer used in our scheme: we find that they disappear at higher angular resolutions. We note that the ray features are emphasized here as it is a slice plot and represents an instantaneous snapshot of the ray-sampling. In practice, the rays are rotated randomly every timestep \citep{Menon_2022_VETTAM}, so errors induced by finite sampling do not accumulate and affect the thermochemical state of the gas -- indeed, this is why the temperature/chemical structure do not show the ray-features. 

Finally, in the bottom row we show the ratio of the total cooling from $\OI$ and $\CII$ fine-structure lines ($\Lambda_{\rm OI}/\Lambda_{\rm CII}$), which act as important observational tracers of PDRs and control their thermal structure \citep{wolfire_photodissociation_2022}. We can see that $\CII$ cooling dominates within the $\HII$ region as $\OI$ is photoionized\footnote{Our chemical network does not track ionization states beyond $\CII$, and therefore we do not obtain higher ionization states of C that should exist in $\HII$ regions with hard ionizing sources. This would overestimate the apparent role of $\CII$ cooling in our plot; this however doesn't affect the temperature structure as our nebular cooling prescription ($\Lambda_{\rm neb}$) approximately accounts for cooling from untracked ionization species.}. In the PDR region, at $\Zsun$, there is a very thin layer just outside the ionization front in which $\OI$ cooling becomes marginally more important than $\CII$ -- this is consistent with the expectation that $\OI$ cooling becomes important at the high FUV radiation fields experienced at the edge of the ionization front. At lower $\Zsun$, this $\OI$-dominated layer becomes thicker and peak values of $\Lambda_{\rm OI}/\Lambda_{\rm CII}$ are higher, revealing that $\OI$ becomes relatively more important in the neutral gas. Indeed at the lowest $Z$ $\OI$ cooling dominates over $\CII$ even in the diffuse gas, as found also in the case without the strong UV radiation source in Section~\ref{sec:turbulent_sim} (Figure~\ref{fig:TurbMetallicityPhase}). 

Taken together, these results demonstrate that the coupling between radiation transport from point sources (stars/clusters) and thermochemistry in our implementation is able to accurately model the dynamical evolution of HII regions/PDRs across a broad range of $Z$.

\section{Discussion}
\label{sec:Discussion}

\subsection{Non-equilibrium chemistry and CI abundances}
\label{sec:ComparisonObservations}

\begin{figure*}
    \centering
    \includegraphics[width=\linewidth]{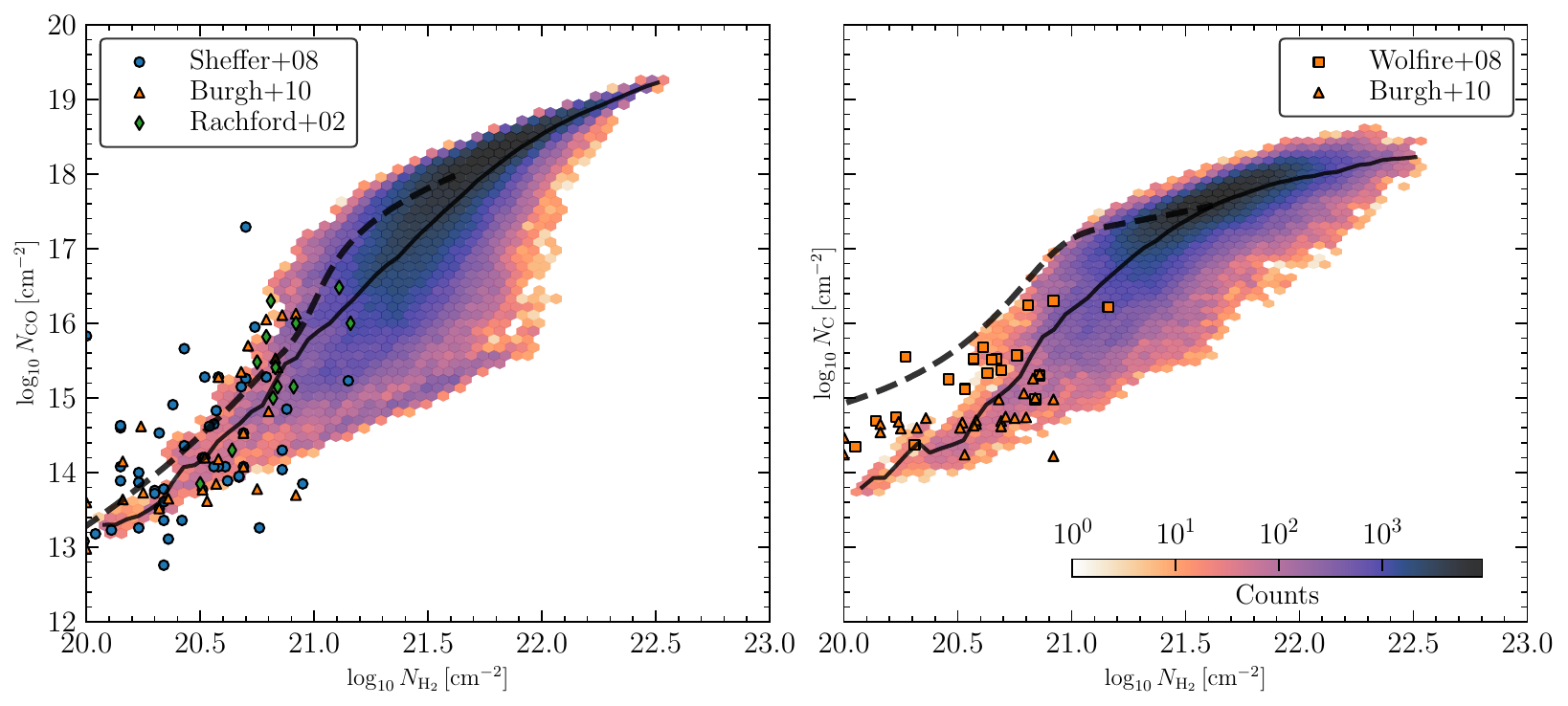}
    \caption{2D histograms of the line integrated column density distributions of CO and $\CI$ with respect to $\Htwo$ obtained in the turbulent patch simulations presented in Section~\ref{sec:turbulent_sim}. Solid black line indicates the binned-average trend of this obtained distribution from the simulation, which can be compared with the dashed line that shows the outcome of a corresponding static 1D PDR with the same \textit{average} $n$ and radiation/CR background as the simulation (see Section~\ref{sec:ComparisonObservations}). Markers indicate a compilation of UV absorption line measurements of these species in the diffuse ISM \citep{Rachford_2002, Sheffer_2008, Wolfire_2008, Burgh_2010}. We can see that our simulations are able to simultaneously reproduce observed $\CI$ and CO columns, whereas the corresponding static model overestimates $\CI$ columns.}
    \label{fig:CarbonColumnsTurb}
\end{figure*}

A long-standing puzzle in PDR modeling of the diffuse ISM is that steady-state, equilibrium calculations tend to overpredict the atomic carbon (\CI) column at a given $N_{\rm H_2}$ (or $A_{\rm V}$) relative to UV/optical absorption-line measurements, even though the corresponding CO columns are reproduced comparably well. \citet{Gong_2017} demonstrated this discrepancy explicitly using a slab PDR model tested against a compilation of observed diffuse and translucent sightlines \citep{Sheffer_2008,Rachford_2002,Burgh_2010}, finding that their equilibrium \CI\ predictions exceeded the observed columns by roughly an order of magnitude at fixed $N_{\rm H_2}$. Varying the incident radiation field, the cosmic-ray ionization rate, or the efficiency of grain-assisted recombination does not resolve the discrepancy without invoking an implausibly large ($\sim10\times$) depletion of gas-phase carbon; \citet{Gong_2017} instead suggested that non-equilibrium, dynamical effects -- gas being cycled between shielded cloud interiors and their unshielded, \CI-forming surfaces on timescales shorter than the relevant chemical/collisional timescales -- were the more likely resolution, though this was not tested directly with time-dependent 3D chemistry in their work.

Our turbulent-patch simulations (Section~\ref{sec:turbulent_sim}) provide a natural test of this hypothesis, since they evolve the coupled non-equilibrium thermochemistry self-consistently in a dynamically evolving, turbulent density field. In Figure~\ref{fig:CarbonColumnsTurb} we compare the line-of-sight integrated CO and \CI\ columns obtained from our $\Zsun$ turbulent-patch simulation against the same compilation of absorption-line measurements used by \citet{Gong_2017}: CO column densities from \citet{Sheffer_2008}, \citet{Burgh_2010}, and \citet{Rachford_2002}, and \CI\ column densities from \citet{Wolfire_2008} and \citet{Burgh_2010}. For reference, we also compute a corresponding static, steady-state 1D PDR model using the same average density and radiation/CR background as our turbulent box, following the same procedure used for our PDR benchmark comparisons in Section~\ref{sec:RolligPDR}.

We find that our turbulent, time-dependent simulation reproduces both the observed CO and \CI\ columns simultaneously to good accuracy across the full range of $N_{\rm H_2}$ probed by the observations. By contrast, the static, steady-state PDR model computed at the same mean density and radiation field noticeably overpredicts the \CI\ column at fixed $N_{\rm H_2}$, mirroring the discrepancy identified by \citet{Gong_2017} for their own equilibrium models. Since our approach differs from a standard static PDR model only in that it self-consistently couples the chemistry to a turbulent, dynamically evolving density and radiation field, this agreement lends direct support to the interpretation offered by \citet{Gong_2017}: turbulent mixing continually cycles gas between well-shielded, \CI$\to$CO-converting regions and the unshielded surface layers where CO is efficiently photodissociated back to \CI\ and CII, preventing the gas from ever fully relaxing to the higher equilibrium \CI\ abundance that a static, one-sided slab calculation would predict at the same mean physical conditions. Follow-up experiments spanning a broader range of physical conditions would shed light on these effects and the impact of non-equilibrium chemistry on other species of interest. Time-dependent chemistry can also have important implications for inferring physical conditions from molecular observations, including CR ionization rates \citep[e.g.,][]{Viti_2017}. More recently, \citet{Ho_2026} demonstrated that assuming chemical equilibrium can lead to systematic overestimates of CR ionization rates inferred from $\rm H_3^+$ observations.

\subsection{Comparison to other networks}

\begin{figure*}
    \centering
    \includegraphics[width=\textwidth]{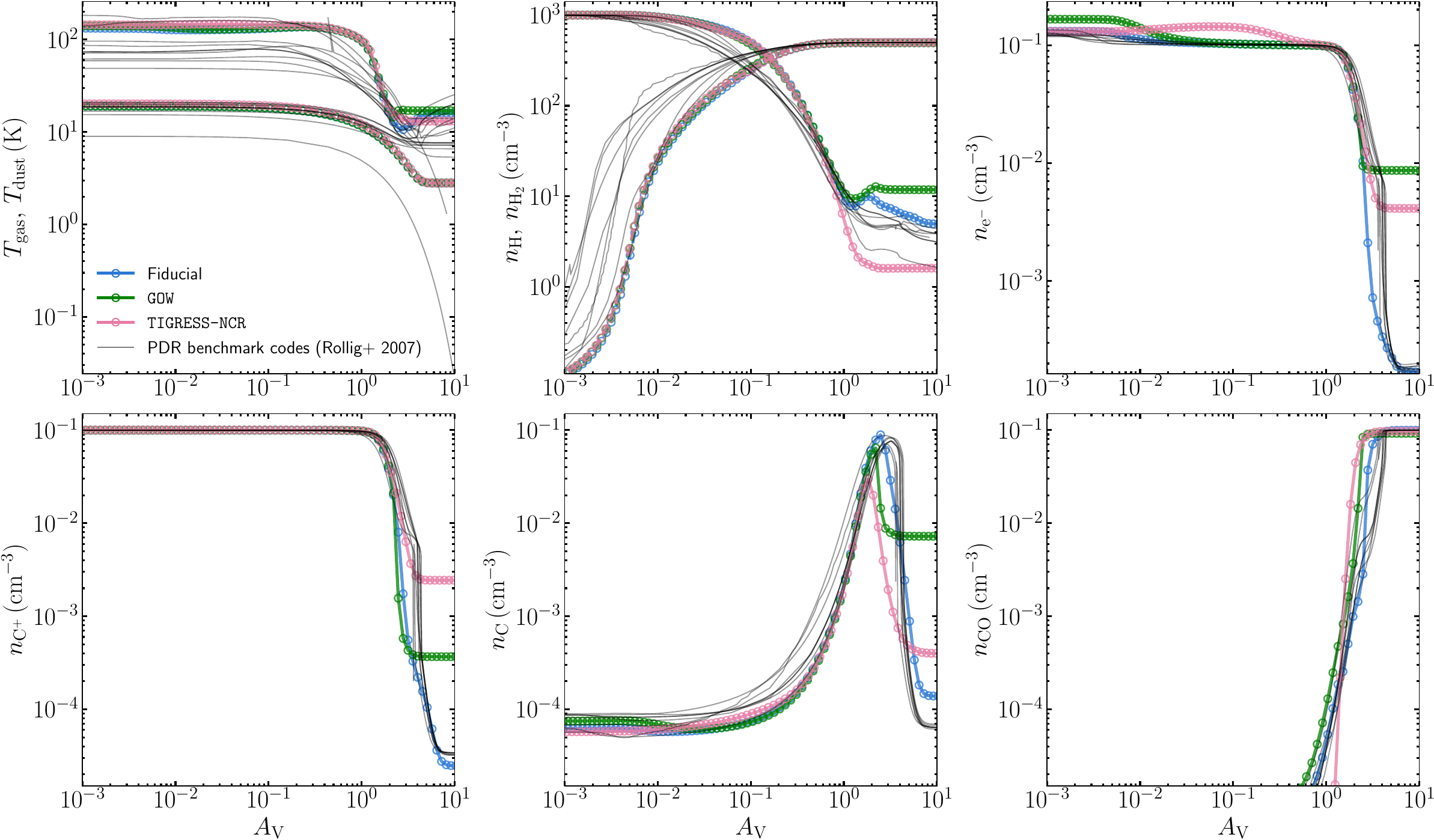}
    \caption{Comparison of the three chemical networks we have implemented -- this work, \texttt{GOW} \citep{Gong_2017}, and \texttt{TIGRESS-NCR} \citep{kim2023_tigress} -- for the V1 configuration of the \citet{Rollig_2007} PDR benchmark ($n = 10^3\,\mathrm{cm^{-3}}$, $\chi_{\rm FUV} = 10$). Panels show, from left to right and top to bottom: gas and dust temperature, H/H$_2$, electron, C$^+$, C, and CO number densities as a function of visual extinction $A_{\rm V}$. Thin grey lines show the published PDR codes from the benchmark (see Figure~\ref{fig:RolligTesta} for sources). All three networks obtain values similar to each other and the published benchmark codes, with noticeable differences seen mostly at high $A_{\rm V}$ due to missing reaction pathways in the reduced networks (see text).}
    \label{fig:NetworkCompare}
\end{figure*}

In addition to our fiducial network, we also implement chemical networks with a reduced number of species, namely the networks presented in \citet[][\texttt{TIGRESS-NCR}]{kim2023_tigress} and \citet[][\texttt{GOW}]{Gong_2017}. The \citet{kim2023_tigress} network contains 3 non-equilibrium species, and was designed with the essential components to accurately capture the thermal evolution of the ISM and photon-mediated $\Hplus$-HI-$\Htwo$ transitions, however, with simplified/equilibrium approaches for the $\CII$-$\CI$-CO cycle. This is similar in level of complexity to the commonly used \citet{Nelson_1997} network for CO formation, although with differences in practice for the treatment of carbon species. \citet{Gong_2017}, on the other hand, contains 18 species and captures the carbon cycle and CO formation pathways explicitly, albeit with the use of pseudo-species for the intermediaries  -- i.e. hydrocarbon species such as $\rm CH$, $\rm CH_2$, $\rm CH^+$, $\rm CH_2^+$ and $\rm CH_3^+$ are combined into a single (pseudo-)species $\rm CH_{x}$, and $\rm OH$, $\rm H_2O$, $\rm OH^+$, $\rm H_2O^+$, and $\rm H_3O^+$ into $\rm OH_{x}$. This network is similar in complexity to the \citet{Nelson_1999} network used in some works, although with improvements that significantly improved the accuracy of CO abundances at higher CR rates. 

Our fiducial network, containing 37 species, explicitly considers the non-equilibrium evolution of intermediary species involved in the CO formation pathways. Therefore, the three networks we have implemented span the range of complexities of chemical networks used in the literature. Since complexity comes with additional computational cost, it is critical to evaluate whether the advantages of more involved networks are worth the additional cost for the problem under consideration. For instance, the explicit treatment of OH and $\rm H_2O$, which, as we established, are important at low $Z$ and dense gas, offers a clear advantage for problems modeling such conditions (e.g., low-$Z$ star formation), or those focused on predicting the abundances of these molecules in the ISM to compare with observations \citep[e.g.,][]{Rawlins_2023}. However, it is also entirely possible that there are non-negligible differences in the $\CII$-$\CI$-CO abundances due to the presence of additional pathways for their formation-destruction in more involved networks. It is therefore informative to compare the outcomes for common species considered by the networks. 

We repeat the V1 test of the \citet{Rollig_2007} PDR benchmark comparison project we performed in Section~\ref{sec:RolligPDR} with the reduced networks. This is useful as the dedicated PDR codes themselves show a noticeable degree of scatter in outcomes, and therefore, only differences in the reduced networks beyond this level of scatter are meaningful. The test setup is identical to that described in Section~\ref{sec:RolligPDR}\footnote{In the \citet{Gong_2017} network, we set the Si abundance to zero since the participating codes in the Rollig PDR benchmark did not include Si for the purposes of the test.}. In Figure~\ref{fig:NetworkCompare}, we compare the gas and dust temperatures, number densities of $\HI$-$\Htwo$, $\elec$, and of $\CII$-$\CI$-CO as a function of $A_{\rm V}$; also overplotted are the curves obtained by the PDR codes that participated in the benchmark. We can see that gas/dust temperatures are virtually identical between the runs, reassuring as this is of primary concern for the coupling to hydrodynamics\footnote{We note however that this may not be the case at lower $Z$ where, as we've discussed (Figure~\ref{fig:CollapsePhysics}), the temperature is more sensitive to the chemical state of the gas and driven by coolants (e.g., $\rm H_2O$, $\rm OH$) not considered by these alternative networks.}. The $\Htwo$ abundances are also virtually identical between the networks. Noticeable differences are seen for the other species, however, especially at high $A_{\rm V}$. We find most of this can be accounted for by missing gas-phase reactions in the reduced networks. For instance, the $\HI$ excess in the \citet{Gong_2017} network arises from the lack of certain gas-phase OH reactions -- the radiative association reaction $\rm O + H \rightarrow OH$ is the \emph{dominant} destruction channel for atomic hydrogen at these depths. Similarly, the $\elec$ at high $A_{\rm V}$ abundance can be explained by differences in the $\HII$ abundances (primary contributor of electrons at high $A_{\rm V}$), driven by the lack of charge transfer destruction channels ($\rm OH + H^+ \rightarrow OH^+ + H$ and $\rm H_2O + H^+ \rightarrow H + H_2O^+$) in the reduced networks. Similarly, for $\CII$ we found that reactions with $\rm O_2$ explain the differences with this work. Nevertheless, these differences occur where these species are \emph{subdominant} and their abundances are relatively small, and should not affect integrated column densities, for instance. 

On the other hand, differences persist in $\CI$ (and CO) abundances, particularly at intermediate/high $A_{\rm V} \sim 1$--3, where it is the dominant carrier of C nuclei. As the network complexity decreases, the transition from $\rm CI$ to CO occurs at increasingly lower $A_{\rm V}$ with increasingly thinner $\rm CI$-dominant layers than the reference PDR models. Corresponding differences are seen in the CO abundances in this layer, which, however, disappear at higher $A_{\rm V} \gtrsim 4$ where CO dominates. Therefore, our analysis indicates that the $\CI$-layer of the PDR is most sensitive to the level of completeness of the chemical network. This is understandable as the $\rm CI$ to CO transition is facilitated by gas-phase reaction pathways involving hydroxyl and hydrocarbon ions, with their rate-limiting steps often involving reactions with $\rm H_3^+$. The reduced networks either omit (\texttt{TIGRESS-NCR}) or simplify (\texttt{GOW}) precisely the reactions involving these species. 

\subsection{Cost-accuracy tradeoff}
\label{sec:cost-accuracy}

In the previous section, we discussed the improved accuracy offered by our fiducial network over reduced networks in the thermal evolution of low-$Z$ dense gas (by capturing OH/$\rm H_2O$ explicitly) and in the abundances of certain PDR tracers (e.g., $\CI$). However, the increased number of non-equilibrium species considered by the network ($N_{\rm species}$) comes with the additional computational cost of solving a larger system of coupled ordinary differential equations (ODEs). Since the stiffness of the system necessitates implicit solvers, the cost can scale steeply with $N_{\rm species}$, with worst-case scenarios implying $\propto N_{\rm species}^3$ for dense Jacobians. However, chemical networks are relatively sparse, which the \texttt{DLSODES} solver utilized by \texttt{KROME} takes advantage of \citep{2013MNRAS.431.1659G}, providing a significant speed-up over commonly used solvers that do not -- for e.g., \texttt{CVODE/DVODE} provided by the \texttt{SUNDIALS} package \citep{sundials}, with the degree of speed-up depending on the sparsity pattern. However, this does not necessarily reflect the absolute efficiency of the two solvers, as tailored implementations or optimizations may yield different levels of efficiency regardless of the specific solver. In addition, the inclusion of some species/pathways may make the network more stable and permit the solver to dynamically choose larger substeps within the implicit update. 

\begin{figure}
\centering
\includegraphics[width=\columnwidth]{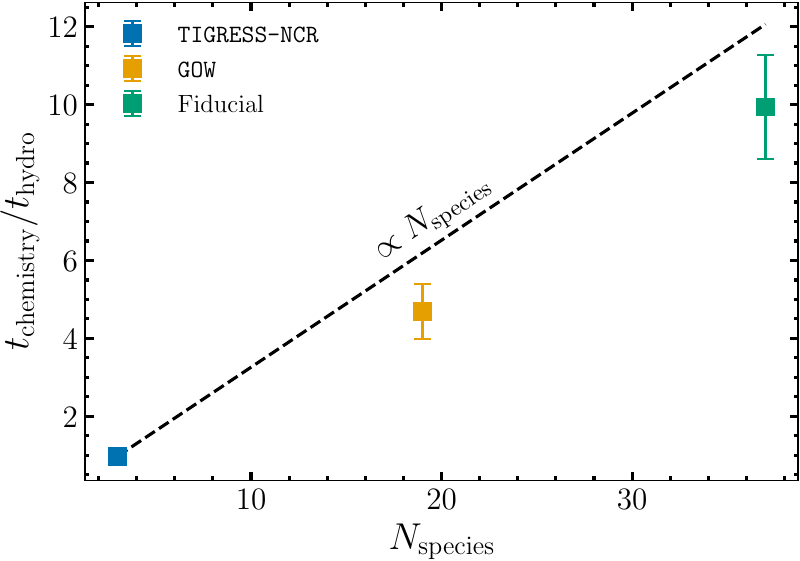}
\caption{Cost incurred by chemistry relative to hydrodynamics per timestep for the 3D turbulent patch simulation of Section~\ref{sec:turbulent_sim}, compared for the chemical networks we implement: \texttt{TIGRESS-NCR} \citep{kim2023_tigress}, \texttt{GOW} \citep{Gong_2017} and the fiducial network presented in this work. Errorbars indicate the $1\sigma$ variation across timesteps. Comparison with the dashed line shows that we obtain (sub-)linear scaling with the number of non-equilibrium species ($N_{\rm species}$).}
\label{fig:CostScaling}
\end{figure}

To quantify the cost-accuracy tradeoff, in Figure~\ref{fig:CostScaling} we compare the cost of solving thermochemistry relative to hydrodynamics for the 3D driven turbulent medium (at $\Zsun$) of Section~\ref{sec:turbulent_sim} with the three different networks. We can see that our fiducial network adds a factor $\sim 10$ cost over hydrodynamics, with the cost progressively lower for \texttt{GOW} ($\sim 5$) and \texttt{TIGRESS-NCR} ($\sim 1$). Comparison with the $\propto N_{\rm species}$ dashed line suggests that the cost scales sub-linearly, which is significantly shallower than the $\mathcal{O}\left(N_{\rm species}^{3}\right)$ expected without sparse factorizations. While this points to the efficiency of our solvers, our fiducial network still incurs a slowdown of a factor of $\sim 10$. Whether this trade-off is worth it depends on the range of conditions realized in the simulation, the underlying scientific questions, and the level of accuracy of a given species required for the problem (e.g., Figure~\ref{fig:NetworkCompare}). This problem-dependence of the cost-accuracy is  precisely our motivation for designing the implementation with a flexible choice of networks.

\section{Summary and Outlook}
\label{sec:Summary}

We present a flexible implementation for the thermal evolution and photon/CR-mediated non-equilibrium chemistry of hydrogen, carbon, and oxygen species relevant in the neutral ISM. We embed the thermochemical network and solve for it using the \texttt{KROME} framework, coupled to the radiation-magnetohydrodynamics \texttt{FLASH} code. Our implementation utilizes \texttt{VETTAM}\footnote{\url{https://github.com/shm-1996/VETTAM}} \citep{Menon_2022_VETTAM}, a variable Eddington tensor-closed radiation hydrodynamics scheme that accurately retains the non-locality of the stellar/diffuse radiation field for photon-mediated reactions and dust cooling \citep{Menon_2022_VETTAM}. Our fiducial chemical network consists of 37 chemical species, designed to capture the formation pathways of \emph{all} relevant species that may be critical to the thermal evolution of (neutral) gas at density ($n<10^{17} \pcm$) and/or metallicity ($\Zsun$ to metal-free): specifically, $\CI$, $\CII$ and $\OI$, CO ($Z\gtrsim 0.01 \Zsun$), OH and $\rm H_2O$ ($Z\lesssim 0.01 \Zsun$) and primordial coolants (including deuterium), along with dust-mediated cooling. In addition to this, we also implement alternate chemical networks with a reduced number of species -- that of \citet{Gong_2017} and \citet{kim2023_tigress} -- that are more cost-effective ($\propto N_{\rm species}$) and may be sufficient for certain problems restricted to a subset of $n$ and/or $Z$ and depending on the problem of interest (Section~\ref{sec:cost-accuracy}). 

We show that our scheme captures the multiphase evolution of neutral gas across $Z$ (Section~\ref{sec:ismphase}), its thermal evolution to prestellar densities ($n \sim 10^{17} \pcm$; Section~\ref{sec:collapse}) and the chemical abundances of the aforementioned important coolants with comparable accuracy as dedicated (static) PDR codes across a wide range of conditions (Section~\ref{sec:RolligPDR}). We also demonstrate the working of our scheme with example 3D applications: i) the evolution of multiphase turbulent media of $\langle n\rangle = 100 \pcm$ at various metallicities ($\Zsun$ -- $10^{-4} \Zsun$), and ii) the expansion of $\HII$  regions and PDRs into these turbulent media driven by a $1000 M_{\sun}$ star cluster. We show that weaker cooling at lower $Z$ imposes higher gas temperatures, less pronounced density fluctuations in turbulent gas, and lower densities in compressed PDR shells (Figure~\ref{fig:TurbHIIRadialProfile}). We also find that, consistent with PDR model expectations, the dominant fine-structure line coolant transitions from $\OI$ in the warmer (inner) edge of the PDR to $\CII$ deeper in the PDR, with the $\OI$-dominant region increasingly more volume-filling at increasingly lower $Z$ until it eventually dominates everywhere at $Z<0.01 \Zsun$. Importantly, we demonstrate from our simulations that the abundances of some of these species are out of chemical equilibrium due to turbulent mixing and dynamics, which we show alleviates the long-standing issue of (static) PDR models overestimating $\CI$ abundances compared to UV absorption-line measurements of the diffuse ISM (Figure~\ref{fig:CarbonColumnsTurb}). 

We expect that the ability of our implementation to handle a broad range of physical conditions, and its modular and flexible design will facilitate a wide range of applications for modeling the dynamical evolution of the ISM across cosmic time. For instance, through the \texttt{POPSICLE} project, we plan to apply it to study the resolved collapse and fragmentation of dense clouds into stars and (potentially) black holes and their interplay with their radiative feedback, with a specific emphasis on the metal-poor Universe \citep{2025MNRAS.540.1745S}. Our scheme also facilitates straightforward predictions for atomic/molecular observational tracers of the ISM while retaining the non-equilibrium nature of their formation/destruction pathways, which we have demonstrated may have implications for them. Looking forward, we envision making modeling improvements to address existing limitations of our implementation: for instance i) replacing the fixed dust-to-gas and composition/grain size model assumption with on-the-fly dust and polycyclic aromatic hydrocarbon (PAH) evolution coupled to thermochemistry \citep[e.g.,][]{CALIMA_2026} -- ii) chemistry of molecular ices that freeze-out on dust grains (e.g., CO and $\rm H_2O$) currently missing in our network with implications for dense, cold gas \citep{Klessen_Glover_2016}, iii) inclusion of nitrogen chemistry relevant for several dense gas tracers in the ISM (e.g., HCN -- \citealt{2026MNRAS.547ag527P}), and iv) explicit CR-injection and transport with a moment-based method \citep[e.g.,][]{Jiang_2018,2020MNRAS.491..993G,Hopkins_2022} instead of fixed backgrounds. 

\begin{acknowledgments}
We thank Amiel Sternberg for extensive discussions that guided our approach to this implementation. We also thank Simon Glover, Lachlan Lancaster, Michael Grudić, Julianne Dalcanton, Shivan Khullar, Jacqueline Hodge, Karin Cescon, Munan Gong, Jeong-Gyu Kim, Shmuel Bialy, Mark Krumholz, Eloy van de Genugten, Marco Padovani, and Danielle Galli for several productive discussions and suggestions. PS is supported by the Leiden University Oort Fellowship and the International Astronomical Union -- Gruber Foundation (TGF) Fellowship. We acknowledge high-performance computing resources provided by the Simons Foundation as part of the Centre for Computational Astrophysics at the Flatiron Institute. The Flatiron Institute is a division of the Simons Foundation. This research has made extensive use of NASA’s Astrophysics Data System (ADS) Bibliographic Services. 
\end{acknowledgments}

\software{\texttt{KROME} \citep{grassi_krome_2014}, \texttt{VETTAM} \citep{Menon_2022_VETTAM}, \texttt{PETSc} \citep{PetscConf,PetscRef}, \texttt{DESPOTIC} \citep{Krumholz_2014}, \texttt{yt} \citep{yt}, \texttt{Matplotlib} \citep{matplotlib}, \texttt{Numpy} \citep{numpy}, \texttt{TurbGen} \citep{Federrath_2022}, \texttt{HEALpix} \citep{Gorski_2005}}


\appendix

\section{Thermodynamics}
\label{sec:appendix_thermodynamics}

We include the relevant heating and cooling mechanisms for the thermal balance in the ISM, and solve the resulting ODE for the gas temperature in a way that it is fully coupled to the rate equations for the chemical species tracked in the network. The ODE for the gas temperature can be expressed as 

\begin{equation}
    \frac{d\Tg}{dt} = \frac{\gamma -1}{nk_{\rm B}}\left(\Gamma_{\rm heat} - \Lambda_{\rm cool} \right),
\end{equation}
where $\Gamma_{\rm heat}$ and $\Lambda_{\rm cool}$ are the total heating and cooling rates per unit volume (in units of $\ergpcm \rm s^{-1}$), $\gamma$ is the ratio of specific heats of the gas, and $n$ the total number density. Below, we delineate the individual heating and cooling terms we include in our scheme.

\subsection{Heating}

The total heating rate is given by
\begin{equation}
    \Gamma_{\rm heat} = \Gamma_{\rm PI} + \Gamma_{\rm LW,H_2} + \Gamma_{\rm PE} + \Gamma_{\rm CR} + \Gamma_{\rm chem}.
\end{equation}
The individual terms are as follows:
\begin{itemize}
    \item Photoionization heating ($\Gamma_{\rm PI}$): Any excess energy in photons with energies $>13.6 \, \rm eV$ beyond the ionization potential of H (13.6 eV) and \Htwo\ (15.4 eV) can be absorbed by free electrons resulting from photoionization, which subsequently thermalise and heat the gas. The heat deposited per ionization event, therefore, depends on the ionizing spectrum to which the gas is exposed. For simplicity, we use a constant value of 3.4 eV and 4.4 eV of thermal energy per ionization event of H and \Htwo\ respectively, corresponding to SED-weighted averages over the lifetime of a fully-IMF-sampled stellar population \citep{kim2023_tigress}; it is however trivial to change this constant value to emulate a different spectrum (like a massive star) or dynamically evolve this value in space and time depending on the instantaneous spectrum to which a gas parcel is exposed.

    \item \Htwo\ LW heating ($\Gamma_{\rm LW,H_2}$): The absorption of LW photons by \Htwo\ can excite the molecule to higher electronic states, which can subsequently either decay to the vibrational continuum (i.e. photodissociation) or cascade down the vibrational level ladder back to the ground state. Each photodissociation event yields 0.4 eV of thermal energy when averaged over the series of possible excitation-decay pathways to the continuum permitted by Lyman-Werner excitations \citep{Black_Dalgarno_1977}. On the other hand, the cascade down the vibrational levels is radiative in nature in low-density gas, but can instead undergo collisional-deexcitation for $n \gtrsim n_{\rm crit} \sim 10^4 \, \rm{cm}^{-3}$ \citep{Sternberg_Dalgarno_1989}, thereby heating the gas, and depositing $\sim$2eV of energy for $n>>n_{\rm crit}$ \citep{Burton_1990}. The total heating rate can then be written as 
    \begin{equation}
    \label{eq:UVPumping}
        \Gamma_{\rm LW,H_2} = \dot{n}_{\rm diss,H_2} \left ( 0.4 + 2f_{\rm pump} (1+n_{\rm crit}/n)^{-1}  \right) \rm eV,
    \end{equation}
    where $\dot{n}_{\rm diss,H_2}$ is the LW photodissociation rate per unit volume, $f_{\rm pump} = 8$ is the inverse ratio of the number of LW-pumped events that lead to photodissociations \citep{Sternberg_2014}. $n_{\rm crit} = A_{\Htwo^*}/(k_{\rm H_2-H} x_{\rm H} + k_{\rm H_2-H_2} x_{\rm H_2})$ is the critical density for the vibrational levels, considering the effective radiative de-excitation rate over the vibrational ladder of $A_{\Htwo^*} = 2 \times 10^{-7} \, \rm s^{-1}$, and $k_{\rm H_2-H}$, $k_{\rm H_2-H_2}$ the collision rates with H and \Htwo\ respectively; these values are taken from Table 7 of \citet{Burton_1990}.

\item Photoelectric heating ($\Gamma_{\rm PE}$): Dust grains in the ISM can absorb FUV photons (6 -- 13.6 eV), ejecting electrons on their surface with non-zero kinetic energies through the photoelectric effect, with this kinetic energy corresponding to the excess energy available in the photon over its work potential. The physics implies that the heating rate depends on the strength of the radiation field and the work potential of the grains, which itself depends on the size, composition and charge of the grains. \citet{Weingartner_2001a} provide characteristic heating rates in the form 
\begin{equation}
    \label{eq:PE_heatingrate}
    \Gamma_{\rm {PE}} = 10^{-26} G_0 \epsilon_{\rm PE}n_{\rm H} Z_{\rm d}^{'},
\end{equation}
where $G_0 = 1.69 \chi_{\rm FUV}$ is the FUV field strength in Habing units and $Z_{\rm d}^{'}$ is the dust-to-gas ratio relative to the Solar neighborhood. $\epsilon_{\rm PE}$ is the heating efficiency that encapsulates the dependence of the work potential on the grain properties and charge. $\epsilon_{\rm PE}$ depends on the so-called grain charging parameter $\psi = G_0\sqrt{\Tg}/n_e$ and $\Tg$, through the functional form described in Eq. 44 of \citet{Weingartner_2001a} ; we use their model that adopts the ratio of visual extinction to reddening $R_{\rm V} = A_{\rm V}/E(B-V) = 3.1$, C-abundance in grains of $b_{\rm C}=4 \times 10^{-5}$ and the ISRF background spectral shape (see their Table 2). Following \citet{kim2023_tigress}, we add a value of $50 \rm K^{1/2} cm^3$ to $\psi$ to avoid numerical issues that arise if $\Psi$ becomes very low in strongly shielded regions. 

\item Cosmic-Ray heating ($\Gamma_{\rm CR}$): Cosmic-rays can effectively ionize atom/molecules, with the excess energy over the ionization potential carried by the free electron, some fraction of which can lead to secondary ionizations through collisions, and the rest thermalizing via Coulomb interactions, thereby heating the gas. In molecular regions, additional pathways of heating apart from Coulomb interactions are available: such as the collisional excitation of $\Htwo$ molecules (similar to UV-pumping) by CR-produced $\elec$, and exothermic chemical reactions induced by CRs \citep{Glassgold_2012}. CR-heating becomes the dominant heating mechanism in the ISM in regions of high $A_{\rm V}$ and even in the diffuse ISM at low-$Z$, due to the lack of FUV photons and dust grains respectively. We adopt a heating rate per primary-ionization event of H\footnote{The primary CR-ionization rate of H, $\xi_{\rm p,H} = \xi_{H}/(1+\phi_s)$ where $\xi_{\rm H}$ is the total CR ionization rate of H and $\phi_s$ the number of secondary ionization events per primary ionization, for which we use Equation 38 in \citet{Bialy_Sternberg_2019}.}, $E_{\rm cr,H}$ from \citet{Bialy_Sternberg_2019} (Equation 39) given by 
\begin{equation}
    E_{\rm cr,H}=6.43\left(1+4.06\left[\frac{x_{\mathrm{e}}}{x_{\mathrm{e}}+0.07}\right]^{1 / 2}\right) \mathrm{eV}.
\end{equation}
To account for the additional pathways of heating in molecular gas, we adopt the density-dependent fit to the heating rate $E_{\rm cr,H_2}$ provided by \citet{Krumholz_2014} using data from \citet{Glassgold_2012}, and adopt an effective total heating rate (per primary ionization of H) of $E_{\rm cr} = E_{\rm cr,H} x_{\rm H} + E_{\rm cr,H_2} x_{\rm \Htwo}$ following several works in the literature \citep{Krumholz_2014,Gong_2017,kim2023_tigress}.

\item Chemical heating ($\Gamma_{\rm chem}$): Certain chemical reactions can be exothermic (endothermic) in nature and release (absorb) energies that are non-negligible to the thermal balance of the gas. The most prominent of these is the endothermic formation of $\Htwo$ from H-atoms catalyzed by dust grains; a large part of the energy released in the reaction goes into exciting the rovibrational modes of the newly formed $\Htwo$ molecule. As described earlier, at higher densities these excited modes can undergo collisional de-excitation and heat the gas; the total amount of energy released per $\Htwo$ molecule formed on dust can be given by \citep{Hollenbach_1979}
\begin{equation}
    E_{\rm \Htwo,form} = \left( 0.2 + \frac{4.2}{1+n_{\rm crit}/n} \right) \rm eV,
\end{equation}
where $n_{\rm crit}$ is the critical density of $\Htwo$ as defined above for Equation~\ref{eq:UVPumping}. We also include the heating from the three-body formation of $\Htwo$, and cooling from the collisional dissociation of $\Htwo$ (both with H and $\Htwo$) both of which primarily operate at very high densities; we assume a constant value of 4.48 eV is released/removed to the thermal energy of the gas per reaction. 
\end{itemize}

\subsection{Cooling}
\label{s:app_a2}
The total cooling rate in our scheme is given by 

\begin{align}
    \Lambda_{\rm cool} = \ & \Lambda_{\rm prim} + \Lambda_{\rm ff} + \Lambda_{\rm H_2} + \Lambda_{\rm HD} + \Lambda_{\rm Compton} + \Lambda_{\rm cont} \nonumber  \\
    &+ (1-f(T))\Lambda_{\rm Z} + f(T)\Lambda_{\rm CIE},
    \label{eq:totalcooling}
\end{align}
where 
\begin{equation}
    \Lambda_{Z} = \Lambda_{\rm CI+CII+OI} + \Lambda_{\rm CO} + \Lambda_{\rm OH} + \Lambda_{\rm H_2O} + \Lambda_{\rm gr,rec} + \Lambda_{\rm gd} + \Lambda_{\rm neb},
    \label{eq:metalcooling}
\end{equation}
is the total cooling rate from metal species explicitly tracked by our network and/or assuming a specific composition of metal ionization states (for nebular cooling). Since our chemical network and the metal species/ionization states we track are designed for explicitly modelling the cold/warm phases of the ISM, we cannot directly obtain cooling in hot gas from our species abundances. Instead, we transition to a tabulated cooling rate obtained by assuming CIE for gas with temperatures $\Tg > 3.5 \times 10^4 \, \rm K$. To transition smoothly to this tabulated rate in hot gas, we follow \citet{kim2023_tigress} and use a sigmoid functional form $f(T)$ given by
\begin{equation}
    f(T) = \frac{1}{1+\exp\left({- 10(\Tg - 0.5 (T_{1} + T_2}))/(T_2-T_1) \right)},
\end{equation}
where $T_1 = 2\times 10^4 \, \rm K$ and $T_2 = 3.5\times 10^4 \, \rm K$ represent the gas temperatures between which the transition occurs smoothly. This design choice ensures that the thermal state of the gas is fully consistent with its dynamically computed chemical state for $T_{\rm gas} \lesssim T_1$, while still offering an approximate approach to modelling the temperature of hot gas.

We describe all of the individual cooling terms below.

\begin{itemize}
    \item Primordial cooling ($\Lambda_{\rm prim}$): These include the relevant cooling pathways for the primordial species tracked by our networks. They include i) cooling via collisional excitation of atomic Hydrogen (Lyman-$\alpha$ cooling\footnote{Note that although the terminology of Lyman-$\alpha$ is used in this context, the cooling includes de-excitations over all energy levels, not just the Lyman-$\alpha$ line (which dominates the cooling) or the Lyman series.}) and Helium (both neutral and singly ionized), ii) radiative recombination of ionized H/He, dielectronic recombination of singly ionized He, and iii) collisional ionization cooling -- i.e. from ionization of H and He by collisions with electrons. We use the cooling rate coefficients for each of these pathways from \citet{Cen_1992}. 

    \item Free-free cooling ($\Lambda_{\rm ff}$): Cooling due to Bremsstrahlung emission induced by the interaction of free electrons with all ions. We use a mean Gaunt correction factor of 1.5, and adopt the cooling rate from \citet{Osterbrock_1989}.

    \item \Htwo\ cooling ($\Lambda_{\rm H_2}$): We consider cooling due to rovibrational transitions of the \Htwo\ molecule excited by collisions with H, \Htwo\ (ortho and para), \Hplus, He, and $\elec$; at high densities the cooling saturates to the value determined by the level populations in local thermodynamic equilibrium (LTE), and we interpolate between the low- and high-density limits using the approach outlined in \citet{Galli_1998}. The collision strengths are obtained from \citet{Glover_Abel_2008} using a constant ortho-to-para ratio of 3:1, while incorporating the updated values for some collision rates reported in \citet{Glover_2015} (Appendix A.1.2); for the high-density limit we use the expression from \citet{Hollenbach_1979} with high-temperature corrections from \citet{Glover_2015}. We also correct for a decrease in the cooling rate due to radiative trapping in the optically-thick limit, for which we use the escape probability formalism with opacities provided by \citet{Omukai_2010}.

    \item HD cooling ($\Lambda_{\rm HD}$): The permanent electric dipole moment of the HD molecule permits higher radiative transition probabilities between rotational levels than $\rm H_{2}$, and therefore can be an important source of cooling for primordial/very-low metallicity gas, especially at intermediate densities ($10^4 \lesssim n \lesssim 10^6 \, \rm cm^{-3}$) and low temperatures ($T \lesssim 300 \rm K$) when the HD abundance can be higher and $\Htwo$ cooling becomes ineffective \citep[e.g.,][]{Galli_1998}. We use the tabulated cooling function provided by \citet{Lipkova_2005} for HD cooling.

    \item Compton cooling ($\Lambda_{\rm Compton}$): We include the cooling (or heating, at very high $z$) of gas due to (Thomson) interactions of free electrons with the background CMB radiation field; this is only relevant at very high $z$, and we include it simply for completeness. We use the cooling rate from \citet{Cen_1992} for the cooling rate.

    \item Fine-structure metal line cooling ($\Lambda_{\rm CI+CII+OI}$): Electron transitions between fine-structure energy levels (from  spin-orbit coupling) of metal species are critical cooling pathways for interstellar gas; we include the most important of these -- i.e. from CI, OI and CII. We follow the approach outlined in \citet{Maio_2007}, i.e. assuming statistical equilibrium to solve the resulting system of equations for the occupation fractions of energy levels, considering CI and OI (CII) as 3-level (2-level) systems respectively, and considering H, He, $\Htwo$ (ortho and para) and $\elec$ as collision partners for excitation/de-excitation; see \citet{Maio_2007} and \citet{Glover_Jappsen_2007} for the relevant atomic data. In regions where the CI, CII, or OI lines become optically thick, we apply optical depth corrections to the cooling rates using the local velocity gradient \citep[LVG,][]{Neufeld_1993} approximation. We use the radiative transfer package \texttt{DESPOTIC} \citep[][]{Krumholz_2014} to precompute the corresponding escape probability correction factors as a function of gas temperature and column density of the species and interpolate on this tabulated data.

    \item CO cooling ($\Lambda_{\rm CO}$): In dense molecular gas, cooling is dominated by the rotational lines of the CO molecule, some of which can be optically thick. We use the LVG approximation to account for this in the cooling rates, the expression for which involves fitted data reported in tabulated form in \citet{Omukai_2010} (Appendix B). Following \citet{Glover_2010} and \citet[][equation 34]{Gong_2017}, we also modify the cooling rate expression to account for rotational lines excited by collisions with H and $\elec$ in less dense gas, whereas \citet{Omukai_2010} considered only collisions with $\Htwo$.

    \item OH cooling ($\Lambda_{\rm OH}$): Rotational transitions of the OH molecule can be important for high density gas, especially at lower metallicities where it becomes the dominant molecule (among metals) over CO \citep[e.g.,][]{Omukai_2010,Bialy_2015}. Similar to CO, we adopt tabulated data from \citet{Omukai_2010} for the cooling rates that include a correction for the optical depth using the LVG approximation, considering rotational lines excited via collisions with $\Htwo$ and H.

    \item H$_2$O cooling ($\Lambda_{\rm H_2O}$): The large number of rovibrational lines accessible to the H$_2$O molecule, and their relatively high critical densities (compared to CO) can make it an important coolant in highly dense gas\footnote{Although, in practice, freeze-out on dust grains limits its abundance in the gas-phase, and therefore its role in the thermal balance of dense gas \citep{Klessen_Glover_2016}} \citep[e.g.,][]{Neufeld_1993}, especially at lower metallicities \citep{Omukai_2010}. We use the cooling rate for H$_2$O adopted by \citet{Omukai_2010}, which uses the rates from \citet{Neufeld_1993} for $T<100$ K and \citet{Neufeld_1995} for $T>100$ K, corrected for high optical depths using the LVG approximation. 

    \item Continuum cooling ($\Lambda_{\rm cont}$): We include an effective cooling rate accounting for a range of cooling processes owed to the gas continuum -- see \citet{Omukai_2000}. This cooling becomes prominent at high densities due to the consideration of continuum cooling imposed by the radiation emitted due to Van der Waals interaction of $\Htwo$ molecules (so-called collision-induced emission), which dominates the cooling at extremely high densities $n \gtrsim 10^{13} \rm cm^{-3}$ \citep{Omukai_2000}.

    \item Grain-facilitated recombination cooling ($\Lambda_{\rm gr,rec}$): Ions colliding with small dust grain and PAH molecules can recombine with free electrons on the surface of the grains, and recombine. Apart from being of equal footing as radiative recombination in changing the ionization state of the gas \citep{Weingartner_2001c}, the associated cooling can be important in warm gas at $T\sim 10^4 \rm K$. We use the cooling rate following \citet{Weingartner_2001c} using the parameters of their model (see Table 3) that adopts the ratio of visual extinction to reddening $R_{\rm V} = A_{\rm V}/E(B-V) = 3.1$, C-abundance in grains of $b_{\rm C}=4 \times 10^{-5}$ and the ISRF background. 

    \item Dust-gas interaction ($\Lambda_{\rm gd}$): Inelastic collisions between dust grains and gas, and the associated energy transfer between them can act as a crucial heating/cooling mechanism for the gas in dense gas ($n \gtrsim 10^5 \rm cm^{-3} Z_{\rm d}^{'}$). We use the cooling rate expression $\Lambda_{\rm gd} = \alpha_{\rm gd} \Tg^{0.5} (\Tg - \Td)Z_{\rm d}^{'}$ where we adopt a constant value of $\alpha_{\rm gd} = 3.2 \times 10^{-34} \rm erg \, cm^3\, s^{-1} \, K^{-3/2}$ for the collision rate coefficient \citep{Krumholz_2014}. We elaborate on our approach to determine $\Td$ from the physical state of the system in Section~\ref{sec:dusttemp} below.

    \item Nebular metal cooling ($\Lambda_{\rm neb}$): The cooling in photoionized gas is largely dominated (at $Z \gtrsim 0.1 \Zsun$) by collisionally excited optical/infrared lines (electronic/fine-structure) from ions of metal species such as O, N, S, among others. The cooling rate therefore depends on the ionization states of these species, which itself depends on the density, temperature and the spectrum of the radiation field. Since the focus of our implementation is the thermochemistry of cold gas, we adopt a simplified approach by using the fitting function for the cooling rate from untracked ions in our chemical network (i.e. apart from CII) given in \citet{kim2023_tigress}\footnote{The fitting function reproduces the cooling rate for a range of gas densities ($1 \lesssim n \lesssim 10^6 \, \rm cm^{-3}$ and temperatures ($2000 \lesssim \Tg \lesssim 5 \times 10^4 \, \rm K$) assuming a fixed fraction of 80\% of O, N, and Ne, and 50\% of S, in their singly ionized forms, and the rest in the doubly ionized forms.}. Our simplified approach here, while convenient and suited to reproducing the broad thermal properties of HII regions, cannot produce the detailed thermal/chemical structure within HII regions associated with the local hardness of the radiation field; we refer the reader to other efforts in the literature where this is attempted \citep[e.g.,][]{Sarkar_2021,Katz_2022}

    \item CIE cooling ($\Lambda_{\rm CIE}$): We include cooling due to collisionally excited cooling from metal ions in hot gas ($\Tg > 2 \times 10^4 \, \rm K$, assuming the metal ions are in CIE). We use the cooling rate reflecting this assumption, tabulated by \citet{Gnat_2012} for the metal ions from \texttt{CLOUDY} calculations \citep{Ferland_1998}. Cooling from primordial ions (i.e. H/He) and free-free ($\elec$) are treated explicitly as mentioned earlier.
\end{itemize}

\subsection{Computing the dust temperature}
\label{sec:dusttemp}

We solve for the dust temperature $T_{\rm dust}$ by assuming they are in instantaneous thermal equilibrium -- a reasonable assumption due to their high heat capacities -- by considering cooling due to thermal dust emission, and heating from absorption of background UV/IR and CMB photons, and gas-dust collisions. This results in
\begin{equation}
\dot{u}_{d, \mathrm{IR}}=\dot{u_{r, \mathrm{UV}}} + \dot{u}_{r, \mathrm{CMB}}+n_{\rm H}^2 \Lambda_{g-d}, 
\label{eq:dusttemp}
\end{equation}
where $\dot{u}_{d, \mathrm{IR}}$ is the cooling due to reprocessed IR radiation energy emitted per unit volume by dust, competed by $\dot{u}_{r, \mathrm{UV}}$, $\dot{u}_{r, \mathrm{CMB}}$, the corresponding volumetric heating rates as the dust absorbs UV, (dust-reprocessed) and CMB photons. $\dot{u}_{d, \mathrm{IR}}$ can be given by $\dot{u}_{d, \mathrm{IR}} = \rho \kappaP c \ar \Td^4Z_d^{'}$, where $\kappaP$ is the Planck-mean opacity of dust per-unit H nucleon and $\ar = 4\sigma_{\rm SB}/c$ is the radiation constant. We use the dust temperature-dependent \citet{Semenov_2003} opacities for $\kappaP$. The term $\Lambda_{g-d}$ represents the energy exchange between dust and gas through inelastic collisions, and is given by the relation
\begin{equation}
\Lambda_{g-d}=\alpha_{g-d} T_{\rm gas }^{1 / 2}\left(\Tg-\Td\right) Z_d^{\prime},
\end{equation}
where $\Td$ is the dust temperature, and $\alpha_{g-d}$ the (inelastic) rate coefficient given by $\alpha_{g-d}=3.2 \times 10^{-34} \mathrm{erg} \mathrm{cm}^3 \mathrm{~s}^{-1} \mathrm{~K}^{-3 / 2}$. Using this in Eq.~\ref{eq:dusttemp}, we obtain a fourth order polynomial equation in $\Td$ of the form
\begin{equation}
    A\Td^4 + B\Td + C = 0
\end{equation}
where
\begin{align}
    A &= \rho \, \kappa_P \, Z_d' \, a_R \, c, \\
    B &= \alpha_{g-d} \, T_g^{1/2} \, Z_d' \, n_{\rm H}^2, \\
    C &= -\left( \dot{u}_{r, \mathrm{UV}} + \dot{u}_{r, \mathrm{CMB}} + \alpha_{g-d} \, T_g^{3/2} \, Z_d' \, n_{\rm H}^2 \right).
\end{align}

The approach above assumes that the dust is optically thin to its own thermal emission; this is however, only true for gas mass surface densities $\Sigma \lesssim 10^3 \, \Msolpc$ ($0.2 \, \rm g \, cm^{-2}$). Columns higher than this limit are easily attained in the cores of star-forming regions and in protostellar disks, and in the molecular clouds/ISM of starburst galaxies. The reprocessed IR radiation is further absorbed by dust in this regime, keeping the dust hotter due to the reduced effective cooling rate, and can lead to additional momentum deposition from the additional absorption/re-emission of the IR radiation field \citep[e.g.,][]{Menon_2022_IR}. We treat these effects explicitly when we track the IR band of radiation explicitly, as outlined in detail in \citet{Menon_2022_VETTAM}.

\section{GOW Network Implementation}
\label{sec:GOWComparison}

\begin{figure*}
    \centering
    \includegraphics[width=\textwidth]{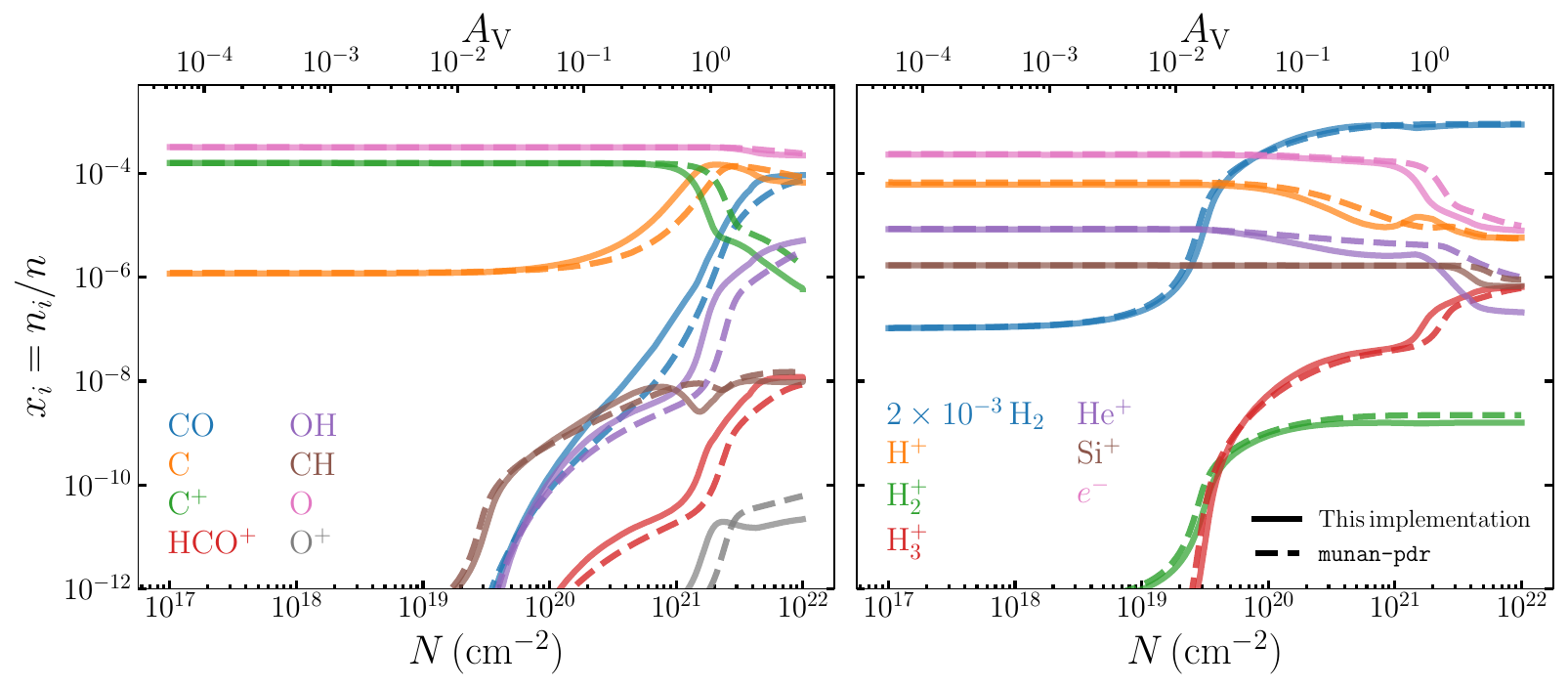}
    \caption{Comparison of equilibrium abundances ($x_i=n_i/n_{\rm H}$) through the PDR column ($N$, $A_{\rm V}$) between our (\texttt{KROME}-based) implementation of the \citet{Gong_2017} (GOW) network (\texttt{this implementation}) and the original, publicly available \citet{Gong_2017} PDR code (\texttt{munan-pdr}). The two implementations agree closely across all species and the full range of $A_{\rm V}$.}
    \label{fig:GOWComparison}
\end{figure*}

As described in Section~\ref{sec:AlternativeNetworks}, in addition to our fiducial network we implement the reduced network of \citet{Gong_2017} along with the modifications suggested in \citet{Hunter_2023} for high-temperature chemistry. To validate our \texttt{KROME}-based implementation of this network, we compare against results obtained with the original, publicly available PDR code\footnote{\url{https://github.com/munan/pdr}}. We reproduce their fiducial one-sided slab PDR test: i.e. a constant $n=100\,\pcm$, $\chi_{\rm FUV}=1$, and $\xi_{\rm H}=2\times10^{-16}\,\rm s^{-1}$ at fixed gas temperature $\Tg = 20 \, \rm K$. We fix the gas temperature to control for differences between our implementation and \texttt{munan-pdr}. In Figure~\ref{fig:GOWComparison} we compare the equilibrium abundances ($x_i = n_i/n_{\rm H}$) through the PDR obtained with our implementation against the original PDR code. The left panel shows the carbon-bearing species (CO, $\CI$, $\CII$, $\rm HCO^+$, OH, CH) together with $\OI$ and $\rm O^+$, while the right panel shows $\Htwo$, $\rm H^+$, $\rm H_2^+$, $\rm H_3^+$, $\rm He^+$, $\rm Si^+$ and $\elec$. We find excellent agreement between the implementations, with residual differences likely attributable to solver tolerances and additional reactions we added from \citet{Hunter_2023}. This demonstrates that our \texttt{KROME}-based implementation of the \citet{Gong_2017} network reproduces the original implementation.

\section{Numerical Tests}

\subsection{\HI\ to \Htwo\ transition}

In this test, we compute the column density at which hydrogen transitions from atomic to molecular and compare against the analytical steady-state calculations of \citet{Bialy_Sternberg_2016}. The \HI-to-\Htwo transition is governed by formation on dust grains competing with the (effective) photodissociation rate after accounting for dust shielding and H$_2$ self-shielding. Although CR dissociation can affect the transition \citep{kim2023_tigress}, we neglect it here to enable a direct comparison with the analytical solution, which includes only photodissociation.

We set up semi-infinite, one-dimensional plane-parallel PDR models discretized into 1000 logarithmically spaced Cartesian cells, with fixed number density $n$ and incident (beamed, unidirectional) FUV intensity $\chi_{\rm FUV}$. The FUV field is attenuated by dust shielding and self-shielding using the cumulative column from the illuminated surface to each cell; in all runs we cover $10^{17} \, \mathrm{cm^{-2}} \leq N \leq 10^{22} \, \mathrm{cm^{-2}}$, so the physical cell size decreases with increasing $n$. For self-shielding we adopt the same prescription used by \citet[][from \citealt{Draine_Bertoldi_1996}]{Bialy_Sternberg_2016}, rather than our default fit from \citet{Wolcott-Green_2011}, to ensure a like-for-like comparison. We also fix the $\Htwo$ formation rate coefficient on dust grains to a constant value of $3 \times 10^{-17} \, \rm cm^3 \, s^{-1}$ to match the analytical setup. Finally, we set $\xi_{\rm H} = 0$ so that only photodissociation operates. We evolve the chemical and thermal state to steady state on a cell-by-cell basis from low to high $N$, ensuring that the updated abundances are used when computing the shielding column from the PDR edge.

\begin{figure}
\centering
\includegraphics[width=\columnwidth]{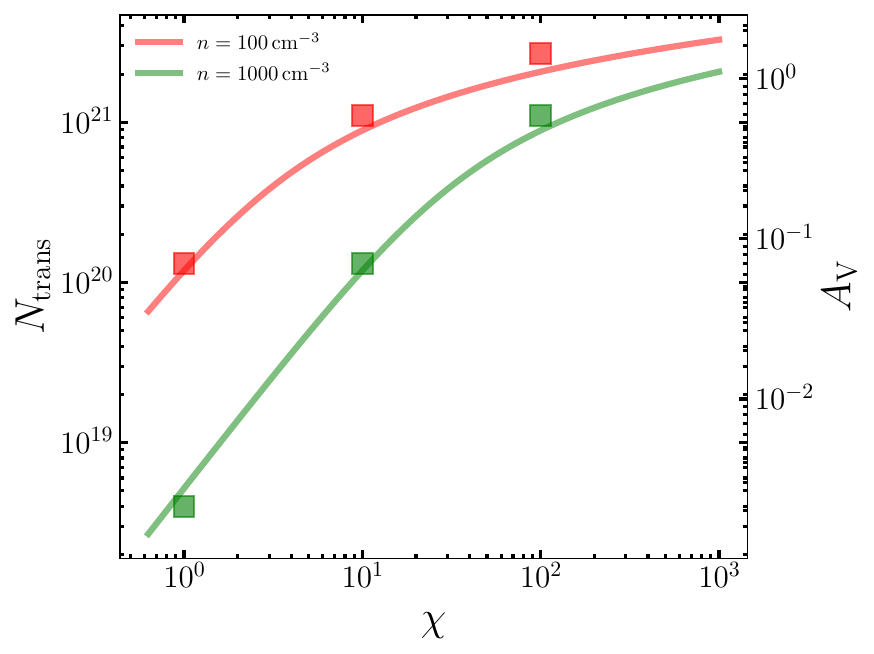}
\caption{Transition column for the \HI-to-\Htwo transition for different radiation-field strengths. Markers show the transition column obtained with our network, while solid lines show the analytical solution from \citet{Bialy_Sternberg_2016}.}
\label{fig:PDR_Ntrans}
\end{figure}

We solve for the PDR state for three radiation-field strengths, $\chi_{\rm FUV} = [1, 10, 100]$, and two densities, $n = [100, 1000] \, \rm cm^{-3}$, at solar metallicity ($Z=\Zsun$). For each parameter pair, we record the \HI-to-\Htwo transition column $N_{\rm trans}$, defined as the value of $N$ where $x_{\rm H} = 2x_{\rm H_2} = 0.5$, i.e. the depth beyond which \Htwo becomes the dominant carrier of H nuclei. In Figure~\ref{fig:PDR_Ntrans} we show $N_{\rm trans}$ from our models (markers) and compare to the analytical approximation of \citet{Bialy_Sternberg_2016}. We find excellent agreement, indicating that our scheme captures the \HI-to-\Htwo transition accurately.

\subsection{R-type Expansion}
\label{sec:Rtype}

In this test, we test the ability of our scheme to reproduce the evolution of an R-type front driven by a constant source of ionising radiation into a uniform medium. The R-type expansion phase of a HII region \citep{Spitzer_1978} refers to the early phase where the ionisation front propagates rapidly (faster than the ionized sound speed) without perturbing gas motions, with the evolution of the front purely driven by ionisation-recombination balance in the gas within the front. The phase ends when the radius of the ionisation front is equal to the Stromgren radius ($R_{\rm S}$) at which point ionisations equal recombinations at the fixed initial density; $R_{\rm S}$  for a dusty medium is given by -
\begin{equation}
    R_{\rm S} = \frac{3f_{\rm ion}Q_{\rm ion}}{4 \pi \alpha_{\rm B}n^2},
\end{equation}
where $Q_{\rm ion}$ is the ionising photon emission rate, $\alpha_{\rm B}$ the case-B recombination coefficient for $\Hplus$, and $n$ the total Hydrogen nuclei number density; $f_{\rm ion}$ refers to the fraction of the total ionising photons absorbed by gas rather than dust, which acts to lower $R_{\rm S}$ from the dust-free case \citep{Petrosian_1972,Draine_2011,Raga_2015}. The time evolution of the ionising front is given by the analytical relation \citep{Spitzer_1978}
\begin{equation}
    R_{\rm ion} = R_{\rm S} \left( 1 - \exp (-t/t_{\rm rec}) \right)^{1/3},
    \label{eq:RType}
\end{equation}
where $t_{\rm rec} = 1/(\alpha_{\rm B}n)$ is the recombination time. $f_{\rm ion}$ can be estimated via the approach outlined in \citet{Petrosian_1972}, which depends primarily on the dust optical depth through the dust-free Stromgren radius; we use a dust opacity $\kappa_{\rm LyC} = 1250 \, \rm cm^2 g^{-1}$ in our test. 

We test the ability of our scheme to reproduce the R-type evolution by setting up a uniform medium in a cubic domain of size $L=(20 \, \rm pc)^3$ with $n_{\rm Hnuclei} = 100 \, \rm cm^{-3}$ and metallicity $\Zsun$, and placing a radiation source at the center, consistent with a star cluster of mass $1000 \, \Msun$ assuming a fully-sampled IMF with SB99; this corresponds to an ionizing photon rate $Q_{\rm ion} = 1.27 \times 10^{50} \, \rm s^{-1}$. The background radiation field and CR flux is set to the Draine ISRF background and $10^{-16} \, \rm s^{-1}$ respectively, while adopting the local Jeans-length shielding prescription for attenuation. The initial thermal and chemical state of the uniform medium is set to the corresponding value in thermochemical equilibrium for the $n_{\rm H}$ we adopt and background CR/radiation fields (see Figure~\ref{fig:ism}) -- i.e. $\Tg = 93 \, \rm K$, 92\% $\Htwo$ fraction, $2 \%$ of C in CO, 57\% in CI and 40\% in CII. We let the system evolve for up to $\sim 100 \, t_{\rm rec}$, and ensure that hydrodynamic motions are turned off. The numerical resolution is set to $256^3$.

A technical consideration is that the expansion speed of R-type fronts are much faster than typical hydrodynamic timescales in the diffuse/dense ISM -- i.e. the ionizing front can propagate multiple computational cells in a hydrodynamic timestep. Capturing this requires the thermochemistry and radiative transfer to be sub-cycled relative to the hydrodynamics with a timestep dictated by the ionization-recombination balance \citep[e.g.,][]{Rosdahl2013Ramses-rt:Context,kim2023_tigress}. We test the sensitivity of the numerical evolution of the front to the timestep by running the test with the substep timesteps limited by a range of values from $\sim t_{\rm rec}/1000$ to $t_{\rm rec}$. 

\begin{figure}
    \centering
    \includegraphics[width=\linewidth]{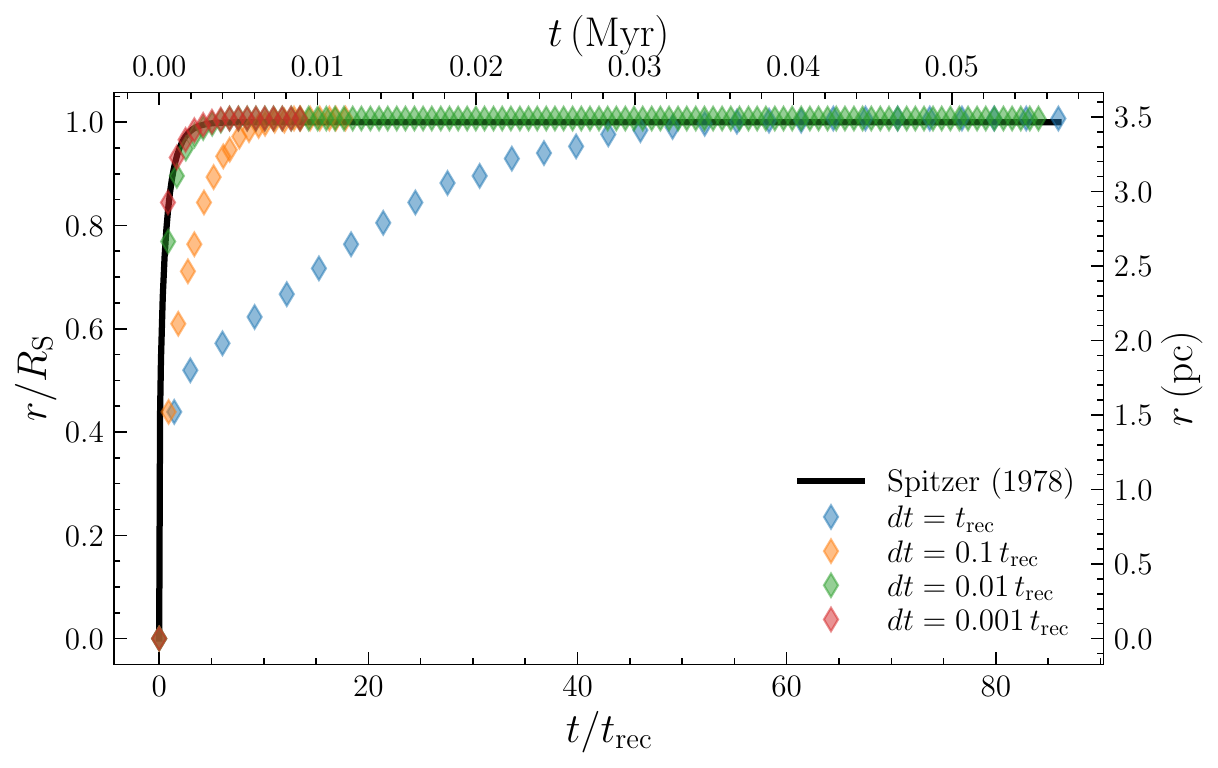}
    \caption{Evolution of the ionisation front radius -- defined as the maximum radius where the fraction $n_{\rm H^+}/n_{\rm Hnuclei} > 0.95$ -- with time for the R-type expansion phase of a HII region. Colored markers indicate the outcome from our implementation for multiple versions where the timestep is restricted by the given value in units of the recombination time ($t_{\rm rec}$). The black line shows the analytical solution (Equation~\ref{eq:RType}) for our given input $n$ and $Q_{\rm ion}$, and where $f_{\rm ion}$ is calculated using the relation provided in \citet{Petrosian_1972}. All our tests asymptotically approach $R_{\rm S}$, however, the timestep needs to be restricted to $\lesssim 0.1 t_{\rm rec}$ to follow the R-type phase accurately.}
    \label{fig:RType}
\end{figure}

In Figure~\ref{fig:RType} we show the evolution of the ionizing front -- defined as the maximum radius where $n_{\rm H^+}/n_{\rm Hnuclei}>0.95$ -- with time obtained with the different timestep constraints, compared with the \citet{Spitzer_1978} analytical solution given by Equation~\ref{eq:RType}. We can see that to capture the time-resolved expansion of the front accurately (to within $\sim 20\%$ at all times), the substeps need to be $\lesssim 0.1 t_{\rm rec}$, with more restrictive timesteps leading to more accurate solutions -- consistent with independent tests conducted in \citet{Deng_2024}. This can be quite restrictive in full 3D applications, especially for our approach that relies on implicit updates for the radiation and thermochemistry, which offers an advantage primarily when performing less frequent, larger timestep updates. However, this is not too much of a concern as the R-type phase is extremely short and not relevant for the dynamical evolution of the gas -- although it is for the ionization state (in this short phase) and applications where the speed of ionization fronts during the R-type phase is important to capture (e.g., reionization of the IGM). More importantly, it is reassuring to note that regardless of the timestep, the ionization front asymptotically approaches the Stromgren radius, which reflects the photon-conserving nature of our scheme. In other words, in the limit of large timesteps relative to the recombination time ($dt/t_{\rm rec} >1$) the ionization front is expected to grow to the Stromgren radius within the timescales resolved in the simulation. 

\subsection{D-type Expansion}
\label{sec:Dtype}

We now test the subsequent (i.e. $r>R_{\rm S}$) D-type expansion of the HII region that is driven by the elevated thermal pressure of the ionized region. The initial conditions are identical to the R-type test for the gas and the radiation source. The domain size is increased to $(80 \, \rm pc)^3$ and the timestep is controlled purely by the hydrodynamics CFL criterion -- the latter is sufficient to capture the expansion of the ionization front as it is the thermal pressure-driven dynamics that facilitates subsequent ionization upstream of the front \citep{Spitzer_1978}. We run the test for metallicities of $Z = \Zsun, 0.1 \Zsun, 0.01 \Zsun$ and set the initial thermochemical state corresponding to that in equilibrium for these metallicities (Figure~\ref{fig:ism}). We quantify the position of the ionization front as the mean radius of cells with $0.1<n_{\rm H^+}/n_{\rm Hnuclei} < 0.9$. 

\begin{figure}
    \centering
    \includegraphics[width=\linewidth]{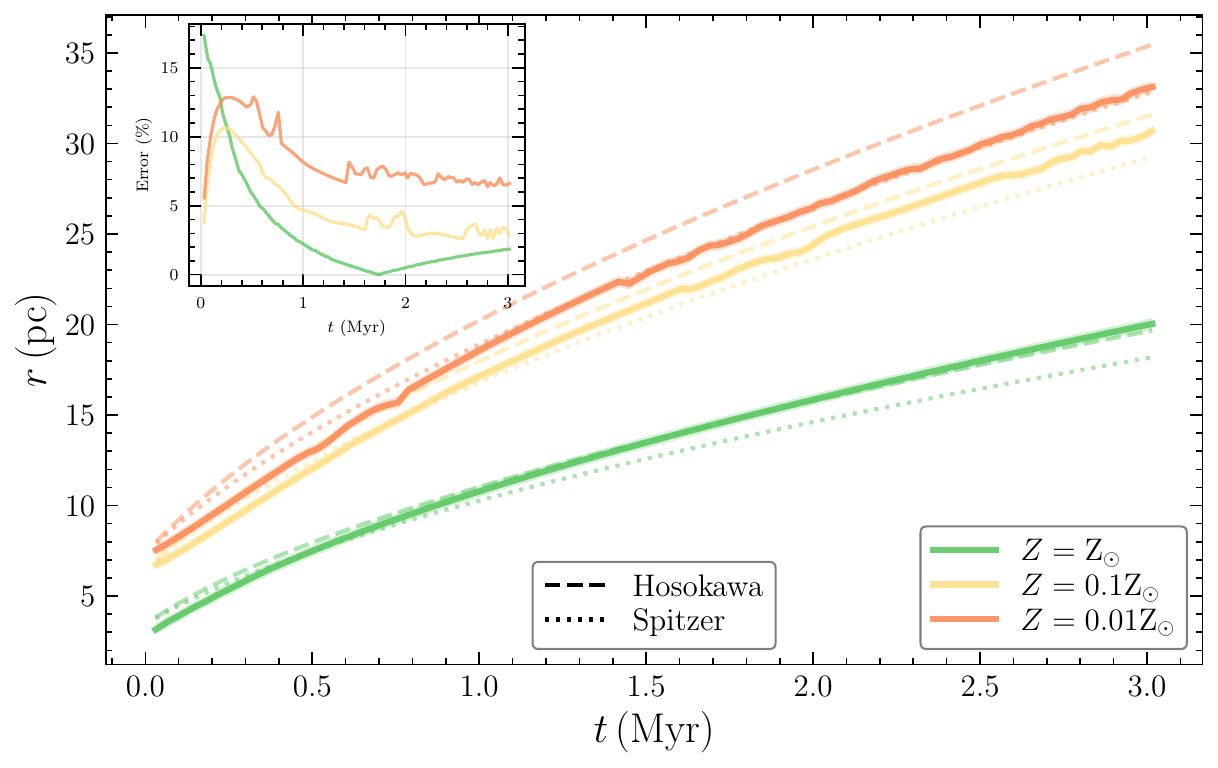}
    \caption{Evolution of the ionisation front radius -- defined as the average radius of cells with $0.1<n_{\rm H^+}/n_{\rm Hnuclei} < 0.9$ -- with time for the D-type expansion phase of a HII region for different metallicities. Dashed and dotted lines indicate the analytical solutions of \citet{Hosokawa_Inutska_2006} and \citet{Spitzer_1978} respectively. The inset denotes the relative error of our implementation compared to the \citet{Hosokawa_Inutska_2006} solution, which is limited to $\lesssim 10\%$ at all times.}
    \label{fig:DType}
\end{figure}

In Figure~\ref{fig:DType} we show the evolution of the ionization front radius, and compare them to the analytical solutions of \citet{Hosokawa_Inutska_2006} and \citet{Spitzer_1978} respectively. The \citet{Hosokawa_Inutska_2006} solution differs from the \citet{Spitzer_1978} solution through a higher scaling factor due to their inclusion of the ram pressure of the dense shell formed ahead of the ionization front. We find decent agreement with the \citet{Hosokawa_Inutska_2006} solution, with stronger agreement at higher metallicities and later times. The weaker agreement at lower metallicities is again due to the delay in the propagation of the ionization front in the R-type test -- i.e. the front has to move one cell at a time limited by the cadence of the radiation update. Nevertheless, the time-averaged error is $<5\%$ and the maximum error is $\lesssim 10\%$; we also verify that the radial momentum of the shell is consistent with the analytical solution. 

\section{Non-equilibrium thermal structure in turbulent clouds}
\label{sec:Turbulence_temperatures}

\begin{figure*}
    \centering
    \includegraphics[width=\linewidth]{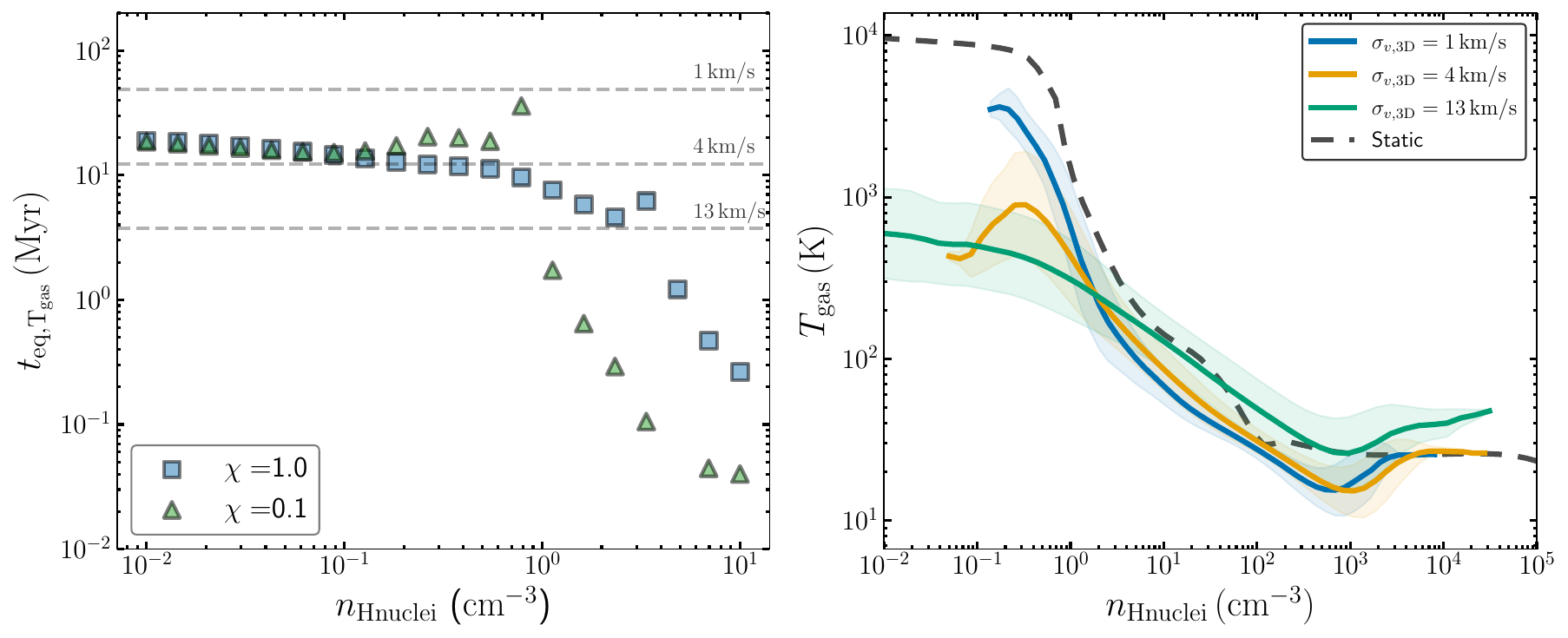}
    \caption{Left: timescales for gas at various number densities ($n_{\rm Hnuclei}$) and radiation backgrounds ($\chi_{\rm FUV}$) to attain its thermal equilibrium value compared to the turbulent eddy turnover timescale ($t_{\rm turb} = L/2\sigma_{v,\rm 3D}$) at the driving scale. We compare these for the $\sigma_{v,\rm 3D}$ used in the simulation in Section~\ref{sec:turbulent_sim}, and for two lower values of $\sigma_{v,\rm 3D}$. We can clearly see that $t_{\rm turb}$ is shorter than $t_{\rm eq, T_{gas}}$ unless $\sigma_{\rm v, 3D} \lesssim 1 \, \kms$. Right: the median gas temperature with shaded regions showing the $1-\sigma$ variation for the turbulent medium simulation of Section~\ref{sec:turbulent_sim}, compared with identical runs with the lower $\sigma_{\rm v,3D}$. Consistent with timescale considerations, we can see that only at the lowest $\sigma_{\rm v,3D} \lesssim 1 \, \kms$ do the temperatures approach the static thermal equilibrium solution for the diffuse ISM.}
    \label{fig:TemperatureTurbulence}
\end{figure*}

In Section~\ref{sec:turbulent_sim} we found that the temperatures in the WNM for the fiducial simulation were noticeably cooler than the expected temperature at thermal equilibrium for these timescales (Figure~\ref{fig:TurbMetallicityPhase}). We attribute this to shorter dynamical (turbulent crossing) timescales in comparison to the timescale for heating CNM gas to WNM temperatures. To compute this timescale, we ran one-zone temporal evolution models of gas at $\Tg = 100 \, \rm K$ and $Z = \Zsun$ with a fixed CR background of $\xi_{\rm H} = 2 \times 10^{-16} \, \rm s^{-1}$, and tested various densities over the range $n_{\rm Hnuclei} \in [0.01,10] \, \pcm$, and fixed radiation backgrounds of $\chi_{\rm FUV} =1$ and $\chi_{\rm FUV} = 0.1$. We then examine the evolution of the temperature with time and measure when the temperature reaches steady state ($t_{\rm eq,\Tg}$). In the left panel of Figure~\ref{fig:TemperatureTurbulence} we compare $t_{\rm eq,\Tg}$ to the turbulent eddy turnover timescale at the driving scale ($t_{\rm turb}=L/(2\sigma_{v,\rm 3D})$), for the fiducial turbulent velocity dispersion used in Section~\ref{sec:turbulent_sim} as well as two lower-$\sigma_{v,\rm 3D}$ cases. The comparison highlights that, for the fiducial run, turbulent motions in diffuse gas ($n_{\rm Hnuclei} \lesssim 1 \, \pcm$) can displace gas on timescales shorter than those required for photoelectric heating (and associated cooling adjustments) to bring the gas to thermal equilibrium -- consistent with the noticeable differences with the equilibrium value in Figure~\ref{fig:TurbMetallicityPhase}. However, it also suggests that at lower $\sigma_{v,\rm 3D}$ the temperature should converge closer to the equilibrium solution. 

To test this picture, we ran simulations identical to that of Section~\ref{sec:turbulent_sim}, but with the lower $\sigma_{v,\rm 3D}$ values. The solid lines in the right panel of Figure~\ref{fig:TemperatureTurbulence} shows the median temperature as a function of $n_{\rm Hnuclei}$ with the shaded region showing the $1\sigma$ scatter about the median, and the dashed lines shows the static equilibrium solution plotted as in Figure~\ref{fig:ism}. We can see that the temperatures progressively get closer to the equilibrium value for lower $\sigma_{v,\rm 3D}$, consistent with expectations from the left panel. We note that the agreement is not perfect even at $\sigma_{v,\rm 3D} = 1 \, \kms$ where the timescale comparison suggests that $t_{\rm eq, T_{gas}} < t_{\rm turb}$. This is because we have compared to $t_{\rm turb}$ at the driving scale ($L/2$); turbulent motions are driven over the range $k/(2\pi/L) \in [2,64]$, and for the turbulence power spectrum we drive ($\propto k^{-2}$), the eddy turnover timescale for a length scale $l$ is $\propto l^{1/2}$ -- i.e. it is shorter at the smaller scales (higher $k$) where the turbulence is driven. Therefore for mixing at the smallest scales where turbulence is (explicitly) driven/forced $t_{\rm turb}$ is shorter by a factor $\sqrt{32} \sim 5$; using this timescale instead would make $t_{\rm turb}$ comparable to $t_{\rm eq, T_{gas}}$ for the diffuse gas. This suggests that the thermal structure of the diffuse gas in our turbulent medium simulations can be reasonably well explained by the competition between turbulent mixing and heating timescales.

\section{Fiducial Chemical network}
\label{sec:appendix_chemistry}

In Table~\ref{tab:network} we show the complete set of reactions considered by our fiducial network and their rates. 

\begin{longtable}{l l l l}
\caption{Rates of chemical reactions included in the fiducial network.}\label{tab:network} \\
\hline
No. & Reaction & Rate coefficient ($\rm cm^3 \, s^{-1}$) & Ref.\\
\hline
\endfirsthead
\hline
No. & Reaction & Rate coefficient & Ref.\\
\hline
\endhead
\hline
\endfoot
\hline
\multicolumn{4}{l}{\textbf{Primordial chemistry}} \\
1 & \ch{H + e- -> H+ + e- + e-} & \parbox[t]{0.65\textwidth}{\raggedright $\begin{aligned}[t] &\operatorname{exp}\Bigl(- 2.03914 \times 10^{-6} \ln^{8}{\Bigl(T_{\rm eV} \Bigr)}  \\ 
 &+ 0.00011 \ln^{7}{\Bigl(T_{\rm eV} \Bigr)} - 0.00263 \ln^{6}{\Bigl(T_{\rm eV} \Bigr)}  \\ 
 &+ 0.03482 \ln^{5}{\Bigl(T_{\rm eV} \Bigr)} - 0.28770 \ln^{4}{\Bigl(T_{\rm eV} \Bigr)}  \\ 
 &+ 1.56315 \ln^{3}{\Bigl(T_{\rm eV} \Bigr)} - 5.73932 \ln^{2}{\Bigl(T_{\rm eV} \Bigr)}  \\ 
 &+ 13.53655 \ln{\Bigl(T_{\rm eV} \Bigr)} - 32.71396 \Bigr)\end{aligned}$} & 1 \\
2 & \ch{H+ + e- -> H} & \parbox[t]{0.65\textwidth}{\raggedright $3.92 \times 10^{-13} \left(\frac{1}{T_{\rm eV}}\right)^{0.6353}\hfill T \leqslant 5500 \, \mathrm{K}$} & 2 \\
 &  & \parbox[t]{0.65\textwidth}{\raggedright $\begin{aligned}[t] &\operatorname{exp}\Bigl(- 3.07113 \times 10^{-9} \ln^{9}{\Bigl(T_{\rm eV} \Bigr)}  \\ 
 &- 1.85676 \times 10^{-8} \ln^{8}{\Bigl(T_{\rm eV} \Bigr)}  \\ 
 &+ 5.75561 \times 10^{-7} \ln^{7}{\Bigl(T_{\rm eV} \Bigr)}  \\ 
 &+ 4.98910 \times 10^{-6} \ln^{6}{\Bigl(T_{\rm eV} \Bigr)}  \\ 
 &- 1.42150 \times 10^{-5} \ln^{5}{\Bigl(T_{\rm eV} \Bigr)}  \\ 
 &- 0.00032 \ln^{4}{\Bigl(T_{\rm eV} \Bigr)}  \\ 
 &- 0.00238 \ln^{3}{\Bigl(T_{\rm eV} \Bigr)}  \\ 
 &- 0.02026 \ln^{2}{\Bigl(T_{\rm eV} \Bigr)}  \\ 
 &- 0.72411 \ln{\Bigl(T_{\rm eV} \Bigr)}  \\ 
 &- 28.61303 \Bigr)\end{aligned}\hfill T > 5500 \, \mathrm{K}$} &  \\
3 & \ch{He + e- -> He+ + e- + e-} & \parbox[t]{0.65\textwidth}{\raggedright $\begin{aligned}[t] &\operatorname{exp}\Bigl(- 3.64916 \times 10^{-6} \ln^{8}{\Bigl(T_{\rm eV} \Bigr)}  \\ 
 &+ 0.00020 \ln^{7}{\Bigl(T_{\rm eV} \Bigr)} - 0.00500 \ln^{6}{\Bigl(T_{\rm eV} \Bigr)}  \\ 
 &+ 0.06795 \ln^{5}{\Bigl(T_{\rm eV} \Bigr)} - 0.56851 \ln^{4}{\Bigl(T_{\rm eV} \Bigr)}  \\ 
 &+ 3.05803 \ln^{3}{\Bigl(T_{\rm eV} \Bigr)} - 10.75323 \ln^{2}{\Bigl(T_{\rm eV} \Bigr)}  \\ 
 &+ 23.91596 \ln{\Bigl(T_{\rm eV} \Bigr)} - 44.09864 \Bigr)\end{aligned}$} & 1 \\
4 & \ch{He+ + e- -> He} & \parbox[t]{0.65\textwidth}{\raggedright $3.92 \times 10^{-13} \left(\frac{1}{T_{\rm eV}}\right)^{0.6353} \hfill T \leqslant 9280 \, \mathrm{K}$} & 3,4 \\
&  & \parbox[t]{0.65\textwidth}{\raggedright $ \frac{1.54\times10^{-9}}{T_{\rm eV}^{3/2}} \left(1+0.3\,e^{-8.09933/T_{\rm eV}}\right)e^{-40.49664/T_{\rm eV}}+   \hfill T > 9280 \, \mathrm{K}$} &  \\
&  & \parbox[t]{0.65\textwidth}{\raggedright $\frac{3.92\times10^{-13}}{T_{\rm eV}^{0.6353}}$} &  \\
5 & \ch{He+ + e- -> He^{++} + e- + e-} & \parbox[t]{0.65\textwidth}{\raggedright $\begin{aligned}[t] &\operatorname{exp}\Bigl(- 3.16558 \times 10^{-6} \ln^{8}{\Bigl(T_{\rm eV} \Bigr)}  \\ 
 &+ 0.00019 \ln^{7}{\Bigl(T_{\rm eV} \Bigr)} - 0.00532 \ln^{6}{\Bigl(T_{\rm eV} \Bigr)}  \\ 
 &+ 0.08113 \ln^{5}{\Bigl(T_{\rm eV} \Bigr)} - 0.76924 \ln^{4}{\Bigl(T_{\rm eV} \Bigr)}  \\ 
 &+ 4.70162 \ln^{3}{\Bigl(T_{\rm eV} \Bigr)} - 18.48066 \ln^{2}{\Bigl(T_{\rm eV} \Bigr)}  \\ 
 &+ 43.93347 \ln{\Bigl(T_{\rm eV} \Bigr)} - 68.71040 \Bigr)\end{aligned}$} & 2 \\
6 & \ch{He^{++} + e- -> He+} &
$\frac{1.891\times10^{-10}}{
\sqrt{\frac{T}{9.37}}
\left(1+\sqrt{\frac{T}{9.37}}\right)^{0.2476}
\left(1+\sqrt{\frac{T}{2.774\times10^6}}\right)^{1.7524}
}$ & 5 \\
7 & \ch{H + e- -> H-} & \parbox[t]{0.65\textwidth}{\raggedright $1.4 \times 10^{-18} T^{0.928} \operatorname{exp}{\left(- 6.17283 \times 10^{-5} T \right)}$} & 6 \\
8 & \ch{H- + H -> H2 + e-} & \parbox[t]{0.65\textwidth}{$\frac{
1.3500\times10^{-9}
\left(
T^{0.09849}
+0.32852\,T^{0.55610}
+2.7710\times10^{-7}\,T^{2.18260}
\right)
}{
1
+0.00619\,T^{1.04610}
+8.9712\times10^{-11}\,T^{3.04240}
+3.2576\times10^{-14}\,T^{3.77410}
}$} & 7 \\
9 & \ch{H + H+ -> H2+} & \parbox[t]{0.65\textwidth}{\raggedright $3.49774 \times 10^{-20}T^{-0.15}\hfill T < 30 \, \mathrm{K}$} & 8 \\
 &  & \parbox[t]{0.65\textwidth}{\raggedright $10^{- 0.2072 \log^{3}{\left(T \right)} + 1.786 \log^{2}{\left(T \right)} - 3.194 \log{\left(T \right)} - 18.2}\hfill T \geqslant 30 \, \mathrm{K}$} &  \\
10 & \ch{H2+ + H -> H2 + H+} & \parbox[t]{0.65\textwidth}{\raggedright $6.4 \times 10^{-10}$} & 9,10 \\
11 & \ch{H2 + H+ -> H2+ + H} & $\begin{aligned}[t] &\Bigl(3.53119 \times 10^{-13} \ln^{7}{\Bigl(T \Bigr)} - 1.81714 \times 10^{-11} \ln^{6}{\Bigl(T \Bigr)}  \\ 
 &+ 3.97315 \times 10^{-10} \ln^{5}{\Bigl(T \Bigr)} - 4.78137 \times 10^{-9} \ln^{4}{\Bigl(T \Bigr)}  \\ 
 &+ 3.41728 \times 10^{-8} \ln^{3}{\Bigl(T \Bigr)} - 1.44913 \times 10^{-7} \ln^{2}{\Bigl(T \Bigr)}  \\ 
 &+ 3.37353 \times 10^{-7} \ln{\Bigl(T \Bigr)} - 3.32321 \times 10^{-7}\Bigr) \operatorname{exp}\Bigl(- \frac{21237.15}{T} \Bigr)\end{aligned} \hfill 100\, \mathrm{K} \leqslant T \leqslant 3 \times 10^{4} \, \mathrm{K}$ & 11,12 \\
12 & \ch{H2 + e- -> H + H-} & \parbox[t]{0.65\textwidth}{\raggedright $35.5 \operatorname{exp}{\left(- \frac{46707}{T} \right)}T^{-2.28}$} & 13 \\
13 & \ch{H2 + e- -> H + H + e-} & \parbox[t]{0.65\textwidth}{\raggedright $4.38 \times 10^{-10} T^{0.35} \operatorname{exp}{\left(- \frac{102000}{T} \right)}$} & 14 \\
14 & \ch{H2 + H -> H + H + H} & \parbox[t]{0.65\textwidth}{\raggedright $k_{\rm H_2,H}$} & 15 \\
15 & \ch{H- + e- -> H + e- + e-} & \parbox[t]{0.65\textwidth}{\raggedright $\begin{aligned}[t] &\operatorname{exp}\Bigl(- 2.63128 \times 10^{-6} \ln^{8}{\Bigl(T_{\rm eV} \Bigr)}  \\ 
 &+ 0.00010 \ln^{7}{\Bigl(T_{\rm eV} \Bigr)} - 0.00165 \ln^{6}{\Bigl(T_{\rm eV} \Bigr)}  \\ 
 &+ 0.01178 \ln^{5}{\Bigl(T_{\rm eV} \Bigr)} - 0.03365 \ln^{4}{\Bigl(T_{\rm eV} \Bigr)}  \\ 
 &+ 0.01623 \ln^{3}{\Bigl(T_{\rm eV} \Bigr)} - 0.28274 \ln^{2}{\Bigl(T_{\rm eV} \Bigr)}  \\ 
 &+ 2.36085 \ln{\Bigl(T_{\rm eV} \Bigr)} - 18.01849 \Bigr)\end{aligned}$} & 1 \\
16 & \ch{H- + H -> H + H + e-} & \parbox[t]{0.65\textwidth}{\raggedright $2.56 \times 10^{-9} T_{\rm eV}^{1.78186} \hfill T \leqslant 1160 \, \mathrm{K}$} & 2 \\
 &  & \parbox[t]{0.65\textwidth}{\raggedright $\begin{aligned}[t] &\operatorname{exp}\Bigl(- 8.06838 \times 10^{-8} \ln^{9}{\Bigl(T_{\rm eV} \Bigr)}  \\ 
 &+ 2.45550 \times 10^{-6} \ln^{8}{\Bigl(T_{\rm eV} \Bigr)} - 2.58500 \times 10^{-5} \ln^{7}{\Bigl(T_{\rm eV} \Bigr)}  \\ 
 &+ 8.66396 \times 10^{-5} \ln^{6}{\Bigl(T_{\rm eV} \Bigr)}  \\ 
 &+ 0.00020 \ln^{5}{\Bigl(T_{\rm eV} \Bigr)} - 0.00143 \ln^{4}{\Bigl(T_{\rm eV} \Bigr)}  \\ 
 &+ 0.00846 \ln^{3}{\Bigl(T_{\rm eV} \Bigr)} - 0.14210 \ln^{2}{\Bigl(T_{\rm eV} \Bigr)}  \\ 
 &+ 1.13944 \ln{\Bigl(T_{\rm eV} \Bigr)} - 20.37260 \Bigr)\end{aligned} \hfill T > 1160 \, \mathrm{K}$} &  \\
17 & \ch{H- + H+ -> H + H} & \parbox[t]{0.65\textwidth}{\raggedright $- 7.77 \times 10^{-13} T + 2.5 \times 10^{-10} \sqrt{\left(T \right)} - 1.73 \times 10^{-9} + \frac{2.96 \times 10^{-6}}{\sqrt{\left(T \right)}} \hfill T \leqslant 10^{5} \, \mathrm{K}$} & 16 \\
18 & \ch{H- + H+ -> H2+ + e-} & \parbox[t]{0.65\textwidth}{\raggedright $1 \times 10^{-8}T^{-0.4}$} & 17 \\
19 & \ch{H2+ + e- -> H + H} & \parbox[t]{0.65\textwidth}{\raggedright $- 5.528 \times 10^{-28} T^{5} + 3.3468 \times 10^{-23} T^{4} - 7.5474 \times 10^{-19} T^{3} + \hfill T \leqslant 23642 \, \mathrm{K}$} & 18 \\
& & \parbox[t]{0.65\textwidth}{\raggedright $7.3428 \times 10^{-15} T^{2} - 2.3088 \times 10^{-11} T + 4.2278 \times 10^{-8}$} & \\
20 & \ch{H2+ + H- -> H + H2} & \parbox[t]{0.65\textwidth}{\raggedright $5 \times 10^{-7} \sqrt{\left(\frac{100}{T} \right)}$} & 19 \\
21 & \ch{H + H + H -> H2 + H} & \parbox[t]{0.65\textwidth}{\raggedright $2 \times 10^{-31}T^{-0.5} + \frac{6 \times 10^{-32}}{T^{0.25}}$} & 20 \\
22 & \ch{H2 + H + H -> H2 + H2} & \parbox[t]{0.65\textwidth}{\raggedright $2.5 \times 10^{-32}T^{-0.5} + \frac{7.5 \times 10^{-33}}{T^{0.25}}$} & 21 \\
23 & \ch{H2 + H2 -> H + H + H2} & \parbox[t]{0.65\textwidth}{\raggedright $k_{\rm H_2, H_2}$} & 15 \\
24 & \ch{He+ + H -> He + H+} & \parbox[t]{0.65\textwidth}{\raggedright $2.88337 \times 10^{-16} T^{0.25}$} & 22 \\
25 & \ch{He + H+ -> He+ + H} & \parbox[t]{0.65\textwidth}{\raggedright $1.26 \times 10^{-9} \operatorname{exp}{\left(- \frac{127500}{T} \right)}T^{-0.75} \hfill T \leqslant 10^{4} \, \mathrm{K}$} & 22 \\
 &  & \parbox[t]{0.65\textwidth}{\raggedright $4 \times 10^{-37} T^{4.74} \hfill 10^{4} < T \leqslant 4.8\times 10^7 \, \mathrm{K}$} &  \\
26 & \ch{He+ + H2 -> H2+ + He} & \parbox[t]{0.65\textwidth}{\raggedright $7.2 \times 10^{-15}$} & 23 \\
27 & \ch{He+ + H2 -> H+ + He + H} & \parbox[t]{0.65\textwidth}{\raggedright $1.26 \times 10^{-13} \operatorname{exp}{\left(- \frac{22.5}{T} \right)}$} & 23 \\
28 & \ch{H2 + D+ -> HD + H+} & \parbox[t]{0.65\textwidth}{\raggedright $- 1.37 \times 10^{-10} \log^{2}{\left(T \right)} + 8.46 \times 10^{-10} \log{\left(T \right)} + 4.17 \times 10^{-10} \hfill T \leqslant 4.3\times 10^6 \, \mathrm{K}$} & 24 \\
29 & \ch{HD + H+ -> H2 + D+} & \parbox[t]{0.65\textwidth}{\raggedright $1 \times 10^{-9} \operatorname{exp}{\left(- \frac{457}{T} \right)}$} & 25 \\
30 & \ch{H2 + D -> HD + H} & \parbox[t]{0.65\textwidth}{\raggedright $7.5 \times 10^{-11} \operatorname{exp}{\left(- \frac{3820}{T} \right)}$} & 25 \\
31 & \ch{HD + H -> H2 + D} & \parbox[t]{0.65\textwidth}{\raggedright $7.5 \times 10^{-11} \operatorname{exp}{\left(- \frac{4240}{T} \right)}$} & 25 \\
32 & \ch{D + H- -> HD + e-} & \parbox[t]{0.65\textwidth}{\raggedright $1.5 \times 10^{-9}(T/300)^{-0.1}$} & 46 \\
33 & \ch{H+ + D -> H + D+} & \parbox[t]{0.65\textwidth}{\raggedright $2 \times 10^{-10} T^{0.402} \operatorname{exp}{\left(- \frac{37.1}{T} \right)} - 3.31 \times 10^{-17} T^{1.48} \hfill T \leqslant 2 \times 10^{5} \, \mathrm{K}$} & 26 \\
 &  & \parbox[t]{0.65\textwidth}{\raggedright $3.44 \times 10^{-10} T^{0.35} \hfill T > 2 \times 10^{5} \, \mathrm{K}$} &  \\
34 & \ch{H + D+ -> H+ + D} & \parbox[t]{0.65\textwidth}{\raggedright $2.03 \times 10^{-9}T^{-0.332} + 2.06 \times 10^{-10} T^{0.396} \operatorname{exp}{\left(- \frac{33}{T} \right)} \hfill T \leqslant 2 \times 10^{5} \, \mathrm{K}$} & 24 \\
35 & \ch{D+ + e- -> D} & \parbox[t]{0.65\textwidth}{\raggedright $2.59503 \times 10^{-10}T^{-0.75}$} & 27,28 \\
36 & \ch{H + D -> HD} & \parbox[t]{0.65\textwidth}{\raggedright $1 \times 10^{-25}$} & 27,28 \\
37 & \ch{HD+ + H -> HD + H+} & \parbox[t]{0.65\textwidth}{\raggedright $6.4 \times 10^{-10}$} & 27,28 \\
38 & \ch{H+ + D -> HD+} & \parbox[t]{0.65\textwidth}{\raggedright ${10}^{- 1.523 \log{\left(T \right)} + 1.118 \log^{2}{\left(T \right)} - 0.1269 \log^{3}{\left(T \right)} - 19.38}$} & 27,28 \\
39 & \ch{H + D+ -> HD+} & \parbox[t]{0.65\textwidth}{\raggedright ${10}^{- 1.523 \log{\left(T \right)} + 1.118 \log^{2}{\left(T \right)} - 0.1269 \log^{3}{\left(T \right)} - 19.38}$} & 27,28 \\
40 & \ch{HD+ + e- -> H + D} & \parbox[t]{0.65\textwidth}{\raggedright $7.2 \times 10^{-8} T^{-0.5}$} & 27,28 \\
41 & \ch{D + e- -> D-} & \parbox[t]{0.65\textwidth}{\raggedright $1.33001 \times 10^{-18} T^{0.95} \operatorname{exp}{\left(- 0.0001 T \right)}$} & 27,28 \\
42 & \ch{D+ + D- -> D + D} & \parbox[t]{0.65\textwidth}{\raggedright $9.87268 \times 10^{-7}T^{-0.5}$} & 27,28 \\
43 & \ch{D- + H+ -> D + H} & \parbox[t]{0.65\textwidth}{\raggedright $7.96743 \times 10^{-7}T^{-0.5}$} & 27,28 \\
44 & \ch{H- + D -> D- + H} & \parbox[t]{0.65\textwidth}{\raggedright $6.17390 \times 10^{-10} T^{0.41}$} & 27,28 \\
45 & \ch{D- + H -> H- + D} & \parbox[t]{0.65\textwidth}{\raggedright $6.17390 \times 10^{-10} T^{0.41}$} & 27,28 \\
46 & \ch{D- + H -> HD + e-} & \parbox[t]{0.65\textwidth}{\raggedright $2.65340 \times 10^{-9}T^{-0.1}$} & 27,28 \\
\hline
\multicolumn{4}{l}{\textbf{Gas-phase metal chemistry}} \\
\hline
47 & \ch{O+ + H -> H+ + O} & \parbox[t]{0.65\textwidth}{\raggedright $7.5 \times 10^{-10}T^{-0.458} + 4.99 \times 10^{-11} T^{0.405}$} & 26 \\
48 & \ch{H+ + O -> O+ + H} & \parbox[t]{0.65\textwidth}{\raggedright $\left(4 \times 10^{-10} T^{0.00669} + 1.1 \times 10^{-11} T^{0.517}\right) \operatorname{exp}{\left(- \frac{227}{T} \right)}$} & 26 \\
49 & \ch{O+ + H2 -> H + OH+} & \parbox[t]{0.65\textwidth}{\raggedright $1.35 \times 10^{-9}$} & 10,29 \\
50 & \ch{OH+ + H2 -> H2O+ + H} & \parbox[t]{0.65\textwidth}{\raggedright $1.27 \times 10^{-9} (T/300)^{0.18}$} & 12,29 \\
51 & \ch{H2O+ + H2 -> H3O+ + H} & \parbox[t]{0.65\textwidth}{\raggedright $9.7 \times 10^{-10}$} & 12,29 \\
52 & \ch{H3+ + H2O -> H3O+ + H2} & \parbox[t]{0.65\textwidth}{\raggedright $5.9 \times 10^{-9}(T/300)^{-0.5}$} & 12,29 \\
53 & \ch{H2O+ + e- -> O + H2} & \parbox[t]{0.65\textwidth}{\raggedright $3.9 \times 10^{-8}(T/300)^{-0.5}$} & 12,29 \\
54 & \ch{H2O+ + e- -> OH + H} & \parbox[t]{0.65\textwidth}{\raggedright $8.6 \times 10^{-8}(T/300)^{-0.5}$} & 12,29 \\
55 & \ch{H2O+ + e- -> H + H + O} & \parbox[t]{0.65\textwidth}{\raggedright $3.05 \times 10^{-7}(T/300)^{-0.5}$} & 29 \\
56 & \ch{H3O+ + e- -> OH + H + H} & \parbox[t]{0.65\textwidth}{\raggedright $3.05 \times 10^{-7}(T/300)^{-0.5}$} & 29 \\
57 & \ch{H3O+ + e- -> O + H2 + H} & \parbox[t]{0.65\textwidth}{\raggedright $5.6 \times 10^{-9}(T/300)^{-0.5}$} & 29 \\
58 & \ch{H3O+ + e- -> H + H2O} & \parbox[t]{0.65\textwidth}{\raggedright $7.09 \times 10^{-8}(T/300)^{-0.5}$} & 29 \\
59 & \ch{H3O+ + e- -> OH + H2} & \parbox[t]{0.65\textwidth}{\raggedright $5.37 \times 10^{-8}(T/300)^{-0.5}$} & 29 \\
60 & \ch{O + H -> OH} & \parbox[t]{0.65\textwidth}{\raggedright $9.9 \times 10^{-19}(T/300)^{-0.38}$} & 29 \\
61 & \ch{O + H2 -> OH + H} & \parbox[t]{0.65\textwidth}{\raggedright $3.14 \times 10^{-13} (T/300)^{2.7} \operatorname{exp}{\left(- \frac{3150}{T} \right)}$} & 30,29,31 \\
62 & \ch{H2 + OH -> H2O + H} & \parbox[t]{0.65\textwidth}{\raggedright $3.6 \times 10^{-16} T^{1.52} \operatorname{exp}{\left(- \frac{1740}{T} \right)}$} & 32 \\
63 & \ch{OH + OH -> H2O + O} & \parbox[t]{0.65\textwidth}{\raggedright $1.65 \times 10^{-12} (T/300)^{1.14} \operatorname{exp}{\left(- \frac{50}{T} \right)}$} & 29 \\
64 & \ch{H2O + O -> OH + OH} & \parbox[t]{0.65\textwidth}{\raggedright $1.85 \times 10^{-11} (T/300)^{0.95} \operatorname{exp}{\left(- \frac{8571}{T} \right)}$} & 29 \\
65 & \ch{OH + H+ -> OH+ + H} & \parbox[t]{0.65\textwidth}{\raggedright $3.74511 \times 10^{-8} \hfill T < 10 \, \mathrm{K}$} & 10 \\
 &  & \parbox[t]{0.65\textwidth}{\raggedright $1.13529 \times 10^{-7} \left(T\right)^{0.5} + 1.55 \times 10^{-9} \hfill T > 10 \, \mathrm{K}$} &  \\
66 & \ch{H2O + H+ -> H + H2O+} & \parbox[t]{0.65\textwidth}{\raggedright $4.20246 \times 10^{-8} \hfill T < 10 \, \mathrm{K}$} & 29 \\
 &  & \parbox[t]{0.65\textwidth}{\raggedright $1.27305 \times 10^{-7} \left(T\right)^{0.5} + 1.767 \times 10^{-9} \hfill T > 10 \, \mathrm{K}$} &  \\
67 & \ch{OH + H -> O + H2} & \parbox[t]{0.65\textwidth}{\raggedright $8.10098 \times 10^{-21} T^{2.8} \operatorname{exp}{\left(- \frac{1950}{T} \right)}$} & 15,33,12 \\
68 & \ch{H2O + H -> H2 + OH} & \parbox[t]{0.65\textwidth}{\raggedright $1.908 \times 10^{-11} (T/300) \operatorname{exp}{\left(- \frac{9610}{T} \right)}$} & 12 \\
69 & \ch{H2O + H -> OH + H + H} & \parbox[t]{0.65\textwidth}{\raggedright $5.8 \times 10^{-9} \operatorname{exp}{\left(- \frac{52900}{T} \right)}$} & 12 \\
70 & \ch{H2O + H2 -> OH + H2 + H} & \parbox[t]{0.65\textwidth}{\raggedright $5.8 \times 10^{-9} \operatorname{exp}{\left(- \frac{52900}{T} \right)}$} & 12 \\
71 & \ch{O + O -> O2} & \parbox[t]{0.65\textwidth}{\raggedright $4.9 \times 10^{-20} (T/300)^{1.58}$} & 12 \\
72 & \ch{H2 + O2 -> OH + OH} & \parbox[t]{0.65\textwidth}{\raggedright $2.82 \times 10^{-11} \operatorname{exp}{\left(- \frac{24200}{T} \right)}$} & 12 \\
73 & \ch{O + OH -> H + O2} & \parbox[t]{0.65\textwidth}{\raggedright $3.69 \times 10^{-11} \operatorname{exp}{\left(- \frac{12.9}{T} \right)}(T/300)^{-0.27}$} & 29 \\
74 & \ch{O2 + H -> OH + O} & \parbox[t]{0.65\textwidth}{\raggedright $3.132 \times 10^{-10} \operatorname{exp}{\left(- \frac{8156}{T} \right)}$} & 12 \\
75 & \ch{O2 + H+ -> O2+ + H} & \parbox[t]{0.65\textwidth}{\raggedright $2 \times 10^{-9}$} & 12 \\
76 & \ch{O2+ + e- -> O + O} & \parbox[t]{0.65\textwidth}{\raggedright $1.95 \times 10^{-7}(T/300)^{-0.7}$} & 29 \\
77 & \ch{CH + O -> CO + H} & \parbox[t]{0.65\textwidth}{\raggedright $6.02 \times 10^{-11} (T/300)^{0.1} \operatorname{exp}{\left(\frac{4.5}{T} \right)}$} & 29 \\
78 & \ch{CH + O -> OH + C} & \parbox[t]{0.65\textwidth}{\raggedright $2.52 \times 10^{-11} \operatorname{exp}{\left(- \frac{2381}{T} \right)}$} & 29 \\
79 & \ch{CH2 + O -> CO + H2} & \parbox[t]{0.65\textwidth}{\raggedright $2.04 \times 10^{-10} \operatorname{exp}{\left(- \frac{270}{T} \right)} \hfill T > 290 \, \mathrm{K}$} & 29 \\
80 & \ch{CH2 + O -> CO + H + H} & \parbox[t]{0.65\textwidth}{\raggedright $3.4 \times 10^{-10} \operatorname{exp}{\left(- \frac{270}{T} \right)} \hfill T > 290 \, \mathrm{K}$} & 29 \\
81 & \ch{C + OH -> H + CO} & \parbox[t]{0.65\textwidth}{\raggedright $7.05076 \times 10^{-11} \hfill T < 10 \, \mathrm{K}$} & 10 \\
 &  & \parbox[t]{0.65\textwidth}{\raggedright $2.25 \times 10^{-11} \operatorname{exp}{\left(\frac{0.108}{T} \right)}(T/300)^{-0.339} \hfill T > 10 \, \mathrm{K}$} &  \\
82 & \ch{O2 + C -> CO + O} & \parbox[t]{0.65\textwidth}{\raggedright $4.7 \times 10^{-11}(T/300)^{-0.34} \hfill T < 1052 \, \mathrm{K}$} & 29 \\
 &  & \parbox[t]{0.65\textwidth}{\raggedright $2.48 \times 10^{-12} (T/300)^{1.54} \operatorname{exp}{\left(\frac{613}{T} \right)} \hfill T > 1052 \, \mathrm{K}$} &  \\
83 & \ch{O2 + C+ -> CO + O+} & \parbox[t]{0.65\textwidth}{\raggedright $4.53 \times 10^{-10}$} & 10 \\
84 & \ch{OH + CO -> CO2 + H} & \parbox[t]{0.65\textwidth}{\raggedright $2.81 \times 10^{-13} \operatorname{exp}{\left(- \frac{176}{T} \right)}$} & 34,29 \\
85 & \ch{C+ + e- -> C} & \parbox[t]{0.65\textwidth}{\raggedright $\begin{aligned}[t] &\Bigl(0.00033 \operatorname{exp}\Bigl(- \frac{214400}{T} \Bigr)  \\ 
 &+ 0.00083 \operatorname{exp}\Bigl(- \frac{120700}{T} \Bigr)  \\ 
 &+ 1.634 \times 10^{-6} \operatorname{exp}\Bigl(- \frac{15230}{T} \Bigr)  \\ 
 &+ 9.793 \times 10^{-9} \operatorname{exp}\Bigl(- \frac{73.8}{T} \Bigr)  \\ 
 &+ 6.346 \times 10^{-9} \operatorname{exp}\Bigl(- \frac{12.17}{T} \Bigr)\Bigr) / \Bigl(T^{1.5}\Bigr)  \\ 
 &+ \Bigl(2.995 \times 10^{-9} \Bigl(\sqrt{\Bigl(5.14668 \times 10^{-7} T \Bigr)}  \\ 
 &+ 1\Bigr)^{- 0.1597 \operatorname{exp}\Bigl(- \frac{49550}{T} \Bigr) - 1.7849} \Bigl(\sqrt{\Bigl(149.92503 T \Bigr)}  \\ 
 &+ 1\Bigr)^{0.1597 \operatorname{exp}\Bigl(- \frac{49550}{T} \Bigr) - 0.2151}\Bigr) / \Bigl(\sqrt{\Bigl(149.92503 T \Bigr)}\Bigr)\end{aligned}$} & 35,36 \\
86 & \ch{OH + C+ -> H + CO+} & \parbox[t]{0.65\textwidth}{\raggedright $1.37071 \times 10^{-8} \hfill T < 10 \, \mathrm{K}$} & 10 \\
 &  & \parbox[t]{0.65\textwidth}{\raggedright $4.15517 \times 10^{-8} \left(T\right)^{0.5} + 5.673 \times 10^{-10} \hfill T > 10 \, \mathrm{K}$} &  \\
87 & \ch{CO+ + H -> CO + H+} & \parbox[t]{0.65\textwidth}{\raggedright $7.5 \times 10^{-10}$} & 29 \\
88 & \ch{C + H -> CH} & \parbox[t]{0.65\textwidth}{\raggedright $1 \times 10^{-17}$} & 29 \\
89 & \ch{C + H2 -> CH + H} & \parbox[t]{0.65\textwidth}{\raggedright $6.64 \times 10^{-10} \operatorname{exp}{\left(- \frac{11700}{T} \right)}$} & 29 \\
90 & \ch{CH + H -> C + H2} & \parbox[t]{0.65\textwidth}{\raggedright $1.31 \times 10^{-10} \operatorname{exp}{\left(- \frac{80}{T} \right)}$} & 29 \\
91 & \ch{CH + H2 -> CH2 + H} & \parbox[t]{0.65\textwidth}{\raggedright $5.46 \times 10^{-10} \operatorname{exp}{\left(- \frac{1943}{T} \right)}$} & 29 \\
92 & \ch{CH2 + H -> CH + H2} & \parbox[t]{0.65\textwidth}{\raggedright $2.2 \times 10^{-10}$} & 10 \\
93 & \ch{CH2 + H2 -> CH3 + H} & \parbox[t]{0.65\textwidth}{\raggedright $5.18 \times 10^{-11} (T/300)^{0.17} \operatorname{exp}{\left(- \frac{6400}{T} \right)}$} & 29 \\
94 & \ch{CH3 + H -> CH2 + H2} & \parbox[t]{0.65\textwidth}{\raggedright $1 \times 10^{-10} \operatorname{exp}{\left(- \frac{7600}{T} \right)}$} & 29 \\
95 & \ch{CH3 + H2 -> CH4 + H} & \parbox[t]{0.65\textwidth}{\raggedright $6.86 \times 10^{-14} (T/300)^{2.74} \operatorname{exp}{\left(- \frac{4740}{T} \right)} \hfill T \leqslant 2\times 10^7 \, \mathrm{K}$} & 29 \\
96 & \ch{CH4 + H -> CH3 + H2} & \parbox[t]{0.65\textwidth}{\raggedright $5.82 \times 10^{-13} (T/300)^{3} \operatorname{exp}{\left(- \frac{4045}{T} \right)} \hfill T \leqslant 3.6 \times 10^6 \, \mathrm{K}$} & 29 \\
97 & \ch{C + H2 -> CH2} & \parbox[t]{0.65\textwidth}{\raggedright $1 \times 10^{-17}$} & 29 \\
98 & \ch{He+ + CO -> C+ + O + He} & \parbox[t]{0.65\textwidth}{\raggedright $1.4 \times 10^{-9}(T/300)^{-0.5}$} & 15 \\
99 & \ch{CH + He+ -> C+ + He + H} & \parbox[t]{0.65\textwidth}{\raggedright $1.90525 \times 10^{-8}T^{-0.5}$} & 15 \\
100 & \ch{He+ + OH -> O+ + He + H} & \parbox[t]{0.65\textwidth}{\raggedright $6.13035 \times 10^{-8}T^{-0.5} + 8.37 \times 10^{-10}$} & 23 \\
101 & \ch{C+ + H2 -> CH2+} & \parbox[t]{0.65\textwidth}{\raggedright $2.31 \times 10^{-16} \operatorname{exp}{\left(- \frac{23}{T} \right)}(T/300)^{-1.3}$} & 23 \\
102 & \ch{C+ + H2 + e- -> C + H + H} & \parbox[t]{0.65\textwidth}{\raggedright $9.9 \times 10^{-17} \operatorname{exp}{\left(- \frac{23}{T} \right)}(T/300)^{-1.3}$} & 23 \\
103 & \ch{CO + H -> C + OH} & \parbox[t]{0.65\textwidth}{\raggedright $1.1 \times 10^{-10} (T/300)^{0.5} \operatorname{exp}{\left(- \frac{77700}{T} \right)}$} & 15 \\
104 & \ch{HCO+ + e- -> CO + H} & \parbox[t]{0.65\textwidth}{\raggedright $9.752 \times 10^{-6}T^{-0.64}$} & 37,23 \\
105 & \ch{HCO+ + e- -> OH + C} & \parbox[t]{0.65\textwidth}{\raggedright $8.48 \times 10^{-7}T^{-0.64}$} & 37,15 \\
106 & \ch{H3+ + CO -> HCO+ + H2} & \parbox[t]{0.65\textwidth}{\raggedright $1.36 \times 10^{-9} \operatorname{exp}{\left(\frac{3.4}{T} \right)}(T/300)^{-0.14}$} & 29 \\
107 & \ch{H2 + CO+ -> HCO+ + H} & \parbox[t]{0.65\textwidth}{\raggedright $7.5 \times 10^{-10}$} & 29 \\
108 & \ch{C + HCO+ -> CO + CH+} & \parbox[t]{0.65\textwidth}{\raggedright $1.1 \times 10^{-9}$} & 38 \\
109 & \ch{H + CO2 -> OH + CO} & \parbox[t]{0.65\textwidth}{\raggedright $2.54 \times 10^{-10} \operatorname{exp}{\left(- \frac{13300}{T} \right)}$} & 39 \\
110 & \ch{O + CO2 -> CO + O2} & \parbox[t]{0.65\textwidth}{\raggedright $2.8 \times 10^{-11} \operatorname{exp}{\left(- \frac{26500}{T} \right)}$} & 39 \\
111 & \ch{H2 + CO2 -> H2O + CO} & \parbox[t]{0.65\textwidth}{\raggedright $2.94 \times 10^{-14} (T/300)^{0.5} \operatorname{exp}{\left(- \frac{7550}{T} \right)}$} & 39 \\
112 & \ch{CO + O2 -> CO2 + O} & \parbox[t]{0.65\textwidth}{\raggedright $4.2 \times 10^{-12} \operatorname{exp}{\left(- \frac{24000}{T} \right)}$} & 39 \\
113 & \ch{C + e- -> C+ + e- + e-} & \parbox[t]{0.65\textwidth}{\raggedright $1.25480 \times 10^{-7} \left(\frac{1}{T_{\rm eV}}\right)^{0.25} \operatorname{exp}{\left(- \frac{11.26}{T_{\rm eV}} \right)}\left[0.193 + \frac{11.26}{T_{\rm eV}}\right]^{-1}$} & 40 \\
114 & \ch{O + e- -> O+ + e- + e-} & \parbox[t]{0.65\textwidth}{\raggedright $8.71962 \times 10^{-8} \left(\frac{1}{T_{\rm eV}}\right)^{0.34} \operatorname{exp}{\left(- \frac{13.6}{T_{\rm eV}} \right)}\left[0.073 + \frac{13.6}{T_{\rm eV}}\right]^{-1}$} & 40 \\
115 & \ch{C + H+ -> C+ + H} & \parbox[t]{0.65\textwidth}{\raggedright $3.9 \times 10^{-16} T^{0.213}$} & 12 \\
116 & \ch{C+ + H -> C + H+} & \parbox[t]{0.65\textwidth}{\raggedright $8.78827 \times 10^{-22} T^{1.96} \operatorname{exp}{\left(- \frac{170000}{T} \right)}$} & 12 \\
117 & \ch{O2 + C+ -> CO+ + O} & \parbox[t]{0.65\textwidth}{\raggedright $3.42 \times 10^{-10}$} & 29 \\
118 & \ch{C + O2+ -> O + CO+} & \parbox[t]{0.65\textwidth}{\raggedright $5.2 \times 10^{-11}$} & 12,10 \\
119 & \ch{C + O2+ -> O2 + C+} & \parbox[t]{0.65\textwidth}{\raggedright $5.2 \times 10^{-11}$} & 12,10 \\
120 & \ch{C+ + O -> CO+} & \parbox[t]{0.65\textwidth}{\raggedright $2.5 \times 10^{-18} \hfill T < 300 \, \mathrm{K}$} & 29 \\
 &  & \parbox[t]{0.65\textwidth}{\raggedright $3.14 \times 10^{-18} \operatorname{exp}{\left(- \frac{68}{T} \right)}(T/300)^{-0.15} \hfill T > 300 \, \mathrm{K}$} &  \\
121 & \ch{OH+ + e- -> O + H} & \parbox[t]{0.65\textwidth}{\raggedright $6.3 \times 10^{-9}(T/300)^{-0.48}$} & \dbDefault{} \\
122 & \ch{CO+ + e- -> C + O} & \parbox[t]{0.65\textwidth}{\raggedright $2.75 \times 10^{-7}(T/300)^{-0.55}$} & 26 \\
123 & \ch{O+ + e- -> O} & \parbox[t]{0.65\textwidth}{\raggedright $1.3 \times 10^{-10}T^{-0.64} \hfill T \leqslant 400 \, \mathrm{K}$} & 12 \\
&  & \parbox[t]{0.65\textwidth}{\raggedright $0.00074 \left(0.062 \operatorname{exp}{\left(- \frac{145000}{T} \right)} + 1\right) \operatorname{exp}{\left(- \frac{175000}{T} \right)}T^{-1.5} +  \hfill T > 400 \, \mathrm{K}$} &  \\
&  & \parbox[t]{0.65\textwidth}{\raggedright $\frac{1.41 \times 10^{-10}}{T^{0.66}}$} &  \\
124 & \ch{CH+ + e- -> C + H} & \parbox[t]{0.65\textwidth}{\raggedright $7 \times 10^{-8}(T/300)^{-0.5}$} & 12 \\
125 & \ch{CH2+ + e- -> CH + H} & \parbox[t]{0.65\textwidth}{\raggedright $1.6 \times 10^{-7}(T/300)^{-0.6}$} & 12 \\
126 & \ch{CH2+ + e- -> C + H2} & \parbox[t]{0.65\textwidth}{\raggedright $7.68 \times 10^{-8}(T/300)^{-0.6}$} & \dbDefault{} \\
127 & \ch{H+ + CH3 -> CH3+ + H} & \parbox[t]{0.65\textwidth}{\raggedright $3.4 \times 10^{-9}$} & 29 \\
128 & \ch{CH3+ + O -> HCO+ + H2} & \parbox[t]{0.65\textwidth}{\raggedright $4 \times 10^{-10}$} & 29 \\
129 & \ch{CH2 + HCO+ -> CH3+ + CO} & \parbox[t]{0.65\textwidth}{\raggedright $8.6 \times 10^{-10}$} & 29 \\
130 & \ch{CH2 + H2O+ -> CH3+ + OH} & \parbox[t]{0.65\textwidth}{\raggedright $4.7 \times 10^{-10}$} & 29 \\
131 & \ch{CH2 + H3O+ -> CH3+ + H2O} & \parbox[t]{0.65\textwidth}{\raggedright $9.4 \times 10^{-10}$} & 29 \\
132 & \ch{H3+ + CH2 -> CH3+ + H2} & \parbox[t]{0.65\textwidth}{\raggedright $1.7 \times 10^{-9}$} & 29 \\
133 & \ch{H2 + CH2+ -> CH3+ + H} & \parbox[t]{0.65\textwidth}{\raggedright $1.6 \times 10^{-9}$} & 29 \\
134 & \ch{H + CH3+ -> CH2+ + H2} & \parbox[t]{0.65\textwidth}{\raggedright $7 \times 10^{-10} \operatorname{exp}{\left(- \frac{10560}{T} \right)}$} & 29 \\
135 & \ch{CH3+ + e- -> CH2 + H} & \parbox[t]{0.65\textwidth}{\raggedright $1.4 \times 10^{-7}(T/300)^{-0.5}$} & 29 \\
136 & \ch{CH3+ + e- -> CH + H2} & \parbox[t]{0.65\textwidth}{\raggedright $4.9 \times 10^{-8}(T/300)^{-0.5}$} & 29 \\
137 & \ch{CH2+ + e- -> C + H + H} & \parbox[t]{0.65\textwidth}{\raggedright $4.03 \times 10^{-7}(T/300)^{-0.6}$} & 29 \\
138 & \ch{CH3+ + e- -> CH + H + H} & \parbox[t]{0.65\textwidth}{\raggedright $5.6 \times 10^{-8}(T/300)^{-0.5}$} & 29 \\
139 & \ch{CH3+ + e- -> C + H2 + H} & \parbox[t]{0.65\textwidth}{\raggedright $1.05 \times 10^{-7}(T/300)^{-0.5}$} & 29 \\
140 & \ch{H2+ + H2 -> H3+ + H} & \parbox[t]{0.65\textwidth}{\raggedright $2.24 \times 10^{-9} (T/300)^{0.042} \operatorname{exp}{\left(- 2.14592 \times 10^{-5} T \right)}$} & 12 \\
141 & \ch{H3+ + H -> H2+ + H2} & \parbox[t]{0.65\textwidth}{\raggedright $7.7 \times 10^{-9} \operatorname{exp}{\left(- \frac{17560}{T} \right)}$} & 29 \\
142 & \ch{C + H2+ -> CH+ + H} & \parbox[t]{0.65\textwidth}{\raggedright $2.4 \times 10^{-9}$} & 29 \\
143 & \ch{C + H3+ -> CH+ + H2} & \parbox[t]{0.65\textwidth}{\raggedright $\begin{aligned}[t] &6.93 \times 10^{-10} (T/300)^{0.0834}  \\ 
 &+ \Bigl(9.93 \times 10^{-5} \operatorname{exp}\Bigl(- \frac{8080}{T} \Bigr)  \\ 
 &+ 7.48 \times 10^{-8} \operatorname{exp}\Bigl(- \frac{41.9}{T} \Bigr)  \\ 
 &- 7.24 \times 10^{-9} \operatorname{exp}\Bigl(- \frac{8.01}{T} \Bigr)  \\ 
 &- 9.07 \times 10^{-10} \operatorname{exp}\Bigl(- \frac{1.92}{T} \Bigr)\Bigr) / \Bigl(T^{1.5}\Bigr)\end{aligned}$} & 41 \\
144 & \ch{C + H3+ -> CH2+ + H} & \parbox[t]{0.65\textwidth}{\raggedright $\begin{aligned}[t] &\frac{3.35 \times 10^{-10}}{(T/300)^{0.189}}  \\ 
 &+ \Bigl(- 0.00019 \operatorname{exp}\Bigl(- \frac{16200}{T} \Bigr)  \\ 
 &+ 7.46 \times 10^{-8} \operatorname{exp}\Bigl(- \frac{19}{T} \Bigr)  \\ 
 &+ 2.73 \times 10^{-8} \operatorname{exp}\Bigl(- \frac{6.49}{T} \Bigr)  \\ 
 &+ 5.58 \times 10^{-9} \operatorname{exp}\Bigl(- \frac{1.3}{T} \Bigr)\Bigr) / \Bigl(T^{1.5}\Bigr)\end{aligned}$} & 41 \\
145 & \ch{O + H3+ -> H2 + OH+} & \parbox[t]{0.65\textwidth}{\raggedright $7.98 \times 10^{-10} \operatorname{exp}{\left(- \frac{1.41}{T} \right)}(T/300)^{-0.156}$} & 10 \\
146 & \ch{O + H3+ -> H + H2O+} & \parbox[t]{0.65\textwidth}{\raggedright $3.42 \times 10^{-10} \operatorname{exp}{\left(- \frac{1.41}{T} \right)}(T/300)^{-0.156}$} & 10 \\
147 & \ch{OH + H3+ -> H2 + H2O+} & \parbox[t]{0.65\textwidth}{\raggedright $2.27703 \times 10^{-8} \hfill T < 10 \, \mathrm{K}$} & 10 \\
 &  & \parbox[t]{0.65\textwidth}{\raggedright $6.90258 \times 10^{-8} \left(T\right)^{0.5} + 9.424 \times 10^{-10} \hfill T > 10 \, \mathrm{K}$} &  \\
148 & \ch{H3+ + e- -> H2 + H} & \parbox[t]{0.65\textwidth}{\raggedright $2.34 \times 10^{-8}(T/300)^{-0.52}$} & 10 \\
149 & \ch{H3+ + e- -> H + H + H} & \parbox[t]{0.65\textwidth}{\raggedright $4.36 \times 10^{-8}(T/300)^{-0.52}$} & 10 \\
\hline
\multicolumn{4}{l}{\textbf{Photoreactions}} \\
\hline
150 & \ch{H -> H+ + e-} & \parbox[t]{0.65\textwidth}{\raggedright $\mathcal{I}_{\rm H}$} & \dbDefault{} \\
151 & \ch{H2 -> H + H} & \parbox[t]{0.65\textwidth}{\raggedright $\mathcal{D}_{\rm H_2}$} & \dbDefault{} \\
152 & \ch{HD -> H + D} & \parbox[t]{0.65\textwidth}{\raggedright $\mathcal{D}_{\rm HD}$} & \dbDefault{} \\
153 & \ch{H- -> H + e-} & \parbox[t]{0.65\textwidth}{\raggedright $7.1 \times 10^{-7} \chi_{0} \operatorname{exp}{\left(- 0.5 A_{\rm V} \right)}$} & 12 \\
154 & \ch{H2+ -> H + H+} & \parbox[t]{0.65\textwidth}{\raggedright $1.1 \times 10^{-9} \chi_{0} \operatorname{exp}{\left(- 1.9 A_{\rm V} \right)}$} & 29 \\
155 & \ch{H3+ -> H2 + H+} & \parbox[t]{0.65\textwidth}{\raggedright $4.9 \times 10^{-13} \chi_{0} \operatorname{exp}{\left(- 1.8 A_{\rm V} \right)}$} & 29 \\
156 & \ch{H3+ -> H2+ + H} & \parbox[t]{0.65\textwidth}{\raggedright $4.9 \times 10^{-13} \chi_{0} \operatorname{exp}{\left(- 2.3 A_{\rm V} \right)}$} & 29 \\
157 & \ch{C -> C+ + e-} & \parbox[t]{0.65\textwidth}{\raggedright $\mathcal{I}_{\rm CI}$} & 29 \\
158 & \ch{CH -> C + H} & \parbox[t]{0.65\textwidth}{\raggedright $8.7 \times 10^{-10} \chi_{0} \operatorname{exp}{\left(- 1.2 A_{\rm V} \right)}$} & 29 \\
159 & \ch{CH -> CH+ + e-} & \parbox[t]{0.65\textwidth}{\raggedright $7.7 \times 10^{-10} \chi_{0} \operatorname{exp}{\left(- 2.8 A_{\rm V} \right)}$} & 29 \\
160 & \ch{CH+ -> C + H+} & \parbox[t]{0.65\textwidth}{\raggedright $2.6 \times 10^{-10} \chi_{0} \operatorname{exp}{\left(- 2.5 A_{\rm V} \right)}$} & 29 \\
161 & \ch{CH2 -> CH + H} & \parbox[t]{0.65\textwidth}{\raggedright $7.1 \times 10^{-10} \chi_{0} \operatorname{exp}{\left(- 1.7 A_{\rm V} \right)}$} & 29 \\
162 & \ch{CH2 -> CH2+ + e-} & \parbox[t]{0.65\textwidth}{\raggedright $5.9 \times 10^{-10} \chi_{0} \operatorname{exp}{\left(- 2.3 A_{\rm V} \right)}$} & 29 \\
163 & \ch{CH2+ -> CH+ + H} & \parbox[t]{0.65\textwidth}{\raggedright $4.6 \times 10^{-10} \chi_{0} \operatorname{exp}{\left(- 1.7 A_{\rm V} \right)}$} & 29 \\
164 & \ch{CH3+ -> CH2+ + H} & \parbox[t]{0.65\textwidth}{\raggedright $1 \times 10^{-9} \chi_{0} \operatorname{exp}{\left(- 1.7 A_{\rm V} \right)}$} & 29 \\
165 & \ch{CH3+ -> CH+ + H2} & \parbox[t]{0.65\textwidth}{\raggedright $1 \times 10^{-9} \chi_{0} \operatorname{exp}{\left(- 1.7 A_{\rm V} \right)}$} & 29 \\
166 & \ch{OH -> O + H} & \parbox[t]{0.65\textwidth}{\raggedright $3.7 \times 10^{-10} \chi_{0} \operatorname{exp}{\left(- 1.7 A_{\rm V} \right)}$} & 29 \\
167 & \ch{OH -> OH+ + e-} & \parbox[t]{0.65\textwidth}{\raggedright $1.6 \times 10^{-12} \chi_{0} \operatorname{exp}{\left(- 3.1 A_{\rm V} \right)}$} & 29 \\
168 & \ch{OH+ -> O + H+} & \parbox[t]{0.65\textwidth}{\raggedright $1 \times 10^{-12} \chi_{0} \operatorname{exp}{\left(- 1.8 A_{\rm V} \right)}$} & 29 \\
169 & \ch{H2O -> OH + H} & \parbox[t]{0.65\textwidth}{\raggedright $6 \times 10^{-10} \chi_{0} \operatorname{exp}{\left(- 1.7 A_{\rm V} \right)}$} & 29 \\
170 & \ch{H2O -> H2O+ + e-} & \parbox[t]{0.65\textwidth}{\raggedright $3.2 \times 10^{-11} \chi_{0} \operatorname{exp}{\left(- 3.9 A_{\rm V} \right)}$} & 29 \\
171 & \ch{O2 -> O2+ + e-} & \parbox[t]{0.65\textwidth}{\raggedright $5.6 \times 10^{-11} \chi_{0} \operatorname{exp}{\left(- 3.7 A_{\rm V} \right)}$} & 29 \\
172 & \ch{O2 -> O + O} & \parbox[t]{0.65\textwidth}{\raggedright $7 \times 10^{-10} \chi_{0} \operatorname{exp}{\left(- 1.8 A_{\rm V} \right)}$} & 29 \\
173 & \ch{CO2 -> CO + O} & \parbox[t]{0.65\textwidth}{\raggedright $9.2 \times 10^{-10} \chi_{0} \operatorname{exp}{\left(- 3.4 A_{\rm V} \right)}$} & 29 \\
174 & \ch{HCO+ -> CO+ + H} & \parbox[t]{0.65\textwidth}{\raggedright $5.39 \times 10^{-12} \chi_{0} \operatorname{exp}{\left(- 3.67 A_{\rm V} \right)}$} & 29 \\
175 & \ch{CO -> C + O} & \parbox[t]{0.65\textwidth}{\raggedright $\mathcal{D}_{\rm CO}$} & \dbDefault{} \\
176 & \ch{H2O+ -> H2+ + O} & \parbox[t]{0.65\textwidth}{\raggedright $5 \times 10^{-11} \chi_{0} f_{\rm sh, H_2O^+}$} & 12 \\
177 & \ch{H2O+ -> H+ + OH} & \parbox[t]{0.65\textwidth}{\raggedright $5 \times 10^{-11} \chi_{0} f_{\rm sh, H_2O^+}$} & 12 \\
178 & \ch{H2O+ -> O+ + H2} & \parbox[t]{0.65\textwidth}{\raggedright $5 \times 10^{-11} \chi_{0} f_{\rm sh, H_2O^+}$} & 12 \\
179 & \ch{H2O+ -> OH+ + H} & \parbox[t]{0.65\textwidth}{\raggedright $1.5 \times 10^{-10} \chi_{0} f_{\rm sh, H_2O^+}$} & 12 \\
\hline
\multicolumn{4}{l}{\textbf{Grain recombination reactions}} \\
180 & \ch{H+ + e- + gr -> H + gr} & \parbox[t]{0.65\textwidth}{\raggedright $\begin{aligned}[t] &\frac{1.225 \times 10^{-13} Z_{\rm d} n_{\rm Hnuclei}}{n_{\rm e} \Bigl(8.074 \times 10^{-6} \psi^{1.378} \Bigl(508.7 T^{0.01586} \psi^{- 1.102 \times 10^{-5} \ln{T} - 0.4723} + 1\Bigr) + 1\Bigr)}\end{aligned}$} & 42,26 \\
181 & \ch{He+ + e- + gr -> He + gr} & \parbox[t]{0.65\textwidth}{\raggedright $\begin{aligned}[t] &\frac{5.572 \times 10^{-14} Z_{\rm d} n_{\rm Hnuclei}}{n_{\rm e} \Bigl(3.185 \times 10^{-7} \psi^{1.512} \Bigl(5115 T^{3.903 \times 10^{-7}} \psi^{- 5.494 \times 10^{-7} \ln{T} - 0.4956} + 1\Bigr) + 1\Bigr)}\end{aligned}$} & 42,26 \\
182 & \ch{O+ + e- + gr -> O + gr} & \parbox[t]{0.65\textwidth}{\raggedright $\begin{aligned}[t] &\frac{3.0625 \times 10^{-14} Z_{\rm d} n_{\rm Hnuclei}}{n_{\rm e} \Bigl(8.074 \times 10^{-6} \psi^{1.378} \Bigl(508.7 T^{0.01586} \psi^{- 1.102 \times 10^{-5} \ln{T} - 0.4723} + 1\Bigr) + 1\Bigr)}\end{aligned}$} & 42,26 \\
183 & \ch{C+ + e- + gr -> C + gr} & \parbox[t]{0.65\textwidth}{\raggedright $\frac{4.558 \times 10^{-13} Z_{\rm d} n_{\rm Hnuclei}}{n_{\rm e} \left(0.00608 \psi^{1.128} \left(433.1 T^{0.04845} \psi^{- 0.00013 \ln{T} - 0.812} + 1\right) + 1\right)}$} & 42,26 \\
\hline
\multicolumn{4}{l}{\textbf{Cosmic ray reactions}} \\
184 & \ch{H ->[CR] H+ + e-} & \parbox[t]{0.65\textwidth}{\raggedright $\xi_{\rm H_2} \left(3.731 n_{\rm H_2}/n_{\rm Hnuclei} + 0.5\right)$} & 43 \\
185 & \ch{D ->[CR] D+ + e-} & \parbox[t]{0.65\textwidth}{\raggedright $\xi_{\rm H_2} \left(3.731 n_{\rm H_2}/n_{\rm Hnuclei} + 0.5\right)$} & 43 \\
186 & \ch{H- ->[CR] H + e-} & \parbox[t]{0.65\textwidth}{\raggedright $924 \xi_{\rm H_2} n_{\rm H_2}/n_{\rm Hnuclei}$} & 43 \\
187 & \ch{H2 ->[CR] H2+ + e-} & \parbox[t]{0.65\textwidth}{\raggedright $0.976 \xi_{\rm H_2}$} & 43 \\
188 & \ch{H2 ->[CR] H + H} & \parbox[t]{0.65\textwidth}{\raggedright $0.7 \xi_{\rm H_2}$} & 43 \\
189 & \ch{H2 ->[CR] H+ + H-} & \parbox[t]{0.65\textwidth}{\raggedright $0.00031 \xi_{\rm H_2}$} & 43 \\
190 & \ch{HD ->[CR] H + D} & \parbox[t]{0.65\textwidth}{\raggedright $0.7 \xi_{\rm H_2}$} & 44,45 \\
191 & \ch{He ->[CR] He+ + e-} & \parbox[t]{0.65\textwidth}{\raggedright $0.55 \xi_{\rm H_2}$} & 43 \\
192 & \ch{CO ->[CR] C + O} & \parbox[t]{0.65\textwidth}{\raggedright $78.175 \xi_{\rm H_2} n_{\rm H_2}/n_{\rm Hnuclei}$} & 43 \\
193 & \ch{CO ->[CR] CO+ + e-} & \parbox[t]{0.65\textwidth}{\raggedright $11.3125 \xi_{\rm H_2}$} & 43 \\
194 & \ch{C ->[CR] C+ + e-} & \parbox[t]{0.65\textwidth}{\raggedright $\xi_{\rm H_2} \left(236.9 n_{\rm H_2}/n_{\rm Hnuclei} + 1.92\right)$} & 43 \\
195 & \ch{O ->[CR] O+ + e-} & \parbox[t]{0.65\textwidth}{\raggedright $\xi_{\rm H_2} \left(2.4355 n_{\rm H_2}/n_{\rm Hnuclei} + 2.84\right)$} & 43 \\
196 & \ch{CH ->[CR] C + H} & \parbox[t]{0.65\textwidth}{\raggedright $450 \xi_{\rm H_2} n_{\rm H_2}/n_{\rm Hnuclei}$} & 43 \\
197 & \ch{O2 ->[CR] O + O} & \parbox[t]{0.65\textwidth}{\raggedright $310 \xi_{\rm H_2} n_{\rm H_2}/n_{\rm Hnuclei}$} & 43 \\
198 & \ch{O2 ->[CR] O2+ + e-} & \parbox[t]{0.65\textwidth}{\raggedright $28 \xi_{\rm H_2} n_{\rm H_2}/n_{\rm Hnuclei}$} & 43 \\
199 & \ch{OH ->[CR] O + H} & \parbox[t]{0.65\textwidth}{\raggedright $267 \xi_{\rm H_2} n_{\rm H_2}/n_{\rm Hnuclei}$} & 43 \\
200 & \ch{CH2 ->[CR] CH + H} & \parbox[t]{0.65\textwidth}{\raggedright $139 \xi_{\rm H_2} n_{\rm H_2}/n_{\rm Hnuclei}$} & 43 \\
201 & \ch{CH3 ->[CR] CH3+ + e-} & \parbox[t]{0.65\textwidth}{\raggedright $104 \xi_{\rm H_2} n_{\rm H_2}/n_{\rm Hnuclei}$} & 43 \\
202 & \ch{H2O ->[CR] OH + H} & \parbox[t]{0.65\textwidth}{\raggedright $468 \xi_{\rm H_2} n_{\rm H_2}/n_{\rm Hnuclei}$} & 43 \\
203 & \ch{H2O ->[CR] H2O+ + e-} & \parbox[t]{0.65\textwidth}{\raggedright $21.5 \xi_{\rm H_2} n_{\rm H_2}/n_{\rm Hnuclei}$} & 43 \\
204 & \ch{CO2 ->[CR] CO + O} & \parbox[t]{0.65\textwidth}{\raggedright $629 \xi_{\rm H_2} n_{\rm H_2}/n_{\rm Hnuclei}$} & 43 \\
205 & \ch{H2 ->[CR] H + H+ + e-} & \parbox[t]{0.65\textwidth}{\raggedright $0.023 \xi_{\rm H_2}$} & 43 \\
\end{longtable}

\textbf{Notes:} $T$ is the same as $T_{\rm gas}$ in the main text. $\chi_0$ indicates the normalized free-space unattenuated FUV (LW+PE bands) radiation field. The units for the rate coefficients are $\rm cm^{3(n_{\rm react}-1)}s^{-1}$, where $\rm n_{\rm react}$ is the number of (gas-phase) reactants in a given reaction. $T_{\rm eV} =  k_{\rm B} T$ where $k_{\rm B}$ is the Boltzmann constant measured in \rm{eV}. $k_{\rm H_2,H}, k_{\rm H_2,H_2}$ are the rate coefficients for collisional dissociation of $\Htwo$ by H and $\Htwo$, respectively, based on \citet{Glover_MacLow_2007}, corrected for high $T$ by \citet{Hunter_2023}. $f_{\rm sh, H_2O^+}$ is the shielding function for the photodissociation of $\rm H_2O^+$ following \citet{Glover_2010}. $\psi = G_0 \sqrt{T}/n_e$ is the grain charging parameter \citep{Weingartner_2001c} where $G_0 = 1.69 \chi_0$. $\xi_{\rm H_2} = 2 \xi_{\rm H}$ is the primary+secondary CR-ionization rate of $\Htwo$ molecules. \textbf{References:} (1) \citet{Janev1987}; (2) \citet{Abel1997}; (3) \citet{Cen_1992}; (4) \citet{AldrovandiPequignot1973}; (5) \citet{Verner1996}; (6) \citet{deJong1972}; (7) \citet{Kreckel2010}; (8) \citet{RamakerPeek1976}; (9) \citet{Karpas1979}; (10) \citet{Wakelam2010}; (11) \citet{Savin2004}; (12) \citet{Glover_2010}; (13) \citet{Capitelli2007}; (14) \citet{MitchellDeveau1983}; (15) \citet{Hunter_2023}; (16) \citet{Stenrup2009}; (17) \citet{Poulaert1978}; (18) \citet{Coppola2011}; (19) \citet{DalgarnoLepp1987}; (20) \citet{Forrey2013}; (21) \citet{Glover_Abel_2008}; (22) \citet{Yoshida2006}; (23) \citet{Gong_2017}; (24) \citet{GalliPalla2002}; (25) \citet{Gay2011}; (26) \citet{Glover_Jappsen_2007}; (27) \citet{Galli_1998}; (28) \citet{StancilLeppDalgarno1998}; (29) \citet{umist24}; (30) \citet{NatarajanRoth1987}; (31) \citet{Rollig_2007}; (32) \citet{Baulch2005}; (33) \citet{umist12}; (34) \citet{Frost1993}; (35) \citet{Badnell2003}; (36) \citet{Badnell2006}; (37) \citet{Geppert2005}; (38) \citet{PrasadHuntress1980}; (39) \citet{Omukai_2000}; (40) \citet{Voronov1997}; (41) \citet{Vissapragada2016}; (42) \citet{Weingartner_2001c}; (43) \citet{Padovani2024}; (44) \citet{Walmsley2004}; (45) \citet{Cazaux2009}; (46) \citet{GloverSavin2009}

%

%

%


\bibliography{references,references_clustersims,TableReferences}{}
\bibliographystyle{aasjournalv7}



\end{document}